\documentclass[12pt]{article}
\usepackage{axodraw}
\usepackage{a4wide,epsfig,psfrag,amsmath,amssymb,cite,scalefnt,color}

\newcommand{\gsim}{\;\rlap{\lower 3.5 pt \hbox{$\mathchar \sim$}} \raise 1pt
 \hbox {$>$}\;}
\newcommand{\lsim}{\;\rlap{\lower 3.5 pt \hbox{$\mathchar \sim$}} \raise 1pt
 \hbox {$<$}\;}
\renewcommand{\arraystretch}{1.2}

\begin{document}


\title{\vskip-3cm{\baselineskip14pt
    \begin{flushleft}
      \normalsize SFB/CPP-11-43\\
      \normalsize TTP11-22
  \end{flushleft}}
  \vskip1.5cm
  Gluino Pair Production at the LHC: The Threshold
}
\author{\small Matthias R. Kauth, Johann H. K\"uhn, Peter Marquard and Matthias Steinhauser\\[1em]
  {\small\it Institut f{\"u}r Theoretische Teilchenphysik
  }\\
  {\small\it Karlsruhe Institute of Technology (KIT)}\\
  {\small\it 76128 Karlsruhe, Germany}}

\date{}

\maketitle

\thispagestyle{empty}

\begin{abstract}

The next-to-leading order analysis of the cross section of hadronic
gluino pair production close to threshold is presented. Within the
framework of non-relativistic QCD a significant enhancement compared to
fixed order perturbation theory is observed which originates from the
characteristic remnant of the $1S$ peak below the nominal pair
threshold. This enhancement is similar to the corresponding one for top
production. However, as a consequence of the larger colour factor of the
QCD potential the effect is significantly enhanced. The analysis
includes all colour configurations of $S$-wave gluino pairs,
i.e. singlet, symmetric and antisymmetric octet, decuplet and
twenty-seven representation. Matching coefficients involving real and
virtual radiation are separately evaluated for all colour and spin
configurations and initial states. We concentrate on the case of gluino
decay rates comparable to the gluino binding energy. The
non-relativistic dynamics of the gluino pair is solved by calculating
the Green's function in NLO. Numerical results for the Large Hadron
Collider at $\sqrt{s}=14\,\mbox{TeV}$ and $7\,\mbox{TeV}$ are presented
for various characteristic scenarios.

\huge
\color{blue}
\vspace{2cm}
\begin{center}
\end{center}
\color{black}
\normalsize

\noindent

\end{abstract}

\thispagestyle{empty}


\newpage

\section{\label{sec:intro}Introduction}
The search for new particles, predicted in supersymmetric models, is one
of the important tasks of the experiments at the Large Hadron Collider
(LHC). The detailed determination of particle masses and couplings will
be crucial for the discrimination between various manifestations of
supersymmetry (SUSY) and alternative models, even more so if one wants
to distinguish between the different variants of supersymmetric models,
to identify the origin of breaking of supersymmetry and to measure the
model parameters.  One of the SUSY signals will be events with missing
energy or missing transverse momentum, resulting from cascade decays of
squarks and gluinos into the lightest supersymmetric particle (LSP)
which escapes detection. The existence of squarks, gluinos and the LSP
is definitely a key prediction of supersymmetry.
\par
Depending on details of the models, in particular the masses of squarks
and the LSP, gluinos with masses up to 3 TeV \cite{Baer:2009dn} could be
detected. The search strategy is quite different if gluinos are heavier
than squarks or if they are lighter. In the former case ($m_{\tilde g} >
m_{\tilde q}$) the two-body decay into a squark and an anti-quark (or its
charge conjugate) dominates, with the subsequent decay of the squark
into a quark plus a chargino or neutralino. In the latter case
($m_{\tilde g} < m_{\tilde q}$) the two-body channel is kinematically
forbidden and the now dominant three-body decay into quark, anti-quark
and neutralino or chargino, mediated by the virtual squark, leads to a
small decay rate. In the limit of extremely heavy squarks, a scenario
denoted Split SUSY, the gluino is quasistable and hadrons composed of
gluinos and gluons or quarks may travel macroscopic distances. The
search strategies will be markedly different in the various cases.
\par 
The importance of squark and gluino searches has motivated a series
of detailed studies of hadroproduction cross sections for squarks and
gluinos. The lowest order has been evaluated long time ago
\cite{Harrison:1982yi,Haber:1984rc,Dawson:1983fw}. Subsequently the
next-to-leading order (NLO) SUSY-QCD corrections were calculated
\cite{Beenakker:1996ch,Beenakker:1996ed,Beenakker:1997ut}, more recently
the effect of soft-gluon resummation \cite{Kulesza:2008jb,Langenfeld:2009eg,
Kulesza:2009kq,Beenakker:2009ha,Beenakker:2010nq,Beneke:2010da,Beenakker:2011fu} 
was included. The present paper will be concerned with gluino-pair
production close to threshold, which exhibits a number of peculiar
features.
\par
As a consequence of their colour-octet representation the production
cross section of gluinos is large and perturbative corrections are
particularly important. Furthermore, the threshold region is strongly
affected by final state interaction, which in leading order is related
to Sommerfeld rescattering corrections and which, compared to the
similar situation in top-anti-top production \cite{Kuhn:1992qw}, is
amplified by the ratio $(C_A/C_F)^2= (9/4)^2$.
\par
Corresponding to the different decay modes and rates, two complementary
scenarios must be considered: For relatively stable gluinos
(corresponding to the case $m_{\tilde g} < m_{\tilde q}$) gluino pairs
may form non-relativistic boundstates, denoted gluinonia, which decay
through gluino pair annihilation into a pair of gluon jets or a quark
plus an anti-quark jet.  This possibility has been described originally in
Refs.\,\cite{Keung:1983wz,Kuhn:1983sc,Goldman:1984mj,Kuhn:1987ty} where
the basic features like boundstate quantum numbers, leading terms in the
potential, spectra and some of the production mechanisms were
investigated.  More recently this aspect has been studied in
Refs.\,\cite{Kartvelishvili:1988tm,Kartvelishvili:1989pp,Chikovani:1996bk}.
Relatively stable gluinos are motivated by the proposal of Split
SUSY~\cite{Giudice:2004tc,ArkaniHamed:2004fb} (see also
Refs.\,\cite{Kilian:2004uj,Cheung:2004ad}), which suggests heavy squarks
and, correspondingly, long-lived gluinos. A detailed study is presented
in \cite{Kauth:2009ud}, which gives the higher order corrections to the
potential, to the boundstate energy spectrum and to the production
cross section for the boundstate in the colour singlet configuration.
\par
The second scenario is relevant for gluinos with a larger decay rate, of
order one GeV or higher. In this case the boundstate decay proceeds
through the decay of the constituents, and for decay rates of several GeV
no well-defined boundstates exist. Even in this case final state
interaction leads to a significant lowering of the effective production
threshold, an enhancement of the cross section and a strong distortion
of the differential cross section, in particular of the distribution in
the invariant mass of the gluino, with details depending on the gluino
mass and decay rate.
\par
This scenario has many similarities with hadronic top quark production
close to threshold \cite{Hagiwara:2008df,Kiyo:2008bv}. In particular the
distribution in the invariant mass of the gluino pair can be treated
with similar methods. In Ref.\,\cite{Hagiwara:2009hq} the dominant
contributions and the leading radiative corrections were classified and
evaluated and a sizable enhancement of the cross section was
observed. However, for a complete NLO evaluation the full ISR and hard
corrections must be included. In the case of the top system these were
available from the literature on the production of non-relativistic
colour singlet and octet quark-anti-quark boundstates
\cite{Kuhn:1992qw,Petrelli:1997ge} and could be directly applied to the
case of unstable top quarks.
\par
In contrast, the situation is more involved in the case of gluino pairs.
On the one hand the production cross section depends in addition on the
squark mass; for gluon induced amplitudes in NLO for quark-induced
amplitudes even at tree level. On the other hand a pair of gluinos can
be combined into boundstates transforming under a variety of
irreducible representations partly with attractive, partly with
repulsive interaction. In Ref.~\cite{Hagiwara:2009hq} the hard
correction was extracted from the NLO result for open gluino
production. However, this continuum cross section corresponds to a
weighted sum of the different contributions. In the present paper these
corrections will be evaluated separately for all the relevant
representations, together with the individual ISR corrections. This
allows to obtain the cross section including the full NLO corrections in
the threshold region, the central topic of this work.
\par
The paper will be organized as follows: For a self-contained treatment
we recall in Section\,\ref{sec:quantum} the quantum numbers of the
boundstates, discuss various SUSY scenarios and present the qualitative
features of threshold production for the case of interest, i.e. for
gluinos with decay rates comparable to the level spacing of the would-be
boundstates.
\par
In Section\,\ref{subsec:Green} we will present the threshold enhancement
(or suppression) for the various colour configurations using Green's
functions in NLO approximation. These will be evaluated similar to those
of the $t\overline{t}$ system discussed in
Refs.~\cite{Hagiwara:2008df,Kiyo:2008bv}. In Section\,\ref{subsec:Short}
the NLO pair production cross section, i.e. real and virtual corrections
will be derived for the different colour and spin configurations. Only
$S$ waves will be considered. Issues of renormalization, in particular
the usage of dimensional reduction (DRED), and the role of virtual gluino
or squark pairs will be discussed, together with the choice of the
proper value of the strong coupling $\alpha_{s}$.
\par
Using this input, the hadronic production cross section can be evaluated
in a straightforward way in Section\,\ref{sec:numerics}. We will limit
the discussion to proton-proton collisions at $7$ and $14$ TeV and give
results for several of the SUSY scenarios discussed in
Section\,\ref{sec:quantum}. Section\,\ref{sec:conclusions} contains our
conclusions.  Appendix~\ref{app:Hyp} contains useful relations for the
generalized hypergeometric function ${}_4F_3$, and some of the longer
formulae are relegated to Appendix~\ref{app:results}.  In
Appendix~\ref{app:3} details of the benchmark scenarios used in this
paper are provided.
\section{\label{sec:quantum}SUSY scenarios, gluino 
boundstates and threshold behaviour}
Let us briefly recall the quantum numbers of gluino pairs in the
threshold region, classified according to their colour, spin and orbital
momentum configurations
\cite{Keung:1983wz,Kuhn:1983sc,Goldman:1984mj}. They differ from those
of quark-anti-quark states due to the restrictions arising from the
Majorana nature of gluinos, and due to their different colour assignment.
\par
Two colour-octet states can be combined into irreducible representations
as follows (see e.g. \cite{MacFarlane:1968vc})
\begin{eqnarray}
8\otimes 8&=&1_{s}\oplus 8_{s}\oplus 8_{a}\oplus
10_{a}\oplus\overline{10}_{a}\oplus 27_{s}
\,,
\label{8times8}
\end{eqnarray}
where the subscript indicates (anti-)symmetry with respect to their
colour index.
\begin{table}[t]
\begin{center}
\begin{tabular}{c|ccccccc}
$^{2S+1}L_{J}$ & $^{1}S_{0}$ & $^{3}S_{1}$ & $^{1}P_{1}$ & $^{3}P_{0}$
& $^{3}P_{1}$ & $^{3}P_{2}$ & $^{1}D_{2}$ \\ \hline
L & $0$ & $0$ & $1$ & $1$ & $1$ & $1$ & $2$ \\
S & $0$ & $1$ & $0$ & $1$ & $1$ & $1$ & $0$ \\
$(\tilde{g}\tilde{g})_{s}$ & $0^{-}$ & $-$ & $-$ & $0^{+}$ & $1^{+}$ & $2^{+}$ & $2^{-}$ \\
$(\tilde{g}\tilde{g})_{a}$ & $-$ & $1^{-}$ & $1^{+}$ & $-$ & $-$ & $-$ & $-$
\end{tabular}
\caption{Lowest-lying states $J^{P}$ of the gluinonium spectrum. $L$
  and $S$ correspond to the angular momentum and spin quantum numbers,
  respectively.}
\label{table2}
\end{center}
\end{table}
Fermi statistics and the Majorana nature of the gluinos lead to
additional restrictions. For the symmetric colour configurations $1_s$,
$8_s$ and $27_s$ antisymmetric spin-angular momentum wave functions,
$(-1)^{L+S}=1$, are required, for the antisymmetric colour
configurations $8_{a}$, $10_{a}$ and $\overline{10}_{a}$ symmetric ones,
$(-1)^{L+S}=-1$. The intrinsic parity of a Majorana particle can be
chosen to be imaginary leading
to negative intrinsic parity of the boundstate. 
For a few lowest orbital angular momenta the
boundstate quantum numbers $J^P$ are listed in Tab.~\ref{table2}.
The transformation of these states under charge conjugation
is more involved.\footnote{We thank Y.~Kats and D.~Kahawala for drawing our attention to this
  issue.}

Since we restrict the discussion to boundstates, only the colour
configurations $1_s$, $8_s$ and $8_a$ with attractive potentials are discussed
in the following. 
Colour quantum numbers now play a non-trivial role. A
self-conjugate Majorana particle (without colour) transforms with $C=+1$ under
charge conjugation, the transformation of coloured constituents and their
boundstates, however, depends on
their colour index. Let us consider boundstates of two Majorana
particles. For the symmetric singlet states with the colour projector
$\delta_{ab}$ the charge parity of both constituents is identical, leading to
overall positive charge conjugation. This is consistent with the fact that the
decay of the pseudoscalar boundstate into two photons is
non-vanishing~\cite{Kauth:2009ud}, and agrees with earlier statements in the
literature~\cite{Keung:1983wz,Kuhn:1983sc,Goldman:1984mj}. The quantum numbers of the
antisymmetric octet ground state ($^3S_1$) with $J^P=1^-$ can be obtained by
considering the wave function of the boundstate 
$\sim \tilde{g}_b\gamma^\mu\tilde{g}_c f_{abc}$
(where $\tilde{g}_b$ and $\tilde{g}_b$ are the Majorana fields) 
with $C$-quantum numbers identical to
those of the gluon. These, in turn, can be obtained from the
relation~\cite{Smolyakov:1980wq,Tyutin:1982fx} between the gluon field
$A^a_\mu$ and its charge conjugate $A^{aC}_\mu$
\begin{eqnarray}
  \Gamma_a A^a_\mu &=& - \Gamma_a^T A_\mu^{aC}
  \,,
\end{eqnarray}
where $\Gamma_a$ are generators of SU(3) in a specific representation,
$\Gamma_a^T$ the corresponding transposed matrices. Specifically this implies
$C=-1$ for $A^a_\mu$ with $a=1,3,4,6,8$ and $C=+1$ for $a=2,5,7$
and correspondingly\footnote{This result
  is at variance with~\cite{Goldman:1984mj,Kauth:2009ud} which find $C=+1$ for
  all colour labels $a$ as well as with~\cite{Kats:2009bv,Kahawala:2011pc}
  which find $C=-1$ for all values of $a$.}
for the boundstates with $J^P=1^-$.
The same assignment is also valid for the $(\tilde{g}\tilde{g})_a$ state with
$J^P=1^+$. 

The charge parity of the pseudoscalar state
$(\tilde{g}\tilde{g})_s$ in the symmetric octet representation which
corresponds to the wave function $\sim \tilde{g}_b\gamma_5\tilde{g}_c d_{abc}$
is given by $C=+1$ for $a=1,3,4,6,8$ and $C=-1$ for $a=2,5,7$ and hence
opposite to the one of the antisymmetric states.\footnote{This result
  is at variance with~\cite{Goldman:1984mj,Kauth:2009ud,Kats:2009bv,Kahawala:2011pc} 
  which find $C=+1$ for all colour labels $a$.}
Note that for this
assignment $C$-parity is conserved in the decay of the boundstate to two
gluons which proceeds through the coupling $\sim d_{abc} A^bA^c$. The same
assignment is valid also for the other states in the symmetric octet
representation.
In total this can be summarized by defining a factor $C_a$ with $C_a=+1$ for
$a=1,3,4,6,8$, $C_a=-1$ for $a=2,5,7$, and assigning charge conjugation
$-C_a$ for the gluon field $A^a_\mu$  charge conjugation
$+C_a$ to the gluino, $-C_a$ to the antisymmetric octet and $+C_a$ to the
symmetric octet boundstate.
Restricting
ourselves now to the near threshold region, only $S$-wave configurations will
be retained.

\begin{table}[t]
\begin{center}
\begin{tabular}{c|ccc}
R & $(F^{R})^2$ & $F^{a,1}\cdot F^{a,2}$ & \mbox{interaction} \\ \hline
$1_{s}$ & $0$ & $-3$ & \mbox{attractive}\\
$8_{s},8_{a}$ & $3$ & $-\frac{3}{2}$ & \mbox{attractive}\\
$10_{a},\overline{10}_{a}$ & $6$ & $0$ & \mbox{neutral}\\
$27_{s}$ & $8$ & $1$ & \mbox{repulsive}
\end{tabular}
\caption{Colour interaction of two $\mbox{SU}(3)$ octets.}
\label{table1}
\end{center}
\end{table}
Depending on the representation $R$ of the boundstate, the interaction
between the two gluinos can be either attractive, repulsive or absent
(in lowest order). In lowest order the coefficient $C^{[R]}$ of the QCD
potential which governs the final state interaction is given by the
expectation value of the product of the colour generators
$F^{a}_{ij}F^{a}_{kl}$, taken between two-particle states in the
respective representation. This product, in turn, can be expressed by
the eigenvalues of the quadratic Casimir operator of the constituents,
$C_{A}=3$, and the boundstate in representation $R$,
$C_{R}=\left(F^{R}\right)^2$:
\begin{eqnarray}
  C^{[R]}\hspace{0.2cm}\equiv\hspace{0.2cm}F^{a,1}\cdot F^{a,2}&=&
  \frac{1}{2}\left[(F^{R})^2-(F^{a,1})^2-(F^{a,2})^2\right]=\frac{1}{2}\left(C_{{R}}-2C_{A}\right)
  \,.
\label{FdotF}
\end{eqnarray}
The results are listed in Tab.~\ref{table1}. For the cases with negative
(positive) coefficients, corresponding to attraction (repulsion), the
cross section will be enhanced (suppressed). Also the NLO correction of
the QCD potential, which will be needed below, is proportional to the
same group theoretical factor \cite{Hagiwara:2009hq}. The classification
described in Tabs.~\ref{table2} and \ref{table1} is of course also
applicable to continuum production and will be important for the
description of final state interaction.
\par
As mentioned in the Introduction and discussed in the literature
\cite{Keung:1983wz,Kuhn:1983sc,Goldman:1984mj,Kauth:2009ud,Hagiwara:2009hq}
the phenomenology of gluino pair production in the threshold region is
governed by the relative size of the decay rate of a gluino,
$\Gamma_{\tilde{g}}$, compared to the rate for gluinonium annihilation
into gluons, $\Gamma_{gg}$, on the one hand, and by the relative size of
$\Gamma_{\tilde{g}}$ and the level spacing $\Delta M$ between the ground
state and the first radial excitation of the colour singlet boundstate
on the other hand. Specifically we adopt for this comparison the
single-gluino decay rate of the boundstate, corresponding to
$2\Gamma_{\tilde{g}}$. The choice of $\Delta M=\left|E_{1}-E_{2}\right|$
is motivated by the fact that the binding energy per se depends
evidently on the choice of the mass definition (pole mass, potential
subtracted mass, \ldots) while $\Delta M$ is convention independent.
\par
If the decay rate $2\Gamma_{\tilde{g}}$ is smaller than the annihilation
rate of the ($S$-wave) boundstate, which in lowest order is
approximately given by $\Gamma_{gg}\approx
C_{A}^2\alpha_{s}^2\left|R(0)\right|^2/(2m_{\tilde{g}}^2)
\approx(C_{A}\alpha_{s})^5m_{\tilde{g}}/4$, the signatures of boundstate
and open gluino production are distinctively different: Boundstates
produced below the pair threshold decay into two gluon jets (no missing
energy), in contrast to the complicated cascade decays of gluinos above
threshold. This case (class A), evidently true in Split SUSY, was
discussed in detail in Ref.~\cite{Kauth:2009ud}, with emphasis on the
production of colour singlet boundstates.
\par 
If $\Gamma_{gg}\lsim 2\Gamma_{\tilde{g}}$, the constituents decay
before they can annihilate and the qualitative decay signatures of bound
and unbound gluinos are practically indistinguishable. Nevertheless, the
distribution in the invariant mass of the gluino pair will still be
modulated and strongly affected by final state interaction. As long as
$2\Gamma_{\tilde{g}}<\Delta M$, this mass distribution will still
reflect the presence of the boundstates, at least the enhancement
resulting from the lowest $1S$ resonance with quantum numbers $0^{-+}$
(class B). For even larger gluino decay rates
i.e. $2\Gamma_{\tilde{g}}\geq\Delta M$ these structures have essentially
disappeared resulting in an unstructured threshold behaviour (class
C). Nevertheless the cross section is still modified by final state
interaction and receives a significant contribution from the region
below threshold as far as colour singlet and octet rates are concerned
and a sizable suppression for the $27$ representation.
\par
\begin{figure}[tbp]
\begin{center}
\begin{tabular}{cc}
\vspace{-0.75cm}

\mbox{(a)}&
\includegraphics[angle=270,width=0.75\textwidth]{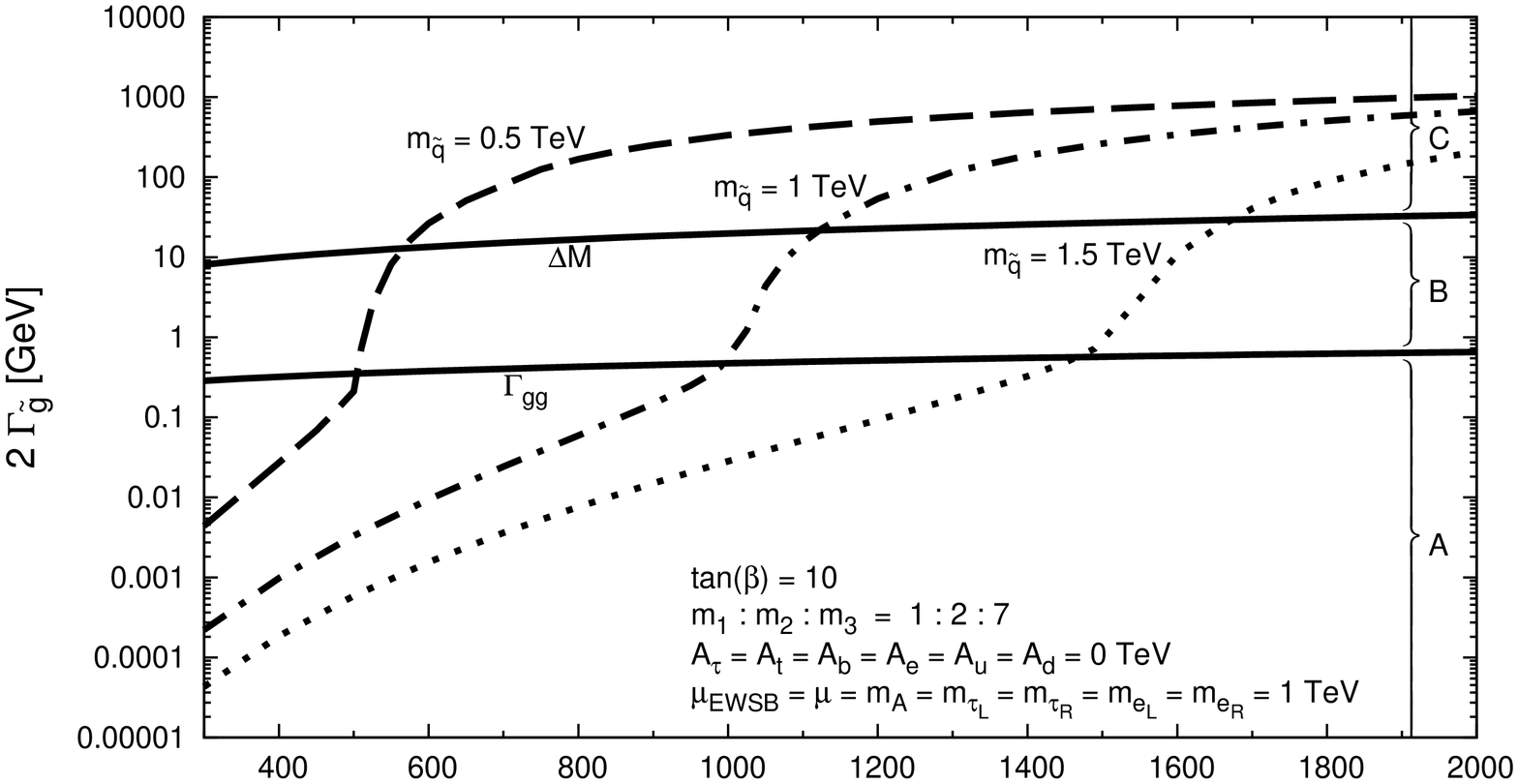}
\\
\mbox{(b)}&
\hspace{-0.1cm}
\includegraphics[angle=270,width=0.75\textwidth]{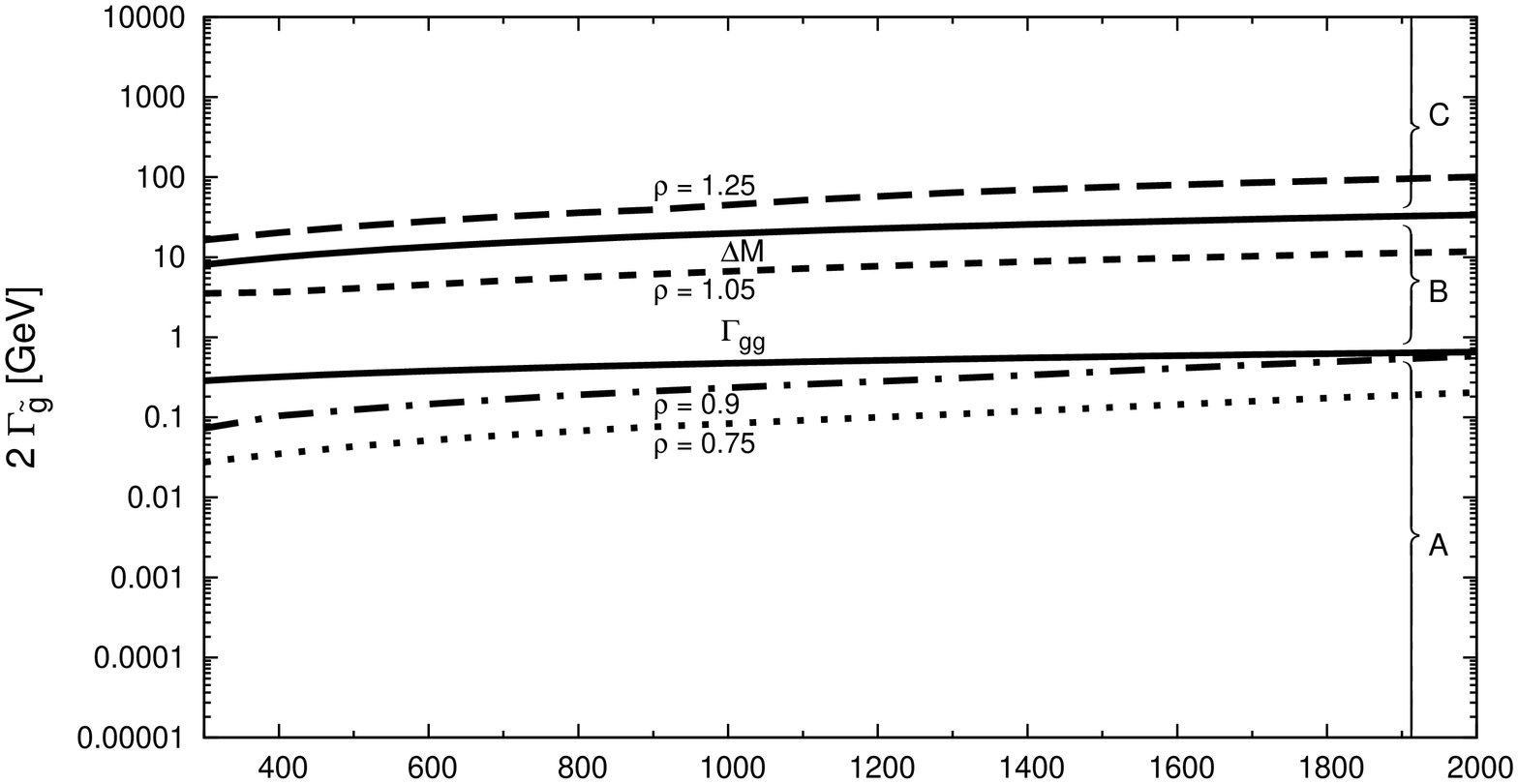}
\vspace{-.75cm}
\\
\mbox{(c)}&
\hspace{-0.25cm}
\includegraphics[angle=270,width=0.75\textwidth]{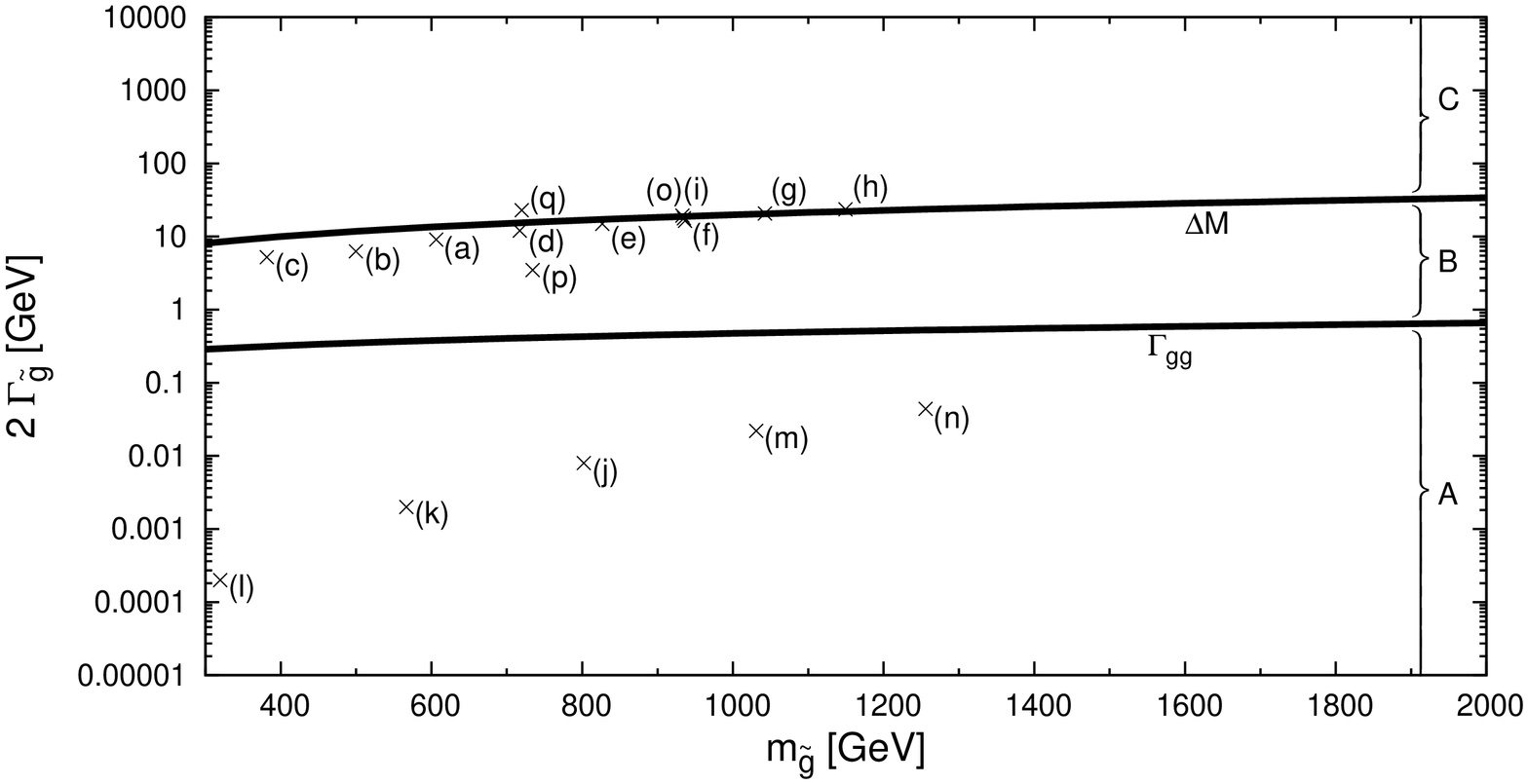}
\end{tabular}
\caption{Level spacing $\Delta M$ and annihilation rates $\Gamma_{gg}$
(solid curves) compared to the gluino decay rates
$2\Gamma_{\tilde{g}}$. Fig.~(a): for three different squark masses
(dashed: $0.5\,\mbox{TeV}$, dash-dotted: $1\,\mbox{TeV}$, dotted:
$1.5\,\mbox{TeV}$). Fig.~(b): for fixed ratios
($\rho=m_{\tilde{g}}/m_{\tilde{q}}$) between squark and gluino
masses. Fig.~(c): for the benchmark points (see Tab.~\ref{SPSmain}).}
\label{GammaGluino}
\end{center}
\end{figure}

The dependence of $\Gamma_{gg}$, $2\Gamma_{\tilde{g}}$ and $\Delta M$ on
$m_{\tilde{g}}$ is illustrated in Fig.~\ref{GammaGluino}. In
Fig.~\ref{GammaGluino}a we compare $\Delta M$ and $\Gamma_{gg}$,
evaluated in NLO \cite{Kauth:2009ud}, to the decay rate
$2\Gamma_{\tilde{g}}$, evaluated for three generic choices of squark
masses $0.5$~TeV, $1.0$~TeV and $1.5$~TeV. In Fig.~\ref{GammaGluino}b
fixed ratios between $m_{\tilde{g}}$ and $m_{\tilde{q}}$
($m_{\tilde{g}}/m_{\tilde{q}}=0.75$, $0.90$, $1.05$, $1.25$) are
adopted. In Tab.~\ref{SPSa} (Appendix~\ref{app:3}) we define 17
benchmark points on the basis of various SPS scenarios introduced in
Ref.~\cite{Allanach:2002nj}. Decay rates and SUSY masses corresponding
to our benchmark points are calculated with the programs {\tt
  SuSpect}~\cite{Djouadi:2002ze} and {\tt
  SDECAY}~\cite{Muhlleitner:2003vg}. The essential information,
i.e. gluino mass, averaged squark mass, level spacing of gluinonium,
gluino decay rate, gluinonium annihilation decay and dominant gluino
decay channels, is listed in Tab.~\ref{SPSmain}. Tabs.~\ref{SPSb}
and~\ref{SPSc} in Appendix~\ref{app:3} contain detailed information on
the squark masses of the benchmark points. Gluino masses and decay
rates, level spacing and annihilation rate of gluinonium for these
benchmark points are shown in Fig.~\ref{GammaGluino}c. It is clear from
this figure and Tab.~\ref{SPSmain}, that all three possibilities,
corresponding to class A, B and C could arise and should be
discussed. Note, that even the three cases which fall into class C are
fairly close to the boundary between B and C, such that a structured
threshold behaviour is typical for all benchmark points discussed in
this paper. Class A has been studied in detail in \cite{Kauth:2009ud},
with emphasis on colour singlet production. The present paper will be
concerned with class B, with results also applicable to class C.
\par

\begin{table}
\begin{center}
\footnotesize
\renewcommand{\arraystretch}{1.3}
\begin{tabular}{c||c|c|c|c|c|c|c}
benchmark&$m_{\tilde{g}}\,\left[\mbox{GeV}\right]$&$\overline{m}_{\tilde{q}}\,\left[\mbox{GeV}\right]$&$\Delta M\,\left[\mbox{GeV}\right]$&$2\Gamma_{\tilde{g}}\,\left[\mbox{GeV}\right]$&$\Gamma_{gg}\,\left[\mbox{GeV}\right]$&class&dominant decay
\\
point&&&&&&&channels
\\\hline\hline
(a)&$606.11$&$541.04$&$13.89$&$9.08$&$0.37$&$B$&$\tilde{g}\rightarrow\tilde{b}_{1}\overline{b},\tilde{b}^{*}_{1}b$
\\\hline
(b)&$493.05$&$443.19$&$12.04$&$6.28$&$0.34$&$B$&$\tilde{g}\rightarrow\tilde{c}_{2}\overline{c},\tilde{c}^{*}_{2}c,$
\\
&&&&&&&\hspace{0.8cm}$\tilde{u}_{2}\overline{u},\tilde{u}^{*}_{2}u$
\\\hline
(c)&$381.45$&$344.04$&$9.88$&$5.20$&$0.30$&$B$&$\tilde{g}\rightarrow\tilde{b}_{1}\overline{b},\tilde{b}^{*}_{1}b$
\\\hline
(d)&$717.12$&$637.39$&$15.75$&$12.00$&$0.40$&$B$&$\tilde{g}\rightarrow\tilde{b}_{1}\overline{b},\tilde{b}^{*}_{1}b$
\\\hline
(e)&$826.71$&$732.62$&$17.52$&$14.92$&$0.43$&$B$&$\tilde{g}\rightarrow\tilde{t}_{1}\overline{t},\tilde{t}^{*}_{1}t$
\\\hline
(f)&$935.18$&$826.94$&$19.22$&$17.84$&$0.46$&$B$&$\tilde{g}\rightarrow\tilde{b}_{1}\overline{b},\tilde{b}^{*}_{1}b$
\\\hline
(g)&$1042.60$&$920.35$&$20.87$&$20.70$&$0.48$&$B$&$\tilde{g}\rightarrow\tilde{t}_{1}\overline{t},\tilde{t}^{*}_{1}t$
\\\hline
(h)&$1149.42$&$1013.25$&$22.47$&$23.52$&$0.51$&$C$&$\tilde{g}\rightarrow\tilde{t}_{1}\overline{t},\tilde{t}^{*}_{1}t$
\\\hline
(i)&$936.42$&$831.88$&$19.24$&$16.38$&$0.46$&$B$&$\tilde{g}\rightarrow\tilde{b}_{1}\overline{b},\tilde{b}^{*}_{1}b$
\\\hline
(j)&$802.21$&$1462.40$&$17.13$&$0.0080$&$0.43$&$A$&$\tilde{g}\rightarrow\tilde{\chi}^{+}_{2}b\overline{t},\tilde{\chi}^{-}_{2}t\overline{b}$
\\\hline
(k)&$566.65$&$1219.50$&$13.21$&$0.0020$&$0.36$&$A$&$\tilde{g}\rightarrow\tilde{\chi}^{0}_{2}b\overline{b}$
\\\hline
(l)&$319.59$&$987.34$&$8.69$&$0.0002$&$0.28$&$A$&$\tilde{g}\rightarrow\tilde{\chi}^{+}_{1}d\overline{u},\tilde{\chi}^{-}_{1}u\overline{d},$
\\
&&&&&&&\hspace{0.5cm}$\tilde{\chi}^{+}_{1}s\overline{c},\tilde{\chi}^{-}_{1}c\overline{s}$
\\\hline
(m)&$1030.98$&$1710.01$&$20.69$&$0.022$&$0.48$&$A$&$\tilde{g}\rightarrow\tilde{\chi}^{+}_{2}b\overline{t},\tilde{\chi}^{-}_{2}t\overline{b}$
\\\hline
(n)&$1255.61$&$1959.83$&$24.04$&$0.044$&$0.53$&$A$&$\tilde{g}\rightarrow\tilde{\chi}^{+}_{2}b\overline{t},\tilde{\chi}^{-}_{2}t\overline{b}$
\\\hline
(o)&$933.03$&$819.58$&$19.19$&$19.26$&$0.46$&$C$&$\tilde{g}\rightarrow\tilde{t}_{1}\overline{t},\tilde{t}^{*}_{1}t$
\\\hline
(p)&$734.11$&$714.46$&$16.02$&$3.48$&$0.41$&$B$&$\tilde{g}\rightarrow\tilde{b}_{1}\overline{b},\tilde{b}^{*}_{1}b$
\\\hline
(q)&$719.66$&$618.86$&$15.79$&$22.92$&$0.40$&$C$&$\tilde{g}\rightarrow\tilde{t}_{1}\overline{t},\tilde{t}^{*}_{1}t$
\\\hline
(X)&$1300.00$&$1360.00$&$26.21$&$3.66$&$1.23$&$B$&$\tilde{g}\rightarrow\tilde{t}_{1}\overline{t},\tilde{t}^{*}_{1}t$\\\hline
(Y)&$1370.00$&$1235.00$&$27.27$&$20.00$&$1.26$&$B$&$\tilde{g}\rightarrow\tilde{t}_{1}\overline{t},\tilde{t}^{*}_{1}t$
\end{tabular}
\caption{Comparison of gluino masses, average squark masses, gluinonium
  level spacing, single decay rate, annihilation rate and dominant
  decay channels for the 19 benchmark points defined in
  Appendix~\ref{app:3}, Tab.~\ref{SPSa}.}
\label{SPSmain}
\end{center}
\end{table}
{From these considerations, from Fig.1 and from the discussion
presented below it is clear that the parameter $\rho\equiv m_{\tilde
  g}/m_{\tilde q}$ which characterizes the relative size of gluino vs.\
squark mass and thus the gluino decay rate is decisive for the
assignment to class A, B, or C. A value close to one is characteristic
for models at the borderline between A and B, a value around 1.14
corresponds to models at the borderline between B and C. Benchmark
points $p$, $a$ and $q$ with $\rho=1.3$, 1.12 and 1.16 respectively,
although excluded by recent LHC results
\cite{Chatrchyan:2011zy,Aad:2011qa,AbdusSalam:2011fc} already\footnote{For the Constraint Minimal Supersymmetric Standard Model
limits of around 1~TeV are quoted for $m_{\tilde{g}}$ and
 $m_{\tilde{q}}$. Note, however, that the calculation performed
in this paper is more general and can be applied to less
restrictive supersymmetric scenarios.}  
nevertheless
provide important insight into the structure of the threshold behaviour
and the relative importance of the different spin and colour
configurations. This will be illustrated by two scenarios X and Y with
the parameters listed in Tabs. 3 and 5 and which are not excluded by LHC
results. Model Y corresponds to point 10.3.3 from
Ref.~\cite{AbdusSalam:2011fc}, in model X the parameters $m_0$ and
$m_{1/2}$ have been chosen such as to ensure a relatively small gluino
decay rate. The phenomenological discussion of gluino pair production
for these two scenarios will be presented at the end of Section 4. We
restrict ourselves to configurations suggested by the Constraint Minimal
Supersymmetric Standard Model, the results, however, depend mainly on
the gluino decay rate only and thus are valid in a wider context.
}
\par
Let us now recall the qualitative aspects of the production
mechanism. Similar to the case of top quark production
\cite{Hagiwara:2008df,Kiyo:2008bv} (see also \cite{Beneke:2009rj}) the
cross section for a bound state $T$, differential in $M$, the invariant mass of the gluino
pair, can be decomposed into a factor representing the hard, short
distance part of the production process, and a factor given by the
imaginary part of the Green's function evaluated at the origin and the
convolution with the luminosity functions
\begin{eqnarray}
M\frac{\mbox{d}\sigma_{PP\rightarrow
    T}}{\mbox{d}M}(S,M^2)&=&\sum_{i,j}\int_{\rho}^{1}\mbox{d}\tau\,\left[\frac{\mbox{d}\mathcal{L}_{ij}}{\mbox{d}\tau}\right](\tau,\mu_F^2)M\frac{\mbox{d}\hat{\sigma}_{ij\rightarrow
T}}{\mbox{d}M}(\hat{s},M^2,\mu_R^2,\mu_F^2)
\,,
\label{master}
\end{eqnarray}
with
\begin{eqnarray}
M\frac{\mbox{d}\hat{\sigma}_{ij\rightarrow
    T}}{\mbox{d}M}(\hat{s},M^2,\mu_R^2,\mu_F^2)&\hspace{-0.1cm}=&\hspace{-0.1cm}\mathcal{F}_{ij\rightarrow T}(\hat{s},M^2,\mu_R^2,\mu_F^2)\frac{1}{m_{\tilde{g}}^2}\mbox{Im}\left\{G^{[1,8,10,27]}(0,M\hspace{-0.1cm}-\hspace{-0.1cm}2m_{\tilde{g}}\hspace{-0.05cm}+\hspace{-0.1cm}i\Gamma_{\tilde{g}})\right\}
\,,
\nonumber\\
\left[\frac{\mbox{d}\mathcal{L}_{ij}}{\mbox{d}\tau}\right](\tau,\mu_F^2)&\hspace{-0.1cm}=&\hspace{-0.1cm}\int_{0}^{1}\mbox{d}x_{1}\int_{0}^{1}\mbox{d}x_{2}f_{i|P}(x_1,\mu_{F}^2)f_{j|P}(x_2,\mu_{F}^2)\delta(\tau-x_1x_2)
\,.
\label{master2}
\end{eqnarray}
As usual $\hat{s}$ and $S$ denote the partonic and the hadronic
center-of-mass energy squared, respectively, and $\tau=\hat{s}/S$. The
lower limit of the $\tau$ integration is given by $\rho=M^2/S$. The
superscript of the Green's function refers to the colour state of $T$
and $\mu_F$ and $\mu_R$ denote the factorization and renormalization
scale.
\par
\begin{figure}[tbp]
\begin{center}
\begin{picture}(420,60)(0,0)
\SetColor{Black}
\Vertex(50,50){1.8}
\Gluon(50,50)(50,10){-2.5}{4}
\Line(50,50)(50,10)
\Vertex(50,10){1.8}
\Gluon(10,60)(50,50){3}{4}
\Gluon(10,0)(50,10){-3}{4}
\Gluon(50,50)(90,60){2.5}{4}
\Line(50,50)(90,60)
\Gluon(50,10)(90,0){-2.5}{4}
\Line(50,10)(90,0)
\Vertex(145,30){1.8}
\Gluon(145,30)(175,30){3}{3}
\Vertex(175,30){1.8}
\Gluon(120,60)(145,30){3}{4}
\Gluon(120,0)(145,30){-3}{4}
\Gluon(200,60)(175,30){-2.5}{4}
\Line(200,60)(175,30)
\Gluon(175,30)(200,0){-2.5}{4}
\Line(175,30)(200,0)
\Vertex(255,30){1.8}
\Gluon(255,30)(285,30){3}{3}
\Vertex(285,30){1.8}
\Line(310,60)(285,30)
\Gluon(310,60)(285,30){-2.5}{4}
\Line(310,0)(285,30)
\Gluon(310,0)(285,30){2.5}{4}
\ArrowLine(255,30)(230,0)
\ArrowLine(230,60)(255,30)
\Vertex(380,50){1.8}
\DashArrowLine(380,50)(380,10){4}
\Vertex(380,10){1.8}
\Line(420,60)(380,50)
\Gluon(420,60)(380,50){-2.5}{4}
\Line(420,0)(380,10)
\Gluon(420,0)(380,10){2.5}{4}
\ArrowLine(380,10)(340,0)
\ArrowLine(340,60)(380,50)
\end{picture}
\caption{Feynman diagrams contributing at LO to
  $gg\rightarrow\tilde{g}\tilde{g}$ and
  $q\overline{q}\rightarrow\tilde{g}\tilde{g}$.}
\label{feynLO}
\end{center}
\end{figure}
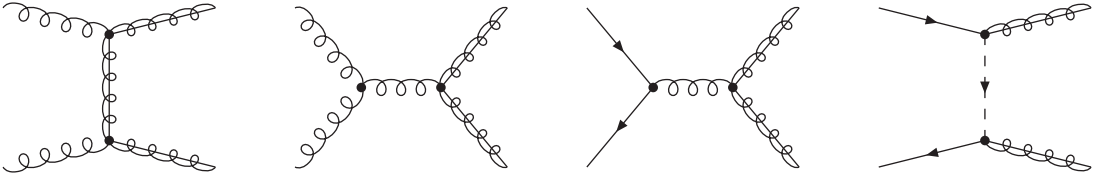

In leading order (LO) gluino pairs can be produced by gluon fusion or
quark-anti-quark annihilation (Fig.~\ref{feynLO}), at NLO also the
quark-gluon channel contributes.
\par
The strength of final state interaction at threshold is determined by
the respective colour representation of the gluino pair. To disentangle
the different representations it is convenient to use the following
projectors \cite{Bartels:1993ih}
\begin{eqnarray}
\mathbb{P}_{1_{S}}^{ab,cd}&=&\frac{1}{8}\delta_{ab}\delta_{cd}\,,
\nonumber\\
\mathbb{P}_{8_{S}}^{ab,cd}&=&\frac{3}{5}d_{abe}d_{cde}\,,
\nonumber\\
\mathbb{P}_{8_{A}}^{ab,cd}&=&\frac{1}{3}f_{abe}f_{cde}\,,
\nonumber\\
\mathbb{P}_{10}^{ab,cd}&=&\frac{1}{2}\left(\delta_{ac}\delta_{bd}-\delta_{ad}\delta_{bc}\right)-\frac{1}{3}f_{abe}f_{cde}\,,
\nonumber\\
\mathbb{P}_{27_{S}}^{ab,cd}&=&\frac{1}{2}\left(\delta_{ac}\delta_{bd}+\delta_{ad}\delta_{bc}\right)-\frac{3}{5}d_{abe}d_{cde}-\frac{1}{8}\delta_{ab}\delta_{cd}\,,
\label{proj}
\end{eqnarray}
where $\mathbb{P}_{10}$ is the projector on $10_a\oplus\overline{10}_a$
and
\begin{eqnarray}
\delta_{ac}\delta_{bd}\mathbb{P}_{X}^{ab,cd}&=&N_{X}\,,
\nonumber\\
\mathbb{P}_{X}^{ab,cd}\mathbb{P}_{Y}^{cd,ef}&=&\delta_{XY}\mathbb{P}_{X}^{ab,ef}\,,
\nonumber\\
\sum_{X}\mathbb{P}_{X}^{ab,cd}&=&\delta_{ac}\delta_{bd}
\,.
\label{proj2}
\end{eqnarray}
The normalization constant $N_{X}$ is given by the dimension of the
representation $X$, where $N_{10}=N_{10_{a}}+N_{\overline{10}_{a}}=20$.
As stated above, colour representation, spin and orbital angular
momentum of the gluino pair are interlocked. $S$-wave gluino pairs in
symmetric and antisymmetric colour configurations combine into spin
singlet and triplet states, respectively. For the quantities
$\mathcal{F}_{ij\rightarrow T}$ (as introduced in Eq.~(\ref{master2}))
describing the hard kernel one finds in Born approximation
\begin{eqnarray}
\mathcal{F}^{(0)}_{ij\rightarrow
  T}&=&\mathcal{N}_{ij}^{[T]}\frac{9\pi^2\alpha_{s}^2(\mu_R)}{4\hat{s}}\delta(1-z)\,
\,,
\label{hardLO}
\end{eqnarray}
with the non-vanishing normalization factors
\begin{eqnarray}
\mathcal{N}_{gg}^{[T]}&=&1,2,3\hspace{0.5cm}\mbox{for}\hspace{0.5cm}T=1_s,8_s,27_s
\, ,
\nonumber\\
\mathcal{N}_{q\overline{q}}^{[8_a]}&=&\frac{128}{27}\left(\frac{r-1}{r+1}\right)^2
\,.
\label{NormHardLO}
\end{eqnarray}
Here $z=M^2/\hat{s}$ and $r=\overline{m}_{\tilde{q}}^2/m_{\tilde{g}}^2$
where $\overline{m}_{\tilde{q}}$ is the averaged squark mass as
introduced in Tab.~\ref{SPSmain}. The Green's function which depends on
the energy $E=M-2m_{\tilde{g}}$ and the decay rate of the gluino
$\Gamma_{\tilde{g}}$ is obtained from the Schr\"odinger equation
\begin{eqnarray}
\left\{\left[\frac{(-i\nabla)^2}{m_{\tilde{g}}}+V_{C}^{[1,8,10,27]}(\vec{r}\,)\right]-(E+i\Gamma_{\tilde{g}})\right\}G^{[1,8,10,27]}(\vec{r},E+i\Gamma_{\tilde{g}})&=&\delta^{(3)}(\vec{r}\,)
\,
\end{eqnarray}
with the ``Coulomb'' potential $V_{C}^{[R]}\left(\vec{r}\,\right)=
-C^{[R]}\alpha_{s}(\mu_R)/r$ (and $C^{[R]}= 3, 3/2, 0,
-1\hspace{0.3cm}\mbox{for}\hspace{0.3cm}R= 1, 8, 10, 27$) in lowest
order. NLO corrections to the hard kernel, the potential and the Green's
function will be discussed in the next section, numerical results will
be presented for a variety of scenarios in
Section~\ref{sec:numerics}. From now on we shall limit our discussion to
three typical benchmark points (p), (a) and (q) (see
Fig.~\ref{GammaGluino} and Tab.~\ref{SPSmain}). For (p), due to the
close proximity of squark and gluino masses, the decay rate of the
boundstate, which is approximately $2\Gamma_{\tilde{g}}$, is
significantly lower than $\Delta M$ and, as a consequence, the
enhancement from the lowest lying resonances is well visible. This is
the case we will discuss in most detail. The other two points serve to
illustrate the case of gluinos with somewhat larger decay rates.
\par
\begin{figure}[tbp]
\begin{center}
\begin{tabular}{cc}
\vspace{-0.1cm}

\mbox{(a)}&
\includegraphics[angle=270,width=0.56\textwidth]{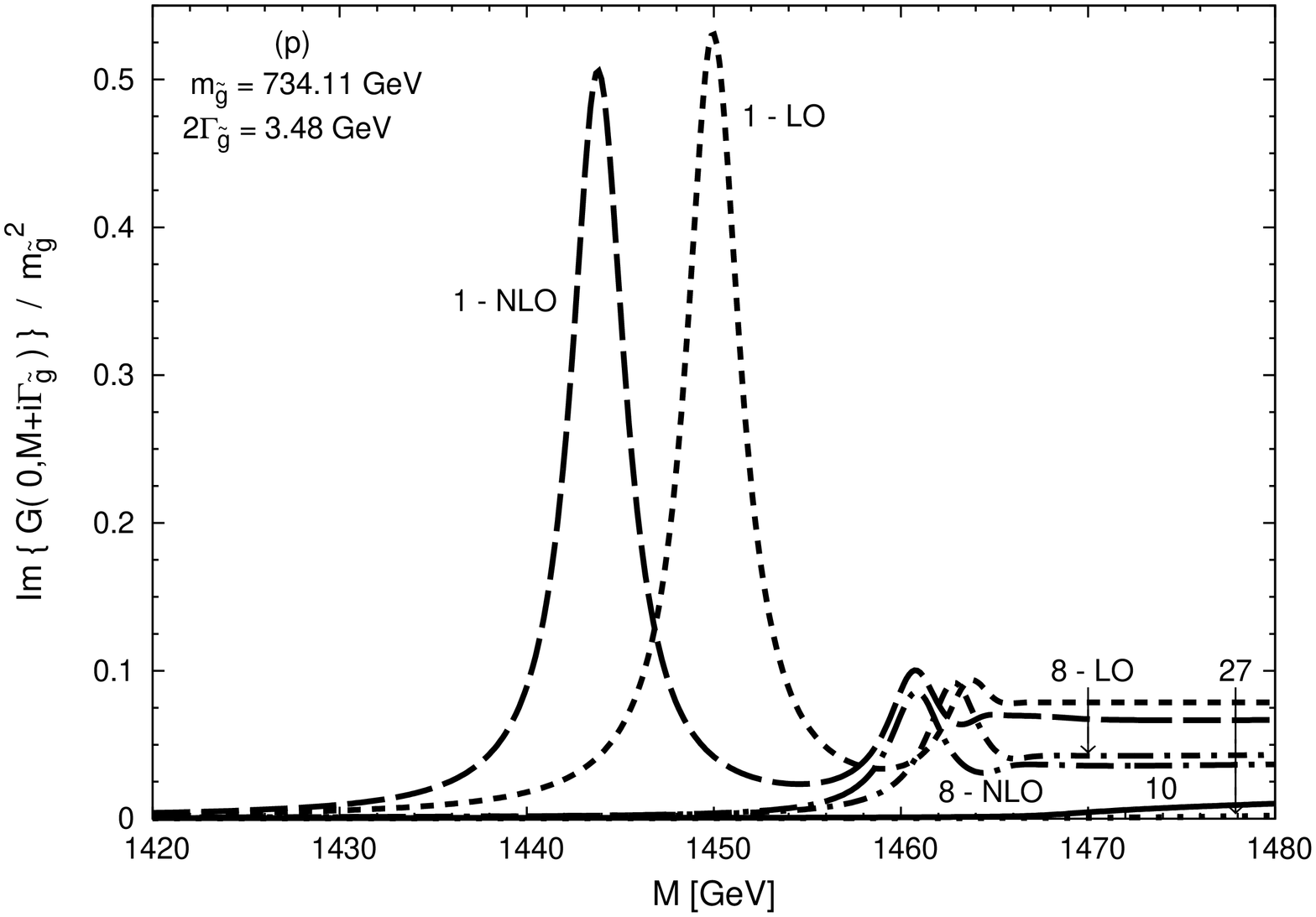}
\vspace{-0.4cm}
\\
\mbox{(b)}&
\hspace{-.3cm}
\includegraphics[angle=270,width=0.57\textwidth]{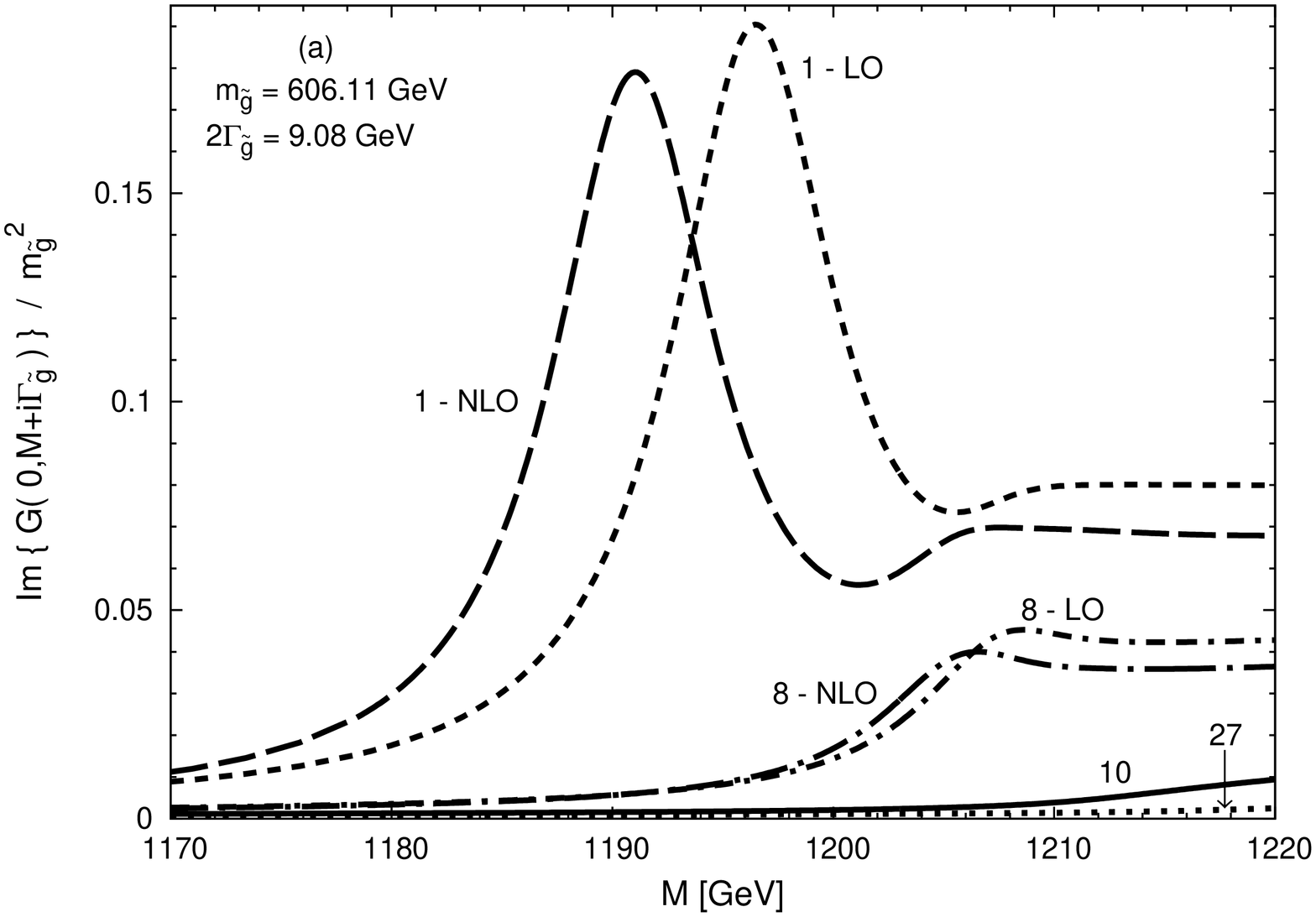}
\vspace{-0.5cm}
\\
\mbox{(c)}&
\hspace{-0.45cm}
\includegraphics[angle=270,width=0.57\textwidth]{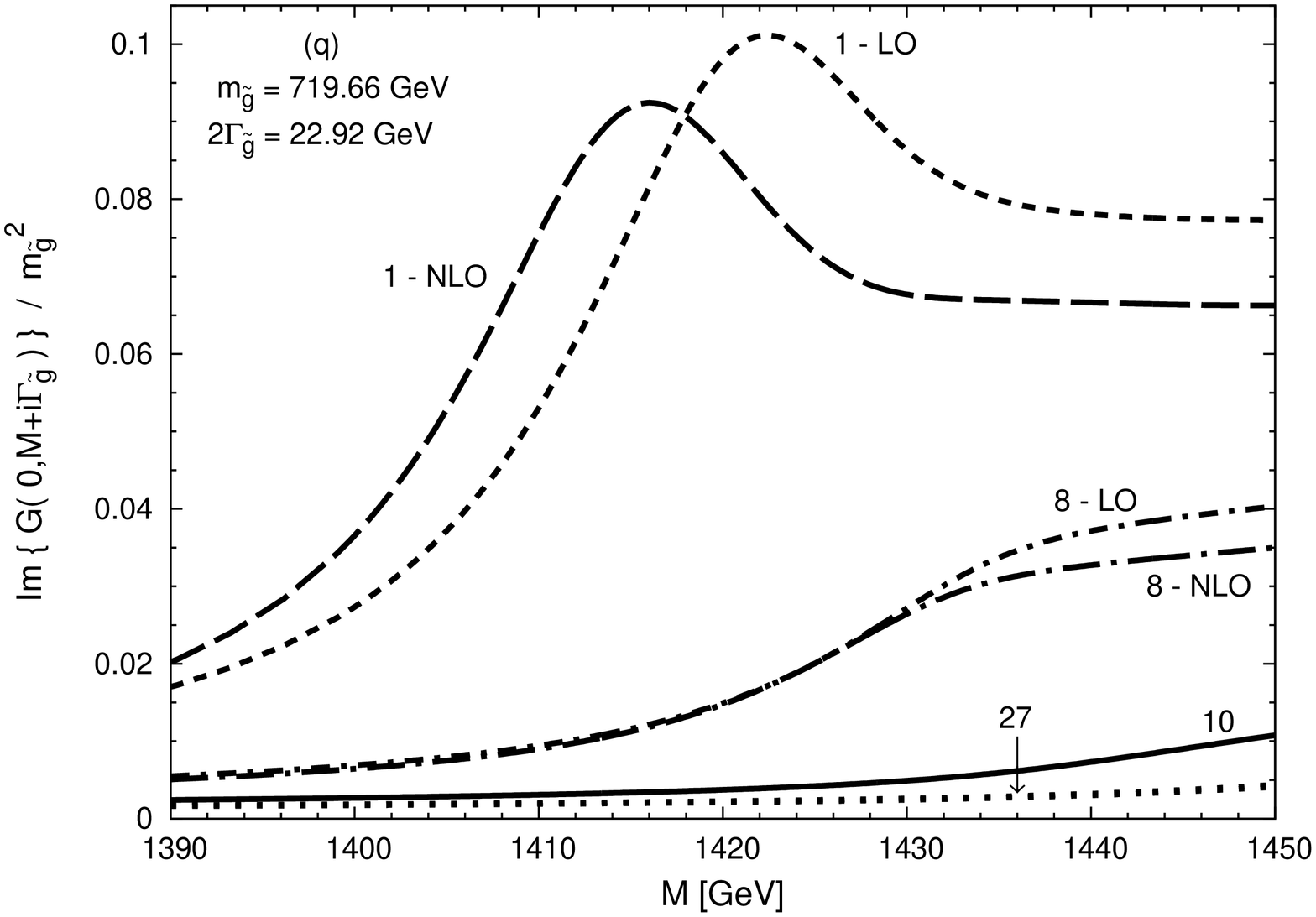}
\end{tabular}
\caption{Imaginary part of the Green's functions for benchmark points
(p), (a) and (q) for the singlet (dashed), octet (dash-dotted), decuplet
(solid) and twenty-seven configuration (dotted). For the decuplet we
show the free Green's function. LO and NLO curves for twenty-seven lie
on top of each other.}
\label{GreenNum}
\end{center}
\end{figure}

The imaginary parts of the LO and NLO Green's functions (for $R=1, 8,
10, 27$) are displayed in Fig.~\ref{GreenNum}(a). We have adopted twice
the Bohr radius as characteristic scale in $\alpha_{s}$ for the
attractive potential, $\alpha_s(M_Z)=0.1202$ as starting value and
two-loop running. This leads to
$\alpha_{s}^{[R]}=\alpha_{s}(C^{[R]}\alpha_{s}^{[R]}m_{\tilde{g}})=
\left\{0.1034,0.1124\right\}$ for $[R]=1,8$ and
$\alpha_{s}^{[27]}=\alpha_{s}(\left|C^{[27]}\right|\alpha_{s}^{[27]}m_{\tilde{g}})
=0.1184$ for the repulsive potential of the twenty-seven
representation. Two further characteristic examples are shown in
Figs.~\ref{GreenNum}(b) and (c) corresponding to benchmark points (a)
and (q) with significantly larger $\Gamma_{\tilde{g}}$.
LO and NLO predictions are shown for the singlet (dashed) and for the
octet configuration (dash-dotted), anticipating the results of
Section~\ref{subsec:Green}. In our approximation the potential of the
decuplet vanishes and thus the free Green's function is shown
(solid). For the repulsive twenty-seven configuration the NLO
corrections are quite small, i.e. LO and NLO nearly coincide. The
accidental near degeneracy of the first radial excitation from $G^{[1]}$
with the lowest enhancement of $G^{[8]}$ follows trivially from the
ratio between the strengths of potential $(C^{[8]}/C^{[1]})^2=1/4$ and
the excitation spectrum for the Coulomb potential, $E_n\sim1/n^2$. For
(p) the lowest resonances are still nicely separated, for (q), which is
close to the border between class B and C (see
Fig.~\ref{GammaGluino}(c)), individual resonances have nearly
disappeared. Nevertheless final state interaction leads to a significant
modification of the threshold behaviour and to marked differences
between the different colour representations. The Green's functions for
symmetric and antisymmetric octet are obviously identical and denoted by
$G^{[8]}$, the difference between the two states is only the spin
configuration.
\begin{figure}[t]
\begin{center}
\begin{minipage}{\textwidth}
\includegraphics[angle=270,width=\textwidth]{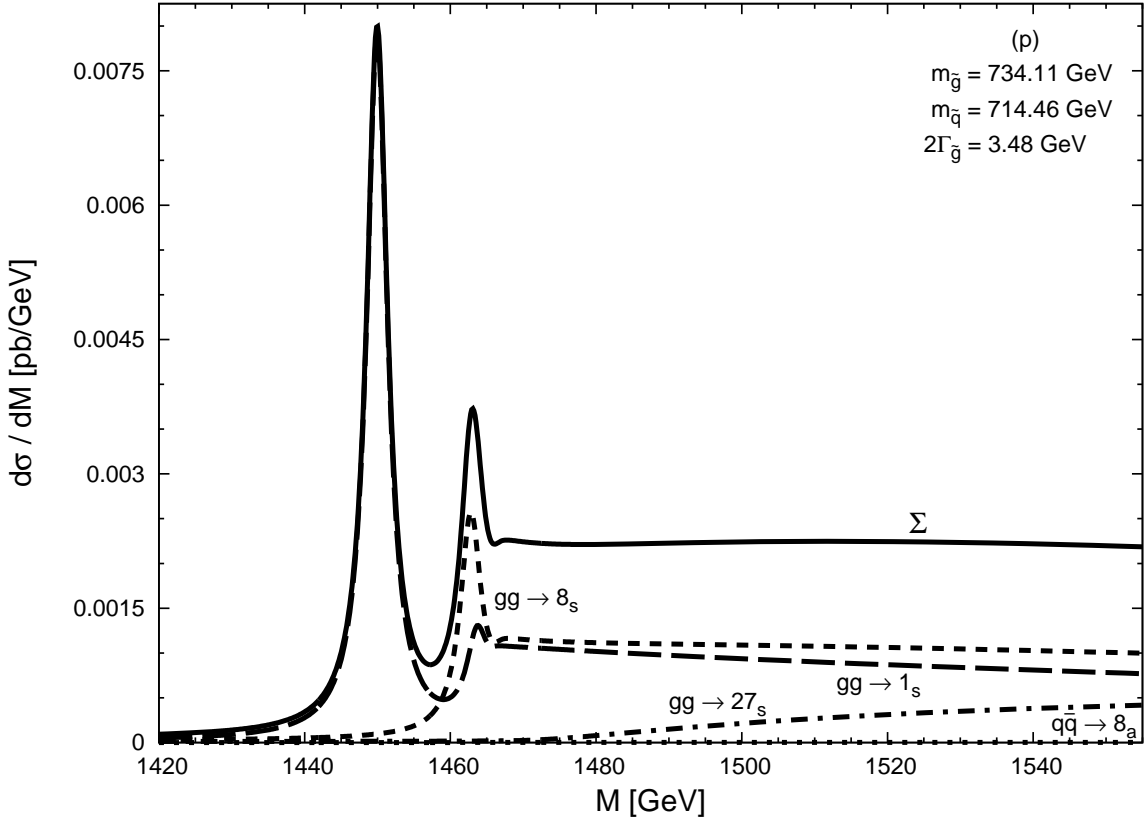}
\end{minipage}
\caption{LO prediction for the invariant mass distribution for scenario
  (p).}
\label{LOcross}
\end{center}
\end{figure}
\par
The LO prediction for the cross section is shown in Fig.~\ref{LOcross}
separately for the four different contributions and the sum.  We have
employed the PDF set MSTW2008LO~\cite{Martin:2009iq} which fixes the value
$\alpha_s(M_Z)=0.1394$. We use two-loop running and one-loop decoupling
(as implemented in the program {\tt RunDec}~\cite{Chetyrkin:2000yt}) to
obtain the strong $\overline{\rm MS}$ coupling at the scale $\mu=2
m_{\tilde{g}}$. In a next step we transform the coupling to the
$\overline{\rm DR}$ scheme in the full SUSY QCD theory with one-loop
approximation (see, e.g., Ref.~\cite{Harlander:2007wh}).  Note, that the
$q\overline{q}$ induced production of $8_{a}$ is strongly suppressed as
a consequence of the destructive interference between the amplitudes
with $s$-channel gluon and $t$- as well as $u$-channel squark exchange
and the near degeneracy of squark and gluino masses
\begin{eqnarray}
\frac{\mathcal{F}_{q\overline{q}\rightarrow 8_{a}}}{\mathcal{F}_{gg\rightarrow 1_{s}}}&=&
\frac{128}{27}\left(\frac{m_{\tilde{g}}^2-m_{\tilde{q}}^2}{m_{\tilde{g}}^2+m_{\tilde{q}}^2}\right)^2
\hspace{0.2cm}
\cong\hspace{0.3cm}3.5\cdot 10^{-3}, 6.1\cdot 10^{-2}, 1.1\cdot 10^{-1}
\,,
\label{eq::FqqLO}
\end{eqnarray}
for the benchmark points (p), (a), (q). Close to threshold the
repulsive final state interaction of $27_{s}$ leads to a strong
suppression of the cross section. For larger energies this effect
disappears quickly as a consequence of the large multiplicity of states
in the $27_{s}$ representation.
\section{\label{sec:NLO}Next-to-leading order corrections}
The NLO corrections to the cross section, whose evaluation is the main
subject of this work, can be separated into those for the Green's
function $\mbox{Im}\left\{G(0,E+i\Gamma_{\tilde{g}})\right\}$ and those
for the hard coefficients $\mathcal{F}_{ij\rightarrow T}$, as defined in
Eq.~(\ref{master2}).
\subsection{\label{subsec:Green}Green's function}
Following the idea of the Green's function formalism developed in
Refs.~\cite{Fadin:1987wz,Fadin:1988fn,Fadin:1990wx} we start with using
the NLO potential in momentum space\footnote{For colour triplet or octet
states combined into singlet boundstates the potential is even known to
$\mbox{NNNLO}$
\cite{Smirnov:2009fh,Smirnov:2008pn,Anzai:2009tm,Anzai:2010td}, for the
combination of colour triplet states into octet representations see
\cite{Kniehl:2004rk}.}
\begin{eqnarray}
\widetilde{V}_{C}^{[R]}\left(\vec{q}\,\right)&=&-C^{[R]}\frac{4\pi\alpha_{s}(\mu_G)}{\vec{q}^{\hspace{0.075cm}2}}\left[1+\frac{\alpha_{s}(\mu_G)}{4\pi}\left(\beta_{0}\ln\frac{\mu_G^2}{\vec{q}^{\hspace{0.075cm}2}}+a_{1}\right)\right]
\,,
\label{potential}
\end{eqnarray}
with~\cite{Goldman:1984mj}
\begin{eqnarray}
C^{[1]}&=&C_A\hspace{0.2cm}=\hspace{0.2cm}3\,,
\nonumber\\
C^{[8]}&=&\frac{1}{2}C_A\hspace{0.2cm}=\hspace{0.2cm}\frac{3}{2}\,,
\nonumber\\
C^{[10]}&=&0\,,
\nonumber\\
C^{[27]}&=&-\frac{1}{3}C_A\hspace{0.2cm}=\hspace{0.2cm}-1
\,,
\label{Cpot}
\end{eqnarray}
and
\begin{eqnarray}
a_{1}&=&\frac{31}{9}C_{A}-\frac{20}{9}T_{F}n_{f}\,,
\nonumber\\
\beta_{0}&=&\frac{11}{3}C_{A}-\frac{4}{3}T_{F}n_{f}
\,.
\label{a1beta0}
\end{eqnarray}
We have checked by an explicit calculation that for all colour
configurations the same coefficient $a_1$ is obtained. The
renormalization scale relevant for potential and Green's function is
denoted by $\mu_G$ and chosen to be the solution of
$\mu_G=\left|C^{[R]}\right|\alpha_s(\mu_G)m_{\tilde{g}}$. The strong
coupling is defined in the $\overline{\mbox{MS}}$ scheme and for five
active flavours. In principle one might include top-quark mass effects,
however, at the present level of precision these terms are still
irrelevant and are neglected in the potential.
\par
The Green's function for the top-anti-top system is known in compact
analytic form \cite{Beneke:1999qg} (see also \cite{Pineda:2006ri}) and
the result is easily applied to the present case
\begin{eqnarray}
G^{[R]}(0,M-2m_{\tilde{g}}+i\Gamma_{\tilde{g}})&=&i\frac{vm_{\tilde{g}}^2}{4\pi}+\frac{C^{[R]}\alpha_{s}(\mu_G)m_{\tilde{g}}^2}{4\pi}\left[g_{\mbox{\scriptsize LO}}+\frac{\alpha_{s}(\mu_G)}{4\pi}g_{\mbox{\scriptsize NLO}}+\,\ldots\,\right]
\,,
\label{Green}
\end{eqnarray}
with
\begin{eqnarray}
g_{\mbox{\scriptsize LO}}&\equiv&L-\psi^{(0)}\,,
\nonumber\\
g_{\mbox{\scriptsize NLO}}&\equiv&\beta_{0}\biggl[\,L^2-2L\left(\psi^{(0)}-\kappa\psi^{(1)}\right)+\kappa\psi^{(2)}+\left(\psi^{(0)}\right)^2-3\psi^{(1)}-2\kappa\psi^{(0)}\psi^{(1)}
\nonumber\\
&&\hspace{0.65cm}+4\,_{4}F_{3}\left(1,1,1,1;2,2,1-\kappa;1\right)\biggr]+a_{1}\biggl[L-\psi^{(0)}+\kappa\psi^{(1)}\biggr]
\,,
\label{Green2}
\end{eqnarray}
and
\begin{eqnarray}
\kappa&\equiv&i\frac{C^{[R]}\alpha_{s}(\mu_G)}{2v}\,,
\nonumber\\
v&\equiv&\sqrt{\frac{M-2m_{\tilde{g}}+i\Gamma_{\tilde{g}}}{m_{\tilde{g}}}}\,,
\nonumber\\
L&\equiv&\ln\frac{i\mu_G}{2m_{\tilde{g}}v}
\,.
\label{Green3}
\end{eqnarray}
The $n$-th derivative $\psi^{(n)}=\psi^{(n)}(1-\kappa)$ of the digamma
function $\psi(z)=\gamma_{E}+(d/dz)\ln\Gamma(z)$ is evaluated at
$(1-\kappa)$ and the Generalized Hypergeometric Function $_4F_3$ is
defined in Appendix~\ref{app:Hyp}.
\par
Solving the Schr\"odinger equation perturbatively induces poles in the
Green's function of the form $\left\{\alpha_{s}E_n^{\mbox{\scriptsize
LO}}/\left[E_n^{\mbox{\scriptsize
LO}}-(E+i\Gamma_{\tilde{g}})\right]\right\}^k$ which become large in the
vicinity of $E=E_n^{\mbox{\scriptsize LO}}$ for a small decay width
$\Gamma_{\tilde{g}}$. To obtain a proper Green's function with single
poles one has to resum the multiple poles as proposed in
\cite{Beneke:1999qg} by adding the term
\begin{eqnarray}
&&\frac{F_n^{\mbox{\scriptsize LO}}\left(1+\alpha_{s}f_{1}\right)}{E_n^{\mbox{\scriptsize LO}}(1+e_{1}\alpha_{s})-(E+i\Gamma_{\tilde{g}})}
\nonumber\\
&&-\Biggl\{\,\frac{F_n^{\mbox{\scriptsize LO}}}{E_n^{\mbox{\scriptsize LO}}-(E+i\Gamma_{\tilde{g}})}+\alpha_{s}\left[-\frac{F_n^{\mbox{\scriptsize LO}}E_n^{\mbox{\scriptsize LO}}e_{1}}{\left(E_n^{\mbox{\scriptsize LO}}-(E+i\Gamma_{\tilde{g}})\right)^2}+\frac{F_n^{\mbox{\scriptsize LO}}f_{1}}{E_n^{\mbox{\scriptsize LO}}-(E+i\Gamma_{\tilde{g}})}\right]\Biggr\}
\,,
\end{eqnarray}
which is of order $\alpha_s^2$. \color{black}The Schr\"odinger wave
function at the origin as well as the binding energy are given by 
$\left|\Psi_{n}(0)\right|^2=F_n^{\mbox{\scriptsize
LO}}\left(1+\alpha_{s}f_{1}+\,\ldots\,\right)$ and
$E_{n}=E_n^{\mbox{\scriptsize
LO}}\left(1+\alpha_{s}e_{1}+\,\ldots\,\right)$. The quantities
$F_n^{\mbox{\scriptsize LO}}$, $E_n^{\mbox{\scriptsize LO}}$, $f_1$ and
$e_1$ can be found in Appendix A of Ref.~\cite{Kauth:2009ud}.
\par
Attention has to be paid to the numerical evaluation of the Generalized
Hypergeometric Function $_4F_3$ in Eq.\,(\ref{Green}). This function has
a branch cut for its last argument on the real positive axis starting
from $1$. Hence the series defining this function converges for
$\mbox{Re}\left(1-\kappa\right)>1$, a condition potentially
violated\footnote{ The condition is fulfilled for
$1>\left|E_1^{\mbox{\scriptsize
LO}}\right|\left(\sqrt{E^2+\Gamma^2}-E\right)/\left[2(E^2+\Gamma^2)\right]$.
Hence this problem does not arise for the $t\overline{t}$ system.} for
small $\left|E+i\Gamma_{\tilde{g}}\right|$. In Appendix\,\ref{app:Hyp}
it is shown how to circumvent this problem by a suitable
transformation.\footnote{We thank Yuichiro Kiyo for communications
concerning this point.}
\par
Due to the factorization, formulated in Eq.~(\ref{master}), the
renormalization scales of Green's function $\mu_G$ and short distance
part $\mu_R$, appearing below, can be chosen independently. For our
choice of $\mu_G$, twice the inverse of the Bohr radius of the two
gluino system, the Coulomb-Green's function has a well-convergent
perturbative series as discussed in Ref.~\cite{Beneke:2005hg} for the
top-anti-top pair in the colour-singlet configuration.
\subsection{\label{subsec:Short}Short distance corrections}
The calculation of the partonic cross sections has been performed in
dimensional regularization (DREG). The Feynman diagrams have been
generated using FeynArts \cite{Hahn:2000kx,Hahn:2001rv}, for the
evaluation we have used in-house FORM \cite{Vermaseren:2000nd} programs
and FormCalc \cite{Hahn:2001rv}. Masses and wave functions are
renormalized on-shell while minimal subtraction ($\overline{\rm MS}$) is
used for the coupling. DREG violates supersymmetry because of a mismatch
of the degrees of freedom of the gauge bosons and their fermionic
superpartners. This problem has been resolved by introducing dimensional
reduction (DRED)~\cite{Siegel:1979wq} as regularization method where the
elegant features of DREG are maintained. While the on-shell masses do
not depend on the choice of scheme the coupling has now to be calculated
in DRED combined with minimal subtraction, the so-called $\overline{\rm
DR}$ scheme. The transition between the $\overline{\rm MS}$ and
$\overline{\rm DR}$ parameters is known up to two
loops~\cite{Martin:1993yx,Mihaila:2009bn}. For our calculation the
one-loop relations are sufficient which we need for the Yukawa coupling,
which appears in the gluino-quark-squark vertices, and the gauge
coupling in the three-gluon, gluon-quark and gluon-gluino vertices.  Our final
results are expressed in terms of
$\alpha_s^{\overline{\mbox{\tiny{DR}}}} \equiv \alpha_s^{\rm (SQCD)}$,
the strong SUSY QCD coupling in the $\overline{\rm DR}$ scheme.
\par
The corresponding amplitudes have to be projected onto the proper spin
configuration \cite{Kuhn:1979bb,Guberina:1980dc} and the corresponding
colour representation (see Eq.~(\ref{proj})). As discussed in
Section~\ref{sec:quantum} the colour-symmetric configurations (with
$L=0$) have spin $S=0$, the antisymmetric ones $S=1$. For the processes,
which are non-vanishing in Born approximation, the hard part of the
partonic cross sections can be written in the form (with the factors
$\mathcal{N}_{ij}^{[T]}$ given in Eq.~(\ref{NormHardLO}))
\begin{eqnarray}
  \mathcal{F}_{ij\rightarrow T}&\hspace{-0.125cm}=\hspace{-0.125cm}&\mathcal{N}_{ij}^{[T]}\frac{9\pi^2\left(\alpha_s^{\overline{\mbox{\tiny{DR}}}}(\mu_R)\right)^2}{4\hat{s}}\left(1+\frac{\alpha_s^{\overline{\mbox{\tiny{DR}}}}(\mu_R)}{\pi}\overline{\mathcal{V}}_{ij}^{[T]}\right)\biggl[\,\delta(1-z)+\frac{\alpha_{s}^{\overline{\mbox{\tiny{DR}}}}(\mu_R)}{\pi}\overline{\mathcal{R}}_{ij}^{[T]}(z)\biggr]\,.
  \nonumber\\
  \label{formulae}
\end{eqnarray}
The remaining ones are given by
\begin{eqnarray}
\mathcal{F}_{ij\rightarrow T}&\hspace{-0.125cm}=\hspace{-0.125cm}&\frac{9\pi\left(\alpha_s^{\overline{\mbox{\tiny{DR}}}}(\mu_R)\right)^3}{4\hat{s}}\overline{\mathcal{R}}_{ij}^{[T]}(z)
\,.
\label{formulae2}
\end{eqnarray}
Furthermore, $\mathcal{F}_{g\overline{q}\rightarrow T}$ is equal to
$\mathcal{F}_{gq\rightarrow T}$. The quantities
$\overline{\mathcal{V}}_{ij}^{[T]}$ and
$\overline{\mathcal{R}}_{ij}^{[T]}$ denote the virtual and real
corrections and are obtained from the results listed below (see
Eqs.~(\ref{virtual}) -- (\ref{realQQ})) by dropping the infrared
singularities, which cancel between $\mathcal{V}_{ij}^{[T]}$ and
$\mathcal{R}_{ij}^{[T]}$. For the Born approximation in
$d=4-2\varepsilon$ dimensions a factor $(1-\varepsilon)(1-2\varepsilon)$
for the pseudoscalar and $(1-\varepsilon)$ for the vector states has been
taken into account. (Note that collinear singularities from
$\mathcal{R}_{ij}^{[T]}$ are already absorbed in the PDFs.)
\par
The only non-vanishing contributions to the hard virtual corrections
read
\begin{subequations}
\begin{eqnarray}
\mathcal{V}_{gg}^{[1_{s}]}&=&\biggl\{-\frac{3}{\varepsilon_{\mbox{\tiny{IR}}}^2}-\frac{11}{2\varepsilon_{\mbox{\tiny{IR}}}}+\frac{n_{f}-1}{3\varepsilon_{\mbox{\tiny{IR}}}}+\frac{1}{3}\ln\frac{m_t^2}{m_{\tilde{g}}^2}-\frac{8}{3}\ln(2)-\frac{25}{2}+2\pi^2+\frac{\beta_0^{\rm (SQCD)}}{2}\ln\left(\frac{\mu_R^2}{M^2}\right)
\nonumber\\
&&\hspace{0.3cm}+n_{f}\biggl[\frac{1}{6}\ln\frac{r}{4}+\frac{9r-1}{9}c_5(r)+\frac{r-1}{2}\Bigl(2b'_1(r)+b_1(r)-b_4(r)\Bigr)\biggr]\biggr\}\,f_{\varepsilon}(M^2)\,,
\nonumber\\
\\
\mathcal{V}_{gg}^{[8_{s}]}&=&\biggl\{-\frac{3}{\varepsilon_{\mbox{\tiny{IR}}}^2}-\frac{7}{\varepsilon_{\mbox{\tiny{IR}}}}+\frac{n_{f}-1}{3\varepsilon_{\mbox{\tiny{IR}}}}+\frac{1}{3}\ln\frac{m_t^2}{m_{\tilde{g}}^2}-\frac{8}{3}\ln(2)-\frac{19}{2}+\frac{13}{8}\pi^2+\frac{\beta_0^{\rm (SQCD)}}{2}\ln\left(\frac{\mu_R^2}{M^2}\right)
\nonumber\\
&&\hspace{0.3cm}+n_{f}\biggl[\frac{1}{6}\ln\frac{r}{4}+\frac{9r-4}{9}c_5(r)+\frac{r-1}{2}\Bigl(2b'_1(r)+b_1(r)-b_4(r)\Bigr)\biggr]\biggr\}\,f_{\varepsilon}(M^2)\,,
\nonumber\\
&&
\\
\mathcal{V}_{q\overline{q}}^{[8_{a}]}&=&\biggl\{-\frac{4}{3\varepsilon_{\mbox{\tiny{IR}}}^2}-\frac{7}{2\varepsilon_{\mbox{\tiny{IR}}}}-\frac{2}{3}+\frac{5}{36}\pi^2+2\ln(2)-\frac{5}{9}n_{f}+\frac{\beta_0^{\rm (SQCD)}}{2}\ln\left(\frac{\mu_R^2}{M^2}\right)
\nonumber\\
&&\hspace{0.3cm}+\mathcal{A}_{q\overline{q}}^{[8_{a}]}(r)\biggr\}\,f_{\varepsilon}(M^2)\,,
\label{virtualC}\\
\mathcal{V}_{gg}^{[27_{s}]}&=&\biggl\{-\frac{3}{\varepsilon_{\mbox{\tiny{IR}}}^2}-\frac{19}{2\varepsilon_{\mbox{\tiny{IR}}}}+\frac{n_{f}-1}{3\varepsilon_{\mbox{\tiny{IR}}}}+\frac{1}{3}\ln\frac{m_t^2}{m_{\tilde{g}}^2}-\frac{8}{3}\ln(2)-\frac{9}{2}+\pi^2+\frac{\beta_0^{\rm (SQCD)}}{2}\ln\left(\frac{\mu_R^2}{M^2}\right)
\nonumber\\
&&\hspace{0.3cm}+n_{f}\biggl[\frac{1}{6}\ln\frac{r}{4}+\frac{r-1}{2}\Bigl(2c_5(r)+2b'_1(r)+b_1(r)-b_4(r)\Bigr)\biggr]\biggr\}\,f_{\varepsilon}(M^2)
\,,
\end{eqnarray}
\label{virtual}
\end{subequations}

where $m_t=172\,\mbox{GeV}$,
$f_{\varepsilon}(Q^2)=(4\pi\mu^2/Q^2)^{\varepsilon}\,\Gamma(1+\varepsilon)$
and $\beta_0^{\rm SQCD}=3C_A-2T_Fn_f$ (with $n_f=6$) is the one-loop
coefficient of the SUSY QCD beta function and the scalar functions
$b_i$, $b_1^\prime$ and $c_5$ are defined in
Appendix~\ref{app:results}. The limit $m_t\rightarrow 0$ has been taken
wherever possible.
\par
The $gg$ initiated processes into $1_{s}, 8_{s}, 27_{s}$ receive
contributions from the diagrams depicted in Fig.~\ref{virOSgg}, which do
not involve squarks and from diagrams with virtual squarks depicted in
Fig.~\ref{virMSgg}. The latter are proportional to $n_{f}$, the number of
quark and squark flavours. For the squarks we assume equal masses. (The
general calculation for different masses for left- and right-handed
squarks and for different generations is straightforward, however, the
formulae are lengthy. Explicit results are presented in
Ref.~\cite{Kauth:2011phd}.) The $q\overline{q}$ initiated process into
$8_{a}$ receives contributions from diagrams without and with squarks in
LO already (see Fig.~\ref{feynLO}). Consequently also the corrections
contain contributions from diagrams without squarks (independent of $r$)
(Fig.~\ref{virOSqq}) and with squarks (Fig.~\ref{virMSqq}). All squark
mass dependent terms are collected in the function
$\mathcal{A}_{q\overline{q}}^{[8_{a}]}$ which is defined in
Appendix~\ref{app:results}.
\par
Born term and virtual corrections are absent for decuplet production
in gluon fusion and quark-anti-quark annihilation (see
Eq.~(\ref{NormHardLO})). The corrections from real radiation as
discussed below give a (small) non-vanishing result.
\begin{figure}[tbp]
\begin{center}
\begin{picture}(330,370)(0,0)
\SetColor{Black}
\Gluon(90,310)(65,320){2.5}{3}
\Line(90,310)(65,320)
\Gluon(90,370)(65,360){-2.5}{3}
\Line(90,370)(65,360)
\Vertex(65,320){1.8}
\Vertex(65,360){1.8}
\Gluon(65,320)(65,360){3}{4}
\Line(25,320)(25,360)
\Gluon(25,320)(65,320){-2.5}{4}
\Line(25,320)(65,320)
\Gluon(65,360)(25,360){-2.5}{4}
\Line(65,360)(25,360)
\Gluon(25,360)(25,320){-2.5}{4}
\Vertex(25,320){1.8}
\Vertex(25,360){1.8}
\Gluon(25,320)(0,310){3}{3}
\Gluon(25,360)(0,370){-3}{3}
\Text(0,340)[r]{($a$)}
\Gluon(210,310)(185,320){2.5}{3}
\Line(210,310)(185,320)
\Gluon(210,370)(185,360){-2.5}{3}
\Line(210,370)(185,360)
\Vertex(185,320){1.8}
\Vertex(185,360){1.8}
\Gluon(185,360)(185,320){-2.5}{4}
\Line(185,360)(185,320)
\Gluon(145,320)(185,320){-3}{4}
\Gluon(185,360)(145,360){-3}{4}
\Gluon(145,320)(145,360){3}{4}
\Vertex(145,320){1.8}
\Vertex(145,360){1.8}
\Gluon(145,320)(120,310){3}{3}
\Gluon(145,360)(120,370){-3}{3}
\Text(120,340)[r]{($b$)}
\Gluon(330,310)(305,320){2.5}{3}
\Line(330,310)(305,320)
\Gluon(330,370)(265,360){-2.5}{6}
\Line(330,370)(265,360)
\Vertex(305,320){1.8}
\Vertex(305,360){1.8}
\Gluon(305,320)(305,360){3}{4}
\Line(265,320)(305,320)
\Gluon(265,320)(305,320){-2.5}{4}
\Gluon(305,360)(265,360){-3}{4}
\Line(265,360)(265,320)
\Gluon(265,360)(265,320){-2.5}{4}
\Vertex(265,320){1.8}
\Vertex(265,360){1.8}
\Gluon(265,320)(240,310){3}{3}
\Gluon(305,360)(240,370){-3}{6}
\Text(240,340)[r]{($c$)}
\Gluon(90,230)(65,240){2.5}{3}
\Line(90,230)(65,240)
\Gluon(90,290)(65,280){-2.5}{3}
\Line(90,290)(65,280)
\Vertex(65,240){1.8}
\Vertex(65,280){1.8}
\Gluon(65,240)(65,280){2.5}{4}
\Line(65,240)(65,280)
\Gluon(65,280)(30,260){-3}{4}
\Gluon(30,260)(65,240){-3}{4}
\Vertex(30,260){1.8}
\Gluon(30,260)(0,230){3}{4}
\Gluon(30,260)(0,290){-3}{4}
\Text(0,260)[r]{($d$)}
\Gluon(210,290)(185,280){-2.5}{3}
\Line(210,290)(185,280)
\Gluon(210,230)(165,235){2.5}{5}
\Line(210,230)(165,235)
\Vertex(165,235){1.8}
\Gluon(165,235)(165,245){2.5}{1}
\Line(165,235)(165,245)
\Vertex(165,245){1.8}
\Gluon(165,245)(185,280){3}{4}
\Line(185,280)(145,280)
\Gluon(185,280)(145,280){-2.5}{4}
\Line(145,280)(165,245)
\Gluon(145,280)(165,245){-2.5}{4}
\Vertex(185,280){1.8}
\Vertex(145,280){1.8}
\Gluon(145,280)(120,290){-3}{3}
\Gluon(165,235)(120,230){3}{5}
\Text(120,260)[r]{($e$)}
\Gluon(330,290)(305,280){-2.5}{3}
\Line(330,290)(305,280)
\Gluon(330,230)(285,235){2.5}{5}
\Line(330,230)(285,235)
\Vertex(285,235){1.8}
\Gluon(285,235)(285,245){2.5}{1}
\Line(285,235)(285,245)
\Vertex(285,245){1.8}
\Line(285,245)(305,280)
\Gluon(285,245)(305,280){2.5}{4}
\Gluon(305,280)(265,280){-3}{3}
\Gluon(265,280)(285,245){-3}{4}
\Vertex(305,280){1.8}
\Vertex(265,280){1.8}
\Gluon(265,280)(240,290){-3}{3}
\Gluon(285,235)(240,230){3}{5}
\Text(240,260)[r]{($f$)}
\Gluon(90,150)(45,155){2.5}{5}
\Line(90,150)(45,155)
\Gluon(90,210)(45,205){-2.5}{5}
\Line(90,210)(45,205)
\Vertex(45,155){1.8}
\Gluon(45,155)(45,165){2.5}{1}
\Line(45,155)(45,165)
\Vertex(45,165){1.8}
\Line(45,165)(45,195)
\Gluon(45,165)(45,195){2.5}{3}
\GlueArc(45,180)(15,270,450){3}{4}
\Vertex(45,195){1.8}
\Gluon(45,195)(45,205){2.5}{1}
\Line(45,195)(45,205)
\Vertex(45,205){1.8}
\Gluon(45,205)(0,210){-3}{5}
\Gluon(45,155)(0,150){3}{5}
\Text(0,180)[r]{($g$)}
\Gluon(210,150)(197.5,151.5){2.5}{1}
\Line(210,150)(197.5,151.5)
\Vertex(197.5,151.5){1.8}
\GlueArc(187.5,152.5)(10,353.5,533.5){3}{3}
\Line(197.5,151.5)(178,153.5)
\Gluon(197.5,151.5)(178,153.5){2.5}{2}
\Vertex(178,153.5){1.8}
\Gluon(178,153.5)(165,155){2.5}{1}
\Line(178,153.5)(165,155)
\Gluon(210,210)(165,205){-2.5}{5}
\Line(210,210)(165,205)
\Vertex(165,155){1.8}
\Gluon(165,155)(165,205){2.5}{6}
\Line(165,155)(165,205)
\Vertex(165,205){1.8}
\Gluon(165,205)(120,210){-3}{5}
\Gluon(165,155)(120,150){3}{5}
\Text(120,180)[r]{($h$)}
\Gluon(330,150)(285,155){2.5}{5}
\Line(330,150)(285,155)
\Gluon(330,210)(285,205){-2.5}{5}
\Line(330,210)(285,205)
\Vertex(285,155){1.8}
\Gluon(285,155)(285,205){2.5}{6}
\Line(285,155)(285,205)
\Vertex(285,205){1.8}
\Gluon(285,205)(240,210){-3}{5}
\Gluon(285,155)(271.5,153.5){3}{1}
\Gluon(253.5,151.5)(240,150){3}{1}
\CCirc(262.5,152.5){10}{Black}{Gray}
\Text(240,180)[r]{($i$)}
\Gluon(90,70)(45,75){2.5}{5}
\Line(90,70)(45,75)
\Gluon(90,130)(45,125){-2.5}{5}
\Line(90,130)(45,125)
\Vertex(45,75){1.8}
\Gluon(45,75)(45,125){2.5}{6}
\Line(45,75)(45,125)
\Vertex(45,125){1.8}
\Gluon(45,125)(0,130){-3}{5}
\Gluon(45,75)(32.44,73.63){3}{1}
\Gluon(12.56,71.37)(0,70){3}{1}
\Vertex(32.44,73.63){1.8}
\ArrowArc(22.5,72.5)(10,6.5,186.5)
\ArrowArc(22.5,72.5)(10,186.5,366.5)
\Vertex(12.56,71.37){1.8}
\Text(0,100)[r]{($j$)}
\Gluon(210,70)(165,75){2.5}{5}
\Line(210,70)(165,75)
\Gluon(210,130)(165,125){-2.5}{5}
\Line(210,130)(165,125)
\Vertex(165,75){1.8}
\Gluon(165,75)(165,125){2.5}{6}
\Line(165,75)(165,125)
\Vertex(165,125){1.8}
\Gluon(165,125)(120,130){-3}{5}
\Gluon(165,75)(152.44,73.63){3}{1}
\Gluon(132.56,71.37)(120,70){3}{1}
\Vertex(152.44,73.63){1.8}
\GlueArc(142.5,72.5)(10,6.5,186.5){2.5}{3}
\GlueArc(142.5,72.5)(10,186.5,366.5){2.5}{3}
\CArc(142.5,72.5)(10,0,360)
\Vertex(132.56,71.37){1.8}
\Text(120,100)[r]{($k$)}
\Text(15,15)[r]{with}
\Gluon(30,15)(45,15){-2.5}{1}
\Gluon(65,15)(80,15){-2.5}{1}
\COval(55,15)(10,10)(0){Black}{Gray}
\Text(100,15)[r]{=}
\Gluon(105,15)(120,15){-2.5}{1}
\Gluon(140,15)(155,15){-2.5}{1}
\GlueArc(130,15)(10,0,180){2.5}{4}
\GlueArc(130,15)(10,180,360){2.5}{4}
\Vertex(120,15){1.6}
\Vertex(140,15){1.6}
\Text(175,15)[r]{+}
\Gluon(180,15)(195,15){-2.5}{1}
\Gluon(215,15)(230,15){-2.5}{1}
\DashCArc(205,15)(10,0,180){1.4}
\DashCArc(205,15)(10,180,360){1.4}
\Vertex(195,15){1.6}
\Vertex(215,15){1.6}
\Text(250,15)[r]{+}
\Gluon(255,15)(280,15){-2.5}{2}
\Gluon(280,15)(305,15){-2.5}{2}
\GlueArc(280,27)(12,270,630){2.5}{10}
\Vertex(280,15){1.6}
\end{picture}
\caption{NLO contributions to the $gg$ initiated processes without
  virtual squarks (dotted lines correspond to ghost fields).}
\label{virOSgg}
\end{center}
\end{figure}
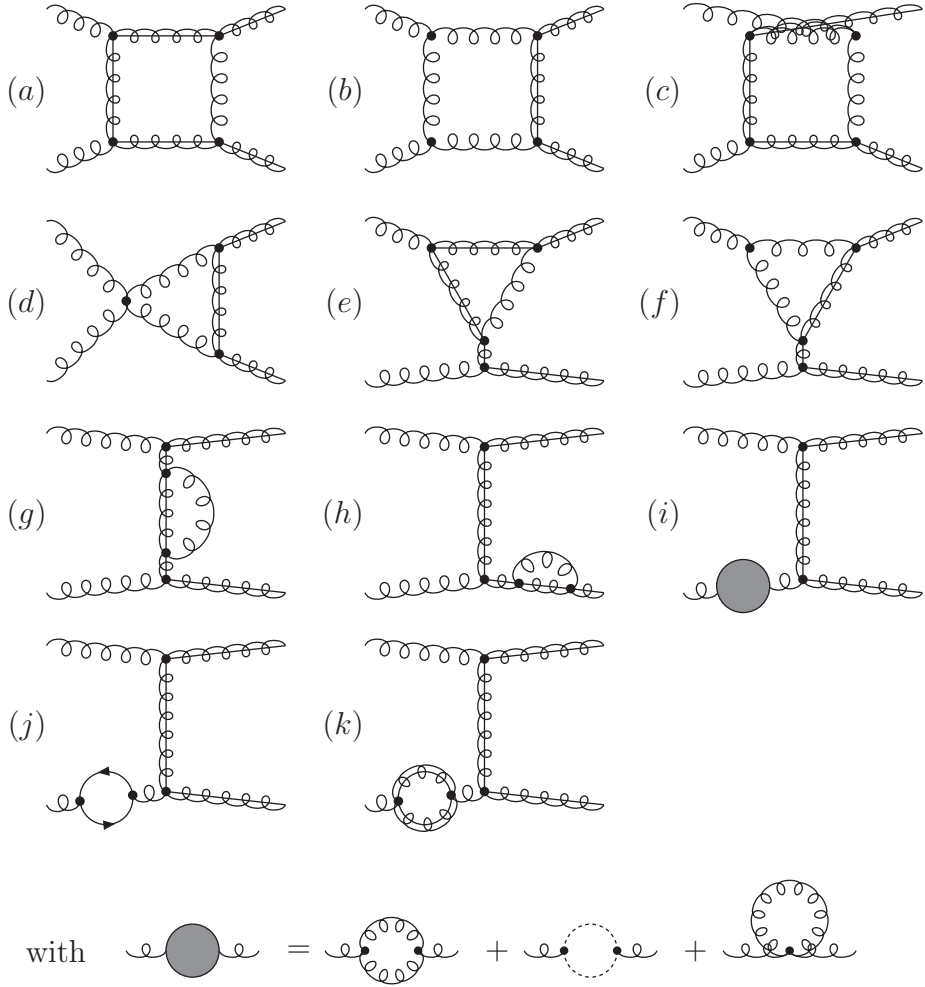
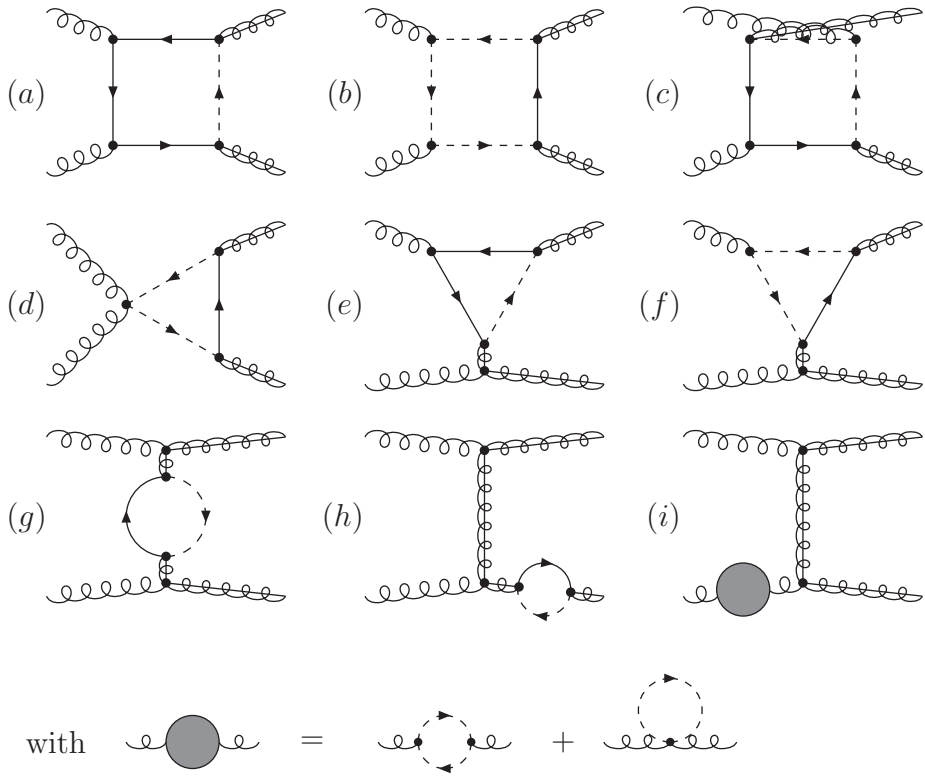
\begin{figure}[tbp]
\begin{center}
\begin{picture}(330,290)(0,0)
\SetColor{Black}
\Gluon(90,230)(65,240){2.5}{3}
\Line(90,230)(65,240)
\Gluon(90,290)(65,280){-2.5}{3}
\Line(90,290)(65,280)
\Vertex(65,240){1.8}
\Vertex(65,280){1.8}
\DashArrowLine(65,240)(65,280){3}
\ArrowLine(25,240)(65,240)
\ArrowLine(65,280)(25,280)
\ArrowLine(25,280)(25,240)
\Vertex(25,240){1.8}
\Vertex(25,280){1.8}
\Gluon(25,240)(0,230){3}{3}
\Gluon(25,280)(0,290){-3}{3}
\Text(0,260)[r]{($a$)}
\Gluon(210,230)(185,240){2.5}{3}
\Line(210,230)(185,240)
\Gluon(210,290)(185,280){-2.5}{3}
\Line(210,290)(185,280)
\Vertex(185,240){1.8}
\Vertex(185,280){1.8}
\ArrowLine(185,240)(185,280)
\DashArrowLine(145,240)(185,240){3}
\DashArrowLine(185,280)(145,280){3}
\DashArrowLine(145,280)(145,240){3}
\Vertex(145,240){1.8}
\Vertex(145,280){1.8}
\Gluon(145,240)(120,230){3}{3}
\Gluon(145,280)(120,290){-3}{3}
\Text(120,260)[r]{($b$)}
\Gluon(330,230)(305,240){2.5}{3}
\Line(330,230)(305,240)
\Gluon(330,290)(265,280){-2.5}{6}
\Line(330,290)(265,280)
\Vertex(305,240){1.8}
\Vertex(305,280){1.8}
\DashArrowLine(305,240)(305,280){3}
\ArrowLine(265,240)(305,240)
\DashArrowLine(305,280)(265,280){3}
\ArrowLine(265,280)(265,240)
\Vertex(265,240){1.8}
\Vertex(265,280){1.8}
\Gluon(265,240)(240,230){3}{3}
\Gluon(305,280)(240,290){-3}{6}
\Text(240,260)[r]{($c$)}
\Gluon(90,150)(65,160){2.5}{3}
\Line(90,150)(65,160)
\Gluon(90,210)(65,200){-2.5}{3}
\Line(90,210)(65,200)
\Vertex(65,160){1.8}
\Vertex(65,200){1.8}
\ArrowLine(65,160)(65,200)
\DashArrowLine(65,200)(30,180){3}
\DashArrowLine(30,180)(65,160){3}
\Vertex(30,180){1.8}
\Gluon(30,180)(0,150){3}{5}
\Gluon(30,180)(0,210){-3}{5}
\Text(0,180)[r]{($d$)}
\Gluon(210,210)(185,200){-2.5}{3}
\Line(210,210)(185,200)
\Gluon(210,150)(165,155){2.5}{5}
\Line(210,150)(165,155)
\Vertex(165,155){1.8}
\Gluon(165,155)(165,165){2.5}{1}
\Line(165,155)(165,165)
\Vertex(165,165){1.8}
\DashArrowLine(165,165)(185,200){3}
\ArrowLine(185,200)(145,200)
\ArrowLine(145,200)(165,165)
\Vertex(185,200){1.8}
\Vertex(145,200){1.8}
\Gluon(145,200)(120,210){-3}{3}
\Gluon(165,155)(120,150){3}{5}
\Text(120,180)[r]{($e$)}
\Gluon(330,210)(305,200){-2.5}{3}
\Line(330,210)(305,200)
\Gluon(330,150)(285,155){2.5}{5}
\Line(330,150)(285,155)
\Vertex(285,155){1.8}
\Gluon(285,155)(285,165){2.5}{1}
\Line(285,155)(285,165)
\Vertex(285,165){1.8}
\ArrowLine(285,165)(305,200)
\DashArrowLine(305,200)(265,200){3}
\DashArrowLine(265,200)(285,165){3}
\Vertex(305,200){1.8}
\Vertex(265,200){1.8}
\Gluon(265,200)(240,210){-3}{3}
\Gluon(285,155)(240,150){3}{5}
\Text(240,180)[r]{($f$)}
\Gluon(90,70)(45,75){2.5}{5}
\Line(90,70)(45,75)
\Gluon(90,130)(45,125){-2.5}{5}
\Line(90,130)(45,125)
\Vertex(45,75){1.8}
\Gluon(45,75)(45,85){2.5}{1}
\Line(45,75)(45,85)
\Vertex(45,85){1.8}
\ArrowArcn(45,100)(15,270,90)
\DashArrowArcn(45,100)(15,450,270){3}
\Vertex(45,115){1.8}
\Gluon(45,115)(45,125){2.5}{1}
\Line(45,115)(45,125)
\Vertex(45,125){1.8}
\Gluon(45,125)(0,130){-3}{5}
\Gluon(45,75)(0,70){3}{5}
\Text(0,100)[r]{($g$)}
\Gluon(210,70)(197.5,71.5){2.5}{1}
\Line(210,70)(197.5,71.5)
\Vertex(197.5,71.5){1.8}
\ArrowArcn(187.5,72.5)(10,173.5,353.5)
\DashArrowArcn(187.5,72.5)(10,353.5,533.5){3}
\Vertex(178,73.5){1.8}
\Gluon(178,73.5)(165,75){2.5}{1}
\Line(178,73.5)(165,75)
\Gluon(210,130)(165,125){-2.5}{5}
\Line(210,130)(165,125)
\Vertex(165,75){1.8}
\Gluon(165,75)(165,125){2.5}{6}
\Line(165,75)(165,125)
\Vertex(165,125){1.8}
\Gluon(165,125)(120,130){-3}{5}
\Gluon(165,75)(120,70){3}{5}
\Text(120,100)[r]{($h$)}
\Gluon(330,70)(285,75){2.5}{5}
\Line(330,70)(285,75)
\Gluon(330,130)(285,125){-2.5}{5}
\Line(330,130)(285,125)
\Vertex(285,75){1.8}
\Gluon(285,75)(285,125){2.5}{6}
\Line(285,75)(285,125)
\Vertex(285,125){1.8}
\Gluon(285,125)(240,130){-3}{5}
\Gluon(285,75)(271.5,73.5){3}{1}
\Gluon(253.5,71.5)(240,70){3}{1}
\CCirc(262.5,72.5){10}{Black}{Gray}
\Text(240,100)[r]{($i$)}
\Text(15,15)[r]{with}
\Gluon(30,15)(45,15){-2.5}{1}
\Gluon(65,15)(80,15){-2.5}{1}
\COval(55,15)(10,10)(0){Black}{Gray}
\Text(105,15)[r]{=}
\Gluon(125,15)(140,15){-2.5}{1}
\Gluon(160,15)(175,15){-2.5}{1}
\DashArrowArcn(150,15)(10,0,180){3}
\DashArrowArcn(150,15)(10,180,360){3}
\Vertex(140,15){1.6}
\Vertex(160,15){1.6}
\Text(200,15)[r]{+}
\Gluon(210,15)(235,15){-2.5}{2}
\Gluon(235,15)(260,15){-2.5}{2}
\DashArrowArcn(235,27)(12,270,-90){3}
\Vertex(235,15){1.6}
\end{picture}
\caption{NLO contributions to the $gg$ initiated processes with
  virtual squarks (dashed lines correspond to squark fields).}
\label{virMSgg}
\end{center}
\end{figure}
\begin{figure}[tbp]
\begin{center}
\begin{picture}(330,370)(0,0)
\SetColor{Black}
\Gluon(90,310)(65,320){2.5}{3}
\Line(90,310)(65,320)
\Vertex(65,320){1.8}
\Gluon(65,320)(65,360){2.5}{4}
\Line(65,320)(65,360)
\Vertex(65,360){1.8}
\Gluon(90,370)(65,360){-2.5}{3}
\Line(90,370)(65,360)
\Gluon(25,320)(65,320){-3}{4}
\Gluon(65,360)(25,360){-3}{4}
\ArrowLine(0,310)(25,320)
\Vertex(25,320){1.8}
\ArrowLine(25,320)(25,360)
\Vertex(25,360){1.8}
\ArrowLine(25,360)(0,370)
\Text(0,340)[r]{($a$)}
\Gluon(210,310)(195,325){2.5}{2}
\Vertex(195,325){1.8}
\Gluon(195,325)(180,340){2.5}{2}
\Line(210,310)(180,340)
\Gluon(210,370)(195,355){-2.5}{2}
\Vertex(195,355){1.8}
\Gluon(195,355)(180,340){-2.5}{2}
\Line(210,370)(180,340)
\GlueArc(180,340)(21.2,315,405){3}{3}
\Vertex(180,340){1.8}
\Gluon(180,340)(140,340){3}{4}
\Vertex(140,340){1.8}
\ArrowLine(120,310)(140,340)
\ArrowLine(140,340)(120,370)
\Text(120,340)[r]{($b$)}
\Gluon(330,310)(315,325){2.5}{2}
\Line(330,310)(315,325)
\Vertex(315,325){1.8}
\Gluon(330,370)(315,355){-2.5}{2}
\Line(330,370)(315,355)
\Vertex(315,355){1.8}
\Gluon(315,325)(315,355){2.5}{3}
\Line(315,325)(315,355)
\Gluon(315,355)(290,340){-3}{3}
\Gluon(315,325)(290,340){3}{3}
\Vertex(290,340){1.8}
\Gluon(290,340)(260,340){3}{3}
\Vertex(260,340){1.8}
\ArrowLine(240,310)(260,340)
\ArrowLine(260,340)(240,370)
\Text(240,340)[r]{($c$)}
\Gluon(90,230)(70,260){2.5}{4}
\Line(90,230)(70,260)
\Gluon(90,290)(70,260){-2.5}{4}
\Line(90,290)(70,260)
\Vertex(70,260){1.8}
\Gluon(70,260)(30,260){3}{4}
\Vertex(30,260){1.8}
\ArrowLine(0,230)(15,245)
\Vertex(15,245){1.8}
\ArrowLine(15,245)(30,260)
\GlueArc(30,260)(21.2,135,225){3}{3}
\ArrowLine(30,260)(15,275)
\Vertex(15,275){1.8}
\ArrowLine(15,275)(0,290)
\Text(0,260)[r]{($d$)}
\Gluon(210,230)(190,260){2.5}{4}
\Line(210,230)(190,260)
\Gluon(210,290)(190,260){-2.5}{4}
\Line(210,290)(190,260)
\Vertex(190,260){1.8}
\Gluon(190,260)(160,260){3}{3}
\Vertex(160,260){1.8}
\Gluon(160,260)(135,245){3}{3}
\Gluon(160,260)(135,275){-3}{3}
\ArrowLine(120,230)(135,245)
\Vertex(135,245){1.8}
\ArrowLine(135,245)(135,275)
\Vertex(135,275){1.8}
\ArrowLine(135,275)(120,290)
\Text(120,260)[r]{($e$)}
\Gluon(330,230)(310,260){2.5}{4}
\Line(330,230)(310,260)
\Gluon(330,290)(310,260){-2.5}{4}
\Line(330,290)(310,260)
\Vertex(310,260){1.8}
\Gluon(310,260)(270,260){3}{4}
\Vertex(270,260){1.8}
\ArrowLine(240,230)(270,260)
\ArrowLine(270,260)(240,290)
\Vertex(262.5,267.5){1.8}
\Vertex(247.5,282.5){1.8}
\GlueArc(255,275)(10.6,315,495){3}{3}
\Text(240,260)[r]{($f$)}
\Gluon(90,150)(82.5,157.5){2.5}{1}
\Line(90,150)(82.5,157.5)
\Vertex(82.5,157.5){1.8}
\Gluon(82.5,157.5)(67.5,172.5){2.5}{2}
\Line(82.5,157.5)(67.5,172.5)
\Vertex(67.5,172.5){1.8}
\Gluon(67.5,172.5)(60,180){2.5}{1}
\Line(67.5,172.5)(60,180)
\Gluon(90,210)(60,180){-2.5}{4}
\Line(90,210)(60,180)
\Vertex(60,180){1.8}
\Gluon(60,180)(20,180){3}{4}
\Vertex(20,180){1.8}
\ArrowLine(0,150)(20,180)
\ArrowLine(20,180)(0,210)
\GlueArc(75,165)(9,315,495){3}{3}
\Text(0,180)[r]{($g$)}
\Gluon(210,150)(190,180){2.5}{4}
\Line(210,150)(190,180)
\Gluon(210,210)(190,180){-2.5}{4}
\Line(210,210)(190,180)
\Vertex(190,180){1.8}
\Gluon(190,180)(175,180){3}{1}
\Vertex(175,180){1.8}
\ArrowArc(165,180)(10,0,180)
\ArrowArc(165,180)(10,180,360)
\Vertex(155,180){1.8}
\Gluon(155,180)(140,180){3}{1}
\Vertex(140,180){1.8}
\ArrowLine(120,150)(140,180)
\ArrowLine(140,180)(120,210)
\Text(120,180)[r]{($h$)}
\Gluon(330,150)(310,180){2.5}{4}
\Line(330,150)(310,180)
\Gluon(330,210)(310,180){-2.5}{4}
\Line(330,210)(310,180)
\Vertex(310,180){1.8}
\Gluon(310,180)(295,180){3}{1}
\CCirc(285,180){10}{Black}{Gray}
\Gluon(275,180)(260,180){3}{1}
\Vertex(260,180){1.8}
\ArrowLine(240,150)(260,180)
\ArrowLine(260,180)(240,210)
\Text(240,180)[r]{($i$)}
\Gluon(90,70)(70,100){2.5}{4}
\Line(90,70)(70,100)
\Gluon(90,130)(70,100){-2.5}{4}
\Line(90,130)(70,100)
\Vertex(70,100){1.8}
\Gluon(70,100)(55,100){3}{1}
\Vertex(55,100){1.8}
\GlueArc(45,100)(10,0,180){2.5}{3}
\CArc(45,100)(10,0,360)
\GlueArc(45,100)(10,180,360){2.5}{3}
\Vertex(35,100){1.8}
\Gluon(35,100)(20,100){3}{1}
\Vertex(20,100){1.8}
\ArrowLine(0,70)(20,100)
\ArrowLine(20,100)(0,130)
\Text(0,100)[r]{($j$)}
\Text(15,15)[r]{with}
\Gluon(30,15)(45,15){-2.5}{1}
\Gluon(65,15)(80,15){-2.5}{1}
\COval(55,15)(10,10)(0){Black}{Gray}
\Text(100,15)[r]{=}
\Gluon(105,15)(120,15){-2.5}{1}
\Gluon(140,15)(155,15){-2.5}{1}
\GlueArc(130,15)(10,0,180){2.5}{4}
\GlueArc(130,15)(10,180,360){2.5}{4}
\Vertex(120,15){1.6}
\Vertex(140,15){1.6}
\Text(175,15)[r]{+}
\Gluon(180,15)(195,15){-2.5}{1}
\Gluon(215,15)(230,15){-2.5}{1}
\DashCArc(205,15)(10,0,180){1.4}
\DashCArc(205,15)(10,180,360){1.4}
\Vertex(195,15){1.6}
\Vertex(215,15){1.6}
\Text(250,15)[r]{+}
\Gluon(255,15)(280,15){-2.5}{2}
\Gluon(280,15)(305,15){-2.5}{2}
\GlueArc(280,27)(12,-90,270){2.5}{10}
\Vertex(280,15){1.6}
\end{picture}
\caption{NLO contributions to the $q\overline{q}$ initiated processes
  without virtual squarks.}
\label{virOSqq}
\end{center}
\end{figure}
\begin{figure}[tbp]
\begin{center}
\begin{picture}(330,540)(0,0)
\SetColor{Black}
\Text(0,510)[r]{($a$)}
\Line(90,480)(65,490)
\Gluon(90,480)(65,490){2.5}{3}
\Line(90,540)(65,530)
\Gluon(90,540)(65,530){-2.5}{3}
\Vertex(65,490){1.8}
\Vertex(65,530){1.8}
\ArrowLine(65,490)(65,530)
\DashArrowLine(65,530)(25,530){3}
\Line(25,530)(25,490)
\Gluon(25,530)(25,490){-2.5}{5}
\DashArrowLine(25,490)(65,490){3}
\Vertex(25,490){1.8}
\Vertex(25,530){1.8}
\ArrowLine(0,480)(25,490)
\ArrowLine(25,530)(0,540)
\Text(120,510)[r]{($b$)}
\Line(210,480)(185,490)
\Gluon(210,480)(185,490){2.5}{3}
\Line(210,540)(185,530)
\Gluon(210,540)(185,530){-2.5}{3}
\Vertex(185,490){1.8}
\Vertex(185,530){1.8}
\DashArrowLine(185,490)(185,530){3}
\ArrowLine(185,530)(145,530)
\Gluon(145,530)(145,490){-3}{5}
\ArrowLine(145,490)(185,490)
\Vertex(145,490){1.8}
\Vertex(145,530){1.8}
\ArrowLine(120,480)(145,490)
\ArrowLine(145,530)(120,540)
\Text(240,510)[r]{($c$)}
\Line(330,480)(305,490)
\Gluon(330,480)(305,490){2.5}{3}
\Line(330,540)(305,530)
\Gluon(330,540)(305,530){-2.5}{3}
\Vertex(305,490){1.8}
\Vertex(305,530){1.8}
\Gluon(305,490)(305,530){-3}{5}
\Line(305,530)(265,530)
\Gluon(305,530)(265,530){-2.5}{5}
\DashArrowLine(265,490)(265,530){3}
\Line(265,490)(305,490)
\Gluon(265,490)(305,490){-2.5}{5}
\Vertex(265,490){1.8}
\Vertex(265,530){1.8}
\ArrowLine(240,480)(265,490)
\ArrowLine(265,530)(240,540)
\Text(0,430)[r]{($d$)}
\Line(90,400)(65,410)
\Gluon(90,400)(65,410){2.5}{3}
\Line(90,460)(25,450)
\Gluon(90,460)(25,450){-2.5}{7}
\Vertex(65,410){1.8}
\Vertex(65,450){1.8}
\Gluon(65,410)(65,450){-3}{5}
\ArrowLine(25,450)(65,450)
\DashArrowLine(25,410)(25,450){3}
\Line(25,410)(65,410)
\Gluon(25,410)(65,410){-2.5}{5}
\Vertex(25,410){1.8}
\Vertex(25,450){1.8}
\ArrowLine(0,400)(25,410)
\Line(65,450)(32.5,455)
\ArrowLine(32.5,455)(0,460)
\Text(120,430)[r]{($e$)}
\Line(210,400)(185,410)
\Gluon(210,400)(185,410){2.5}{3}
\Line(210,460)(145,450)
\Gluon(210,460)(145,450){-2.5}{7}
\Vertex(185,410){1.8}
\Vertex(185,450){1.8}
\DashArrowLine(185,410)(185,450){3}
\Line(145,450)(185,450)
\Gluon(145,450)(185,450){2.5}{5}
\Gluon(145,410)(145,450){3}{5}
\ArrowLine(145,410)(185,410)
\Vertex(145,410){1.8}
\Vertex(145,450){1.8}
\ArrowLine(120,400)(145,410)
\Line(185,450)(152.5,455)
\ArrowLine(152.5,455)(120,460)
\Text(240,430)[r]{($f$)}
\Line(330,400)(310,410)
\Gluon(330,400)(310,410){2.5}{3}
\Line(330,460)(310,450)
\Gluon(330,460)(310,450){-2.5}{3}
\Vertex(310,410){1.8}
\Vertex(310,450){1.8}
\DashArrowLine(310,410)(310,450){3}
\ArrowLine(310,450)(278,430)
\ArrowLine(278,430)(310,410)
\Vertex(278,430){1.8}
\Gluon(278,430)(260,430){-3}{2}
\Vertex(260,430){1.8}
\ArrowLine(240,400)(260,430)
\ArrowLine(260,430)(240,460)
\Text(0,350)[r]{($g$)}
\Line(90,320)(70,330)
\Gluon(90,320)(70,330){2.5}{3}
\Line(90,380)(70,370)
\Gluon(90,380)(70,370){-2.5}{3}
\Vertex(70,330){1.8}
\Vertex(70,370){1.8}
\ArrowLine(70,330)(70,370)
\DashArrowLine(70,370)(38,350){3}
\DashArrowLine(38,350)(70,330){3}
\Vertex(38,350){1.8}
\Gluon(38,350)(20,350){-3}{2}
\Vertex(20,350){1.8}
\ArrowLine(0,320)(20,350)
\ArrowLine(20,350)(0,380)
\Text(120,350)[r]{($h$)}
\Line(210,320)(190,350)
\Gluon(210,320)(190,350){2.5}{4}
\Line(210,380)(190,350)
\Gluon(210,380)(190,350){-2.5}{4}
\Vertex(190,350){1.8}
\Gluon(190,350)(172,350){-3}{2}
\Vertex(172,350){1.8}
\Line(172,350)(140,370)
\Gluon(172,350)(140,370){-2.5}{4}
\Line(172,350)(140,330)
\Gluon(172,350)(140,330){2.5}{4}
\DashArrowLine(140,330)(140,370){3}
\Vertex(140,330){1.8}
\Vertex(140,370){1.8}
\ArrowLine(140,370)(120,380)
\ArrowLine(120,320)(140,330)
\Text(240,350)[r]{($i$)}
\Line(330,320)(310,350)
\Gluon(330,320)(310,350){2.5}{4}
\Line(330,380)(310,350)
\Gluon(330,380)(310,350){-2.5}{4}
\Vertex(310,350){1.8}
\Gluon(310,350)(292,350){-3}{2}
\Vertex(292,350){1.8}
\DashArrowLine(292,350)(260,370){3}
\DashArrowLine(260,330)(292,350){3}
\Line(260,330)(260,370)
\Gluon(260,330)(260,370){2.5}{4}
\Vertex(260,330){1.8}
\Vertex(260,370){1.8}
\ArrowLine(260,370)(240,380)
\ArrowLine(240,320)(260,330)
\Text(0,270)[r]{($j$)}
\Line(90,300)(65,295)
\Gluon(90,300)(65,295){-2.5}{3}
\ArrowLine(25,295)(0,300)
\Vertex(65,295){1.8}
\Vertex(25,295){1.8}
\ArrowLine(45,266)(65,295)
\DashArrowLine(65,295)(25,295){3}
\Line(25,295)(45,266)
\Gluon(25,295)(45,266){-2.5}{4}
\Vertex(45,266){1.8}
\DashArrowLine(45,250)(45,266){3}
\Vertex(45,250){1.8}
\ArrowLine(0,240)(45,250)
\Line(90,240)(45,250)
\Gluon(90,240)(45,250){2.5}{5}
\Text(120,270)[r]{($k$)}
\Line(210,300)(185,295)
\Gluon(210,300)(185,295){-2.5}{3}
\ArrowLine(145,295)(120,300)
\Vertex(185,295){1.8}
\Vertex(145,295){1.8}
\Line(165,266)(185,295)
\Gluon(165,266)(185,295){-2.5}{4}
\Gluon(185,295)(145,295){-3}{4}
\ArrowLine(165,266)(145,295)
\Vertex(165,266){1.8}
\DashArrowLine(165,250)(165,266){3}
\Vertex(165,250){1.8}
\ArrowLine(120,240)(165,250)
\Line(210,240)(165,250)
\Gluon(210,240)(165,250){2.5}{5}
\Text(240,270)[r]{($l$)}
\Line(330,300)(305,295)
\Gluon(330,300)(305,295){-2.5}{3}
\ArrowLine(265,295)(240,300)
\Vertex(305,295){1.8}
\Vertex(265,295){1.8}
\DashArrowLine(285,266)(305,295){3}
\ArrowLine(305,295)(265,295)
\Gluon(265,295)(285,266){-3}{4}
\Vertex(285,266){1.8}
\DashArrowLine(285,250)(285,266){3}
\Vertex(285,250){1.8}
\ArrowLine(240,240)(285,250)
\Line(330,240)(285,250)
\Gluon(330,240)(285,250){2.5}{5}
\Text(0,190)[r]{($m$)}
\Line(90,220)(65,215)
\Gluon(90,220)(65,215){-2.5}{3}
\ArrowLine(25,215)(0,220)
\Vertex(65,215){1.8}
\Vertex(25,215){1.8}
\Gluon(45,186)(65,215){-3}{4}
\Line(65,215)(25,215)
\Gluon(65,215)(25,215){-2.5}{4}
\DashArrowLine(45,186)(25,215){3}
\Vertex(45,186){1.8}
\DashArrowLine(45,170)(45,186){3}
\Vertex(45,170){1.8}
\ArrowLine(0,160)(45,170)
\Line(90,160)(45,170)
\Gluon(90,160)(45,170){2.5}{5}
\Text(120,190)[r]{($n$)}
\Line(210,220)(190,190)
\Gluon(210,220)(190,190){-2.5}{4}
\Line(210,160)(190,190)
\Gluon(210,160)(190,190){2.5}{4}
\Vertex(190,190){1.8}
\Gluon(190,190)(165,190){3}{3}
\DashArrowArcn(165,205)(15,630,270){3}
\Vertex(165,190){1.8}
\Gluon(165,190)(140,190){3}{3}
\Vertex(140,190){1.8}
\ArrowLine(120,160)(140,190)
\ArrowLine(140,190)(120,220)
\Text(240,190)[r]{($o$)}
\Line(330,220)(310,190)
\Gluon(330,220)(310,190){-2.5}{4}
\Line(330,160)(310,190)
\Gluon(330,160)(310,190){2.5}{4}
\Vertex(310,190){1.8}
\Gluon(310,190)(300,190){3}{1}
\Vertex(300,190){1.8}
\DashArrowArcn(285,190)(15,0,180){3}
\DashArrowArcn(285,190)(15,180,360){3}
\Vertex(270,190){1.8}
\Gluon(270,190)(260,190){3}{1}
\Vertex(260,190){1.8}
\ArrowLine(240,160)(260,190)
\ArrowLine(260,190)(240,220)
\Text(0,110)[r]{($p$)}
\Line(90,140)(45,140)
\Gluon(90,140)(45,140){-2.5}{5}
\ArrowLine(45,140)(0,140)
\Vertex(45,140){1.8}
\DashArrowLine(45,125)(45,140){3}
\Vertex(45,125){1.8}
\ArrowArc(45,110)(15,270,450)
\GlueArc(45,110)(15,90,270){2.5}{5}
\CArc(45,110)(15,90,270)
\Vertex(45,95){1.8}
\DashArrowLine(45,80)(45,95){3}
\Vertex(45,80){1.8}
\ArrowLine(0,80)(45,80)
\Line(90,80)(45,80)
\Gluon(90,80)(45,80){2.5}{5}
\Text(120,110)[r]{($q$)}
\Line(210,140)(165,140)
\Gluon(210,140)(165,140){-2.5}{5}
\ArrowLine(165,140)(120,140)
\Vertex(165,140){1.8}
\DashArrowLine(165,125)(165,140){3}
\Vertex(165,125){1.8}
\GlueArc(165,110)(15,90,270){3}{5}
\DashArrowArc(165,110)(15,270,90){3}
\Vertex(165,95){1.8}
\DashArrowLine(165,80)(165,95){3}
\Vertex(165,80){1.8}
\ArrowLine(120,80)(165,80)
\Line(210,80)(165,80)
\Gluon(210,80)(165,80){2.5}{5}
\Text(240,110)[r]{($r$)}
\Line(330,140)(285,130)
\Gluon(330,140)(285,130){-2.5}{5}
\ArrowLine(285,130)(240,140)
\Vertex(285,130){1.8}
\DashArrowLine(285,110)(285,130){3}
\DashArrowArcn(270,110)(15,360,0){3}
\Vertex(285,110){1.8}
\DashArrowLine(285,90)(285,110){3}
\Vertex(285,90){1.8}
\ArrowLine(240,80)(285,90)
\Line(330,80)(285,90)
\Gluon(330,80)(285,90){2.5}{5}
\Text(120,30)[r]{($t$)}
\Line(90,60)(70,40)
\Gluon(90,60)(70,40){-2.5}{3}
\Line(90,0)(70,20)
\Gluon(90,0)(70,20){2.5}{3}
\COval(60,30)(15,15)(0){Black}{Gray}
\ArrowLine(0,0)(50,20)
\ArrowLine(50,40)(40,44)
\Vertex(40,44){1.8}
\GlueArc(25,50)(16,-22,158){2.5}{5}
\DashArrowArc(25,50)(16,158,338){3}
\CArc(25,50)(16,-22,158)
\Vertex(10,56){1.8}
\ArrowLine(10,56)(0,60)
\Text(61,30)[c]{LO}
\Text(0,30)[r]{($s$)}
\Line(210,60)(200,56)
\Gluon(210,60)(200,56){-2.5}{1}
\Vertex(200,56){1.8}
\DashArrowArc(185,50)(16,202,382){3}
\ArrowArc(185,50)(16,382,562)
\Vertex(170,44){1.8}
\Line(170,44)(160,40)
\Gluon(170,44)(160,40){-2.5}{1}
\Line(210,0)(160,20)
\Gluon(210,0)(160,20){2.5}{8}
\COval(150,30)(15,15)(0){Black}{Gray}
\ArrowLine(120,0)(140,20)
\ArrowLine(140,40)(120,60)
\Text(151.5,30)[c]{LO}
\end{picture}
\caption{NLO contributions to the $q\overline{q}$ initiated processes
  with virtual squarks.}
\label{virMSqq}
\end{center}
\end{figure}
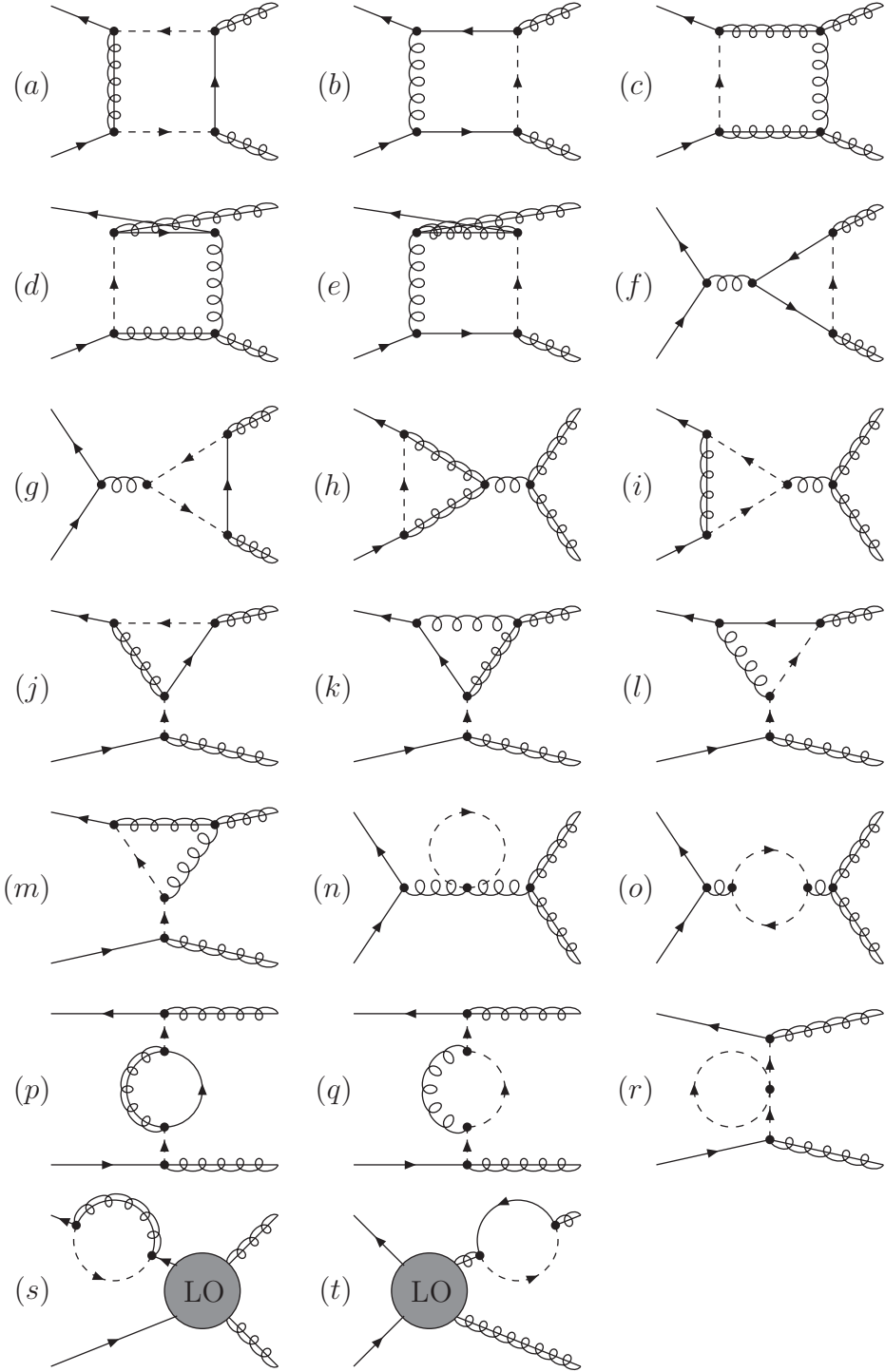
\par
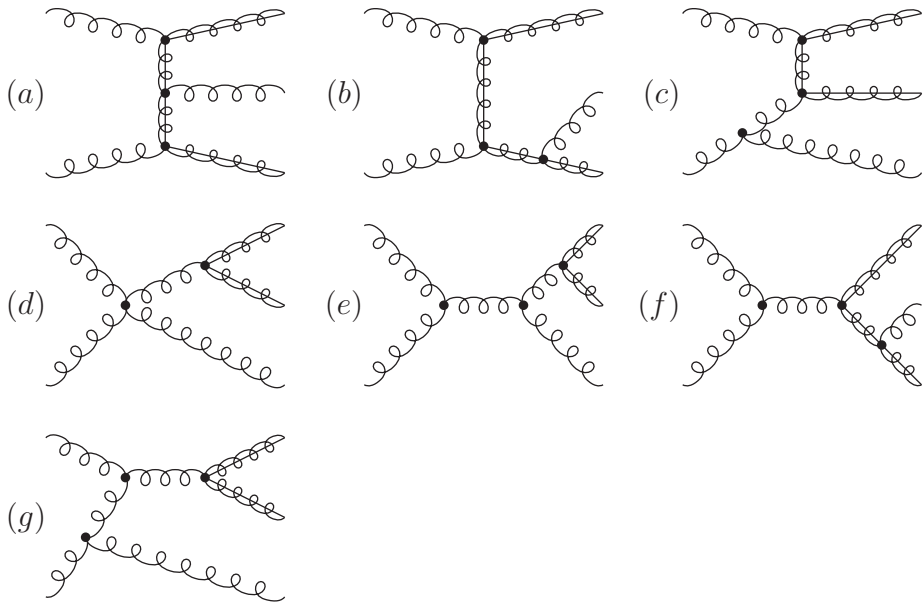
\begin{figure}[tbp]
\begin{center}
\begin{picture}(360,220)(0,0)
\SetColor{Black}
\Text(0,190)[r]{($a$)}
\Vertex(45,210){1.8}
\Gluon(45,210)(0,220){-3}{4}
\Vertex(45,190){1.8}
\Gluon(45,170)(0,160){3}{4}
\Line(45,190)(45,210)
\Gluon(45,190)(45,210){2.5}{2}
\Vertex(45,170){1.8}
\Gluon(45,210)(90,220){2.5}{4}
\Line(45,210)(90,220)
\Gluon(45,190)(45,170){-2.5}{2}
\Line(45,190)(45,170)
\Gluon(45,170)(90,160){-2.5}{4}
\Line(45,170)(90,160)
\Gluon(45,190)(90,190){3}{4}
\Text(120,190)[r]{($b$)}
\Vertex(165,210){1.8}
\Gluon(165,210)(165,170){-2.5}{4}
\Line(165,210)(165,170)
\Vertex(165,170){1.8}
\Line(210,220)(165,210)
\Gluon(210,220)(165,210){-2.5}{4}
\Line(187.5,165)(165,170)
\Gluon(187.5,165)(165,170){2.5}{2}
\Vertex(187.5,165){1.8}
\Line(210,160)(187.5,165)
\Gluon(210,160)(187.5,165){2.5}{2}
\Gluon(165,210)(120,220){-3}{4}
\Gluon(165,170)(120,160){3}{4}
\Gluon(187.5,165)(210,190){3}{3}
\Text(240,190)[r]{($c$)}
\Gluon(240,160)(262.5,175){-3}{2}
\Vertex(262.5,175){1.8}
\Gluon(262.5,175)(285,190){-3}{2}
\Gluon(240,220)(285,210){3}{4}
\Vertex(285,210){1.8}
\Gluon(262.5,175)(330,160){-3}{7}
\Line(285,210)(330,220)
\Gluon(285,210)(330,220){2.5}{4}
\Line(285,190)(330,190)
\Gluon(285,190)(330,190){-2.5}{4}
\Vertex(285,190){1.8}
\Line(285,190)(285,210)
\Gluon(285,190)(285,210){2.5}{2}
\Text(0,110)[r]{($d$)}
\Vertex(30,110){1.8}
\Gluon(30,110)(0,140){-3}{4}
\Gluon(30,110)(0,80){3}{4}
\Gluon(30,110)(90,80){-3}{7}
\Vertex(60,125){1.8}
\Gluon(60,125)(90,140){2.5}{3}
\Line(60,125)(90,140)
\Gluon(60,125)(90,110){-2.5}{3}
\Line(60,125)(90,110)
\Gluon(30,110)(60,125){3}{3}
\Text(120,110)[r]{($e$)}
\Gluon(120,140)(150,110){3}{4}
\Gluon(120,80)(150,110){-3}{4}
\Vertex(150,110){1.8}
\Gluon(150,110)(180,110){3}{3}
\Vertex(180,110){1.8}
\Gluon(180,110)(195,125){3}{2}
\Gluon(180,110)(210,80){-3}{4}
\Vertex(195,125){1.8}
\Line(195,125)(210,140)
\Gluon(195,125)(210,140){2.5}{2}
\Line(195,125)(210,110)
\Gluon(195,125)(210,110){-2.5}{2}
\Text(240,110)[r]{($f$)}
\Gluon(240,80)(270,110){-3}{4}
\Gluon(240,140)(270,110){3}{4}
\Vertex(270,110){1.8}
\Gluon(270,110)(300,110){3}{3}
\Vertex(300,110){1.8}
\Line(300,110)(330,140)
\Gluon(300,110)(330,140){2.5}{4}
\Line(300,110)(315,95)
\Gluon(300,110)(315,95){-2.5}{2}
\Vertex(315,95){1.8}
\Line(315,95)(330,80)
\Gluon(315,95)(330,80){-2.5}{2}
\Gluon(315,95)(330,110){3}{2}
\Text(0,30)[r]{($g$)}
\Gluon(0,0)(15,22.5){-3}{2}
\Vertex(15,22.5){1.8}
\Gluon(15,22.5)(30,45){-3}{2}
\Gluon(0,60)(30,45){3}{3}
\Vertex(30,45){1.8}
\Gluon(30,45)(60,45){3}{3}
\Vertex(60,45){1.8}
\Line(60,45)(90,60)
\Gluon(60,45)(90,60){2.5}{4}
\Line(60,45)(90,30)
\Gluon(60,45)(90,30){-2.5}{4}
\Gluon(15,22.5)(90,0){-3}{7}
\end{picture}
\caption{NLO contributions to the $gg$ initiated processes.}
\label{realgg}
\end{center}
\end{figure}

The quantities $\mathcal{R}_{ij}^{[T]}$ result from the real corrections
to the processes and contain the subtraction terms which originate from
the renormalization of the PDFs in the $\overline{\mbox{MS}}$
scheme. Thus the remaining infra-red divergences exactly cancel the ones
present in the virtual corrections of Eq.~(\ref{virtual}).  Let us start
with the process $gg\rightarrow Tg$ which does not involve squarks
(see Fig.~\ref{realgg}). For the individual colour configurations we
obtain
\begin{subequations}
\begin{eqnarray}
  \mathcal{R}_{gg}^{[1_{s}]}(z)&=&\mathcal{R}_{gg}(z)+\frac{11z^5+11z^4+13z^3+19z^2+6z-12}{2z(1+z)^2}
  \nonumber\\
  &&+\frac{3}{1-z}\left[\frac{2z\left(z^3-z+2\right)\left(z^3-2z^2-3z-2\right)}{(1+z)^3}\frac{\ln(z)}{1-z}-3\right]\,,
  \\
  &&
  \nonumber\\
  \mathcal{R}_{gg}^{[8_{s}]}(z)&=&\mathcal{R}_{gg}(z)+3\left\{\delta(1-z)\left(\frac{1}{2\varepsilon_{\mbox{\tiny{IR}}}}+1\right)-\left[\frac{1}{1-z}\right]_{+}\right\}
  \nonumber\\
  &&+\frac{23z^5+29z^4+43z^3+43z^2+18z-12}{2z(1+z)^2}
  \nonumber\\
  &&+\frac{3}{1-z}\left[\frac{2z\left(z^6-2z^5-6z^4+2z^3-3z^2-4z-4\right)}{(1+z)^3}\frac{\ln(z)}{1-z}-4\right]\,,
  \\
  \mathcal{R}_{gg}^{[8_{a}]}(z)&=&\frac{4(1-z)}{z(1-z^2)^3}\Bigl[\,(1-z^2)(21z^5+88z^4+42z^3+92z^2+17z+12)
  \nonumber\\
  &&\hspace{4.cm}+4z(11z^5+25z^4+64z^3+12z^2+21z+3)\ln(z)\Bigr]\,,
  \nonumber\\
  &&\\
\mathcal{R}_{gg}^{[10]}(z)&=&\frac{80z(1-z)}{(1-z^2)^3}\Bigl[(1-z^2)(z^3+4z^2+z+2)+2z^2(z^2+2z+5)\ln(z)\Bigr]\,,
\end{eqnarray}
\end{subequations}
\setcounter{equation}{21}
\begin{subequations}
\setcounter{equation}{4}
\begin{eqnarray}
  \mathcal{R}_{gg}^{[27_{s}]}(z)&=&\mathcal{R}_{gg}(z)+8\left\{\delta(1-z)\left(\frac{1}{2\varepsilon_{\mbox{\tiny{IR}}}}+1\right)-\left[\frac{1}{1-z}\right]_{+}\right\}
  \nonumber\\
  &&+\frac{43z^5+59z^4+93z^3+83z^2+38z-12}{2z(1+z)^2}
  \nonumber\\
  &&+\frac{1}{1-z}\left[\frac{2z\left(3z^6-6z^5-28z^4+6z^3-19z^2-12z-12\right)}{(1+z)^3}\frac{\ln(z)}{1-z}-17\right]
  \,,
  \nonumber\\
\end{eqnarray}
  \label{realGG}
\end{subequations}
with
\begin{eqnarray}
\mathcal{R}_{gg}(z)&=&(1-z)\mathcal{P}_{gg}(z)\left\{2\left[\frac{\ln(1-z)}{1-z}\right]_{+}-\left[\frac{1}{1-z}\right]_{+}\ln\left(\frac{\mu_F^2}{M^2}\right)\right\}+\biggl\{\frac{3}{\varepsilon_{\mbox{\tiny{IR}}}^2}+\frac{11}{2\varepsilon_{\mbox{\tiny{IR}}}}
\nonumber\\
&&-\frac{n_{f}-1}{3\varepsilon_{\mbox{\tiny{IR}}}}-\left(\frac{11}{2}-\frac{n_f-1}{3}\right)\ln\left(\frac{\mu_F^2}{M^2}\right)-\pi^2\biggr\}f_{\varepsilon}(M^2)\,\delta(1-z)
\,,
\label{realGG2}
\end{eqnarray}
where $n_f=6$. The functions $\mathcal{P}_{ij}$ are listed at the end of
this Section. The conventional plus-distribution\footnote{The
plus-distribution follows the prescription
$\int_0^1dz\,\left[\frac{\ln^n(1-z)}{1-z}\right]_{+}f(z)
\equiv\int_0^1dz\,\frac{\ln^n(1-z)}{1-z}\left[f(z)-f(1)\right]$ for
$n=0,1,\ldots$ and any test function $f(z)$. If the lower integration
boundary is given by $1>\rho>0$ the plus distribution can be replaced by
the $\rho$-description via
$\left[\frac{\ln^n(1-z)}{1-z}\right]_{+}\rightarrow\frac{\ln^{n+1}(1-\rho)}{n+1}\delta(1-z)+\left[\frac{\ln^n(1-z)}{1-z}\right]_{\rho}$
where the latter is defined through
$\int_\rho^1dz\,\left[\frac{\ln^n(1-z)}{1-z}\right]_{\rho}f(z)
\equiv\int_\rho^1dz\,\frac{\ln^n(1-z)}{1-z} \left[f(z)-f(1)\right]$.}
is employed to regularize the singularity at $z=1$. Note, that the cross
section for the decuplet configuration (plus a gluon) is non-vanishing,
albeit small.
\begin{figure}[tbp]
\begin{center}
\begin{picture}(330,220)(0,0)
\SetColor{Black}
\Text(0,190)[r]{($a$)}
\ArrowLine(0,220)(45,210)
\ArrowLine(45,210)(90,220)
\Vertex(45,210){1.8}
\Gluon(45,210)(45,190){-3}{2}
\Vertex(45,190){1.8}
\Line(45,190)(45,170)
\Gluon(45,190)(45,170){-2.5}{2}
\Vertex(45,170){1.8}
\Gluon(0,160)(45,170){-3}{4}
\Gluon(45,190)(90,190){2.5}{4}
\Line(45,190)(90,190)
\Gluon(45,170)(90,160){-2.5}{4}
\Line(45,170)(90,160)
\Text(120,190)[r]{($b$)}
\ArrowLine(120,220)(165,210)
\ArrowLine(165,210)(210,220)
\Vertex(165,210){1.8}
\Gluon(165,210)(165,170){-3}{4}
\Gluon(120,160)(165,170){-3}{4}
\Vertex(165,170){1.8}
\Gluon(165,170)(187.5,165){-3}{2}
\Vertex(187.5,165){1.8}
\Gluon(187.5,165)(210,190){2.5}{3}
\Line(187.5,165)(210,190)
\Gluon(187.5,165)(210,160){-2.5}{2}
\Line(187.5,165)(210,160)
\Text(240,190)[r]{($c$)}
\Gluon(240,160)(270,190){-3}{4}
\ArrowLine(240,220)(270,190)
\Vertex(270,190){1.8}
\ArrowLine(270,190)(300,190)
\Vertex(300,190){1.8}
\ArrowLine(300,190)(330,220)
\Gluon(300,190)(315,175){-3}{2}
\Vertex(315,175){1.8}
\Line(315,175)(330,160)
\Gluon(315,175)(330,160){-2.5}{2}
\Line(315,175)(330,190)
\Gluon(315,175)(330,190){2.5}{2}
\Text(0,110)[r]{($d$)}
\Vertex(45,130){1.8}
\ArrowLine(0,140)(45,130)
\Vertex(45,90){1.8}
\Gluon(45,90)(0,80){3}{4}
\Line(45,90)(45,110)
\Gluon(45,90)(45,110){2.5}{2}
\Vertex(45,110){1.8}
\Line(45,130)(90,140)
\Gluon(45,130)(90,140){2.5}{4}
\DashArrowLine(45,130)(45,110){3}
\Gluon(45,90)(90,80){-2.5}{4}
\Line(45,90)(90,80)
\ArrowLine(45,110)(90,110)
\Text(120,110)[r]{($e$)}
\Vertex(165,130){1.8}
\DashArrowLine(165,130)(165,90){3}
\Vertex(165,90){1.8}
\Line(210,140)(165,130)
\Gluon(210,140)(165,130){-2.5}{4}
\DashArrowLine(165,90)(187.5,85){3}
\Vertex(187.5,85){1.8}
\Line(210,80)(187.5,85)
\Gluon(210,80)(187.5,85){2.5}{2}
\ArrowLine(120,140)(165,130)
\Gluon(165,90)(120,80){3}{4}
\ArrowLine(187.5,85)(210,110)
\Text(240,110)[r]{($f$)}
\Gluon(240,80)(270,110){-3}{4}
\ArrowLine(240,140)(270,110)
\Vertex(270,110){1.8}
\ArrowLine(270,110)(300,110)
\Vertex(300,110){1.8}
\Line(300,110)(330,140)
\Gluon(300,110)(330,140){2.5}{4}
\DashArrowLine(300,110)(315,95){3}
\Vertex(315,95){1.8}
\Line(315,95)(330,80)
\Gluon(315,95)(330,80){-2.5}{2}
\ArrowLine(315,95)(330,110)
\Text(0,30)[r]{($g$)}
\Line(45,10)(45,50)
\Gluon(45,10)(45,50){2.5}{4}
\Gluon(0,0)(45,10){-3}{4}
\Vertex(45,10){1.8}
\ArrowLine(0,60)(45,50)
\Vertex(45,50){1.8}
\DashArrowLine(45,50)(67.5,55){3}
\Line(45,10)(90,0)
\Gluon(45,10)(90,0){-2.5}{4}
\Vertex(67.5,55){1.8}
\ArrowLine(67.5,55)(90,30)
\Line(67.5,55)(90,60)
\Gluon(67.5,55)(90,60){2.5}{2}
\Text(120,30)[r]{($h$)}
\ArrowLine(120,60)(165,50)
\Vertex(165,50){1.8}
\Line(165,50)(210,60)
\Gluon(165,50)(210,60){2.5}{4}
\Gluon(120,0)(165,10){-3}{4}
\Vertex(165,10){1.8}
\DashArrowLine(165,50)(165,30){3}
\Vertex(165,30){1.8}
\ArrowLine(165,30)(165,10)
\Line(165,30)(210,30)
\Gluon(165,30)(210,30){-3}{4}
\ArrowLine(165,10)(210,0)
\end{picture}
\caption{NLO contributions from $gq$ initiated processes.}
\label{realgq}
\end{center}
\end{figure}
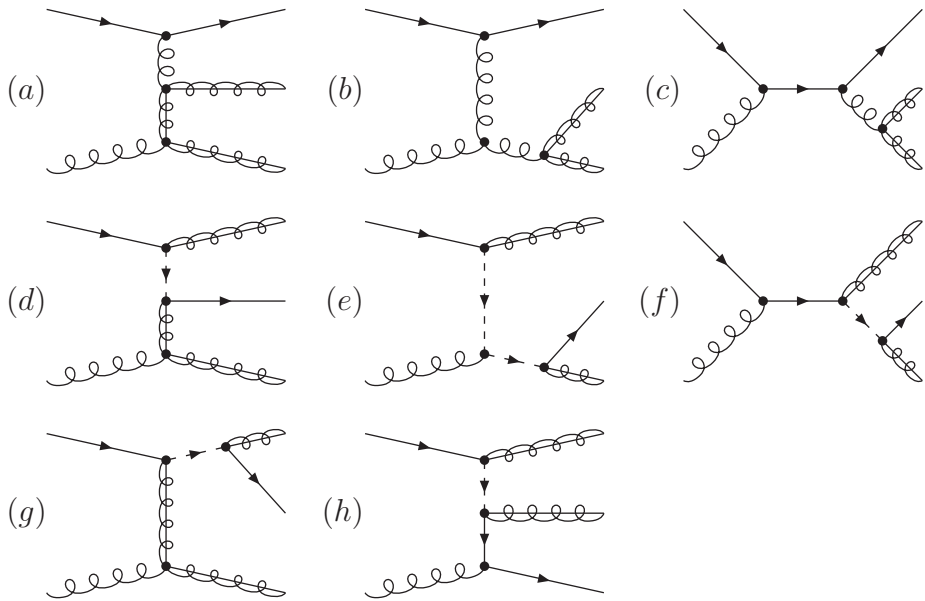
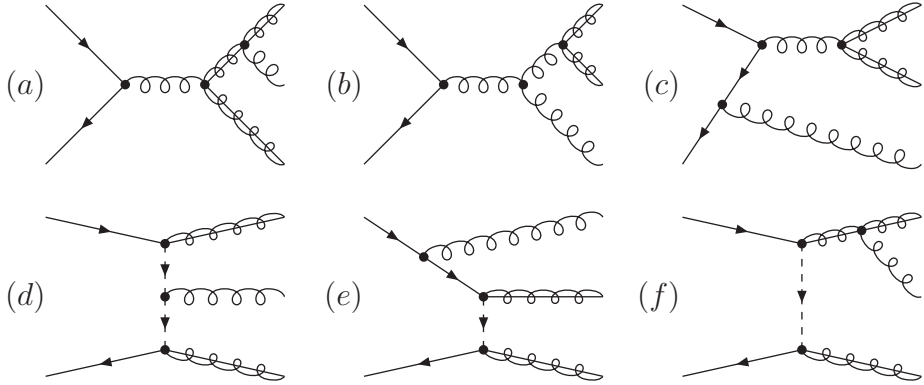
\begin{figure}[tbp]
\begin{center}
\begin{picture}(330,140)(0,0)
\SetColor{Black}
\Text(0,110)[r]{($a$)}
\ArrowLine(0,140)(30,110)
\ArrowLine(30,110)(0,80)
\Vertex(30,110){1.8}
\Gluon(30,110)(60,110){3}{3}
\Vertex(60,110){1.8}
\Line(60,110)(90,80)
\Gluon(60,110)(90,80){-2.5}{4}
\Line(60,110)(90,140)
\Gluon(60,110)(75,125){2.5}{2}
\Vertex(75,125){1.8}
\Gluon(75,125)(90,140){2.5}{2}
\Gluon(75,125)(90,110){-3}{2}
\Text(120,110)[r]{($b$)}
\ArrowLine(120,140)(150,110)
\ArrowLine(150,110)(120,80)
\Vertex(150,110){1.8}
\Gluon(150,110)(180,110){3}{3}
\Vertex(180,110){1.8}
\Gluon(180,110)(210,80){-3}{4}
\Gluon(180,110)(195,125){3}{2}
\Vertex(195,125){1.8}
\Line(195,125)(210,140)
\Gluon(195,125)(210,140){2.5}{2}
\Line(195,125)(210,110)
\Gluon(195,125)(210,110){-2.5}{2}
\Text(240,110)[r]{($c$)}
\ArrowLine(270,125)(255,102.5)
\Vertex(255,102.5){1.8}
\ArrowLine(255,102.5)(240,80)
\ArrowLine(240,140)(270,125)
\Vertex(270,125){1.8}
\Gluon(270,125)(300,125){3}{3}
\Vertex(300,125){1.8}
\Line(300,125)(330,140)
\Gluon(300,125)(330,140){2.5}{3}
\Line(300,125)(330,110)
\Gluon(300,125)(330,110){-2.5}{3}
\Gluon(255,102.5)(330,80){-3}{8}
\Text(0,30)[r]{($d$)}
\ArrowLine(0,60)(45,50)
\Vertex(45,50){1.8}
\DashArrowLine(45,50)(45,30){3}
\Vertex(45,30){1.8}
\DashArrowLine(45,30)(45,10){3}
\Vertex(45,10){1.8}
\ArrowLine(45,10)(0,0)
\Line(45,50)(90,60)
\Gluon(45,50)(90,60){2.5}{4}
\Line(45,10)(90,0)
\Gluon(45,10)(90,0){-2.5}{4}
\Gluon(45,30)(90,30){3}{4}
\Text(120,30)[r]{($e$)}
\ArrowLine(120,60)(142.5,45)
\Vertex(142.5,45){1.8}
\ArrowLine(142.5,45)(165,30)
\Vertex(165,30){1.8}
\DashArrowLine(165,30)(165,10){3}
\Vertex(165,10){1.8}
\ArrowLine(165,10)(120,0)
\Line(165,10)(210,0)
\Gluon(165,10)(210,0){-2.5}{4}
\Line(165,30)(210,30)
\Gluon(165,30)(210,30){2.5}{4}
\Gluon(142.5,45)(210,60){3}{7}
\Text(240,30)[r]{($f$)}
\ArrowLine(240,60)(285,50)
\Vertex(285,50){1.8}
\DashArrowLine(285,50)(285,10){3}
\Vertex(285,10){1.8}
\ArrowLine(285,10)(240,0)
\Line(285,50)(330,60)
\Gluon(285,50)(307.5,55){2.5}{2}
\Vertex(307.5,55){1.8}
\Gluon(307.5,55)(330,60){2.5}{2}
\Line(285,10)(330,0)
\Gluon(285,10)(330,0){-2.5}{4}
\Gluon(307.5,55)(330,30){-2.5}{3}
\end{picture}
\caption{NLO contributions to the $q\overline{q}$ initiated processes.}
\label{realqq}
\end{center}
\end{figure}
\par
The results for the $gq$ and $g\overline{q}$ initiated processes
$gq\rightarrow Tq$ and $g\overline{q}\rightarrow T\overline{q}$ are
identical and we only list the former. The corresponding function
$\mathcal{R}_{qg}^{[T]}$ can be split into a term
$\mathcal{H}_{gq}^{[T]}$, which originates from the diagrams depicted in
Fig.~\ref{realgq}(a)-(c), and is independent of the squark mass and a
second term $\mathcal{F}_{gq}^{[T]}$ from the diagrams depicted in
Fig.~\ref{realgq}(d)-(h) and their interference with those of
(a)-(c). The dependence on the factorization scale $\mu_F$ is contained
in $\mathcal{H}_{gq}^{[T]}$. One finds
\begin{subequations}
\begin{eqnarray}
  \mathcal{R}_{gq}^{[T]}(z)&=&\mathcal{H}_{gq}^{[T]}(z)+\mathcal{F}_{gq}^{[T]}(z,r)\,,
  \\
  \mathcal{H}_{gq}^{[T]}(z)&=&\mathcal{N}_{gg}^{[T]}\left\{-\frac{1}{2}\mathcal{P}_{gq}(z)\ln\left(\frac{\mu_F^2z}{M^2(1-z)^2}\right)-\frac{4(1-z)}{3z}\left[1-\ln(z)\right]+\frac{2z}{3}\right\}\,,
  \nonumber\\
  &&\mbox{for}\hspace{0.6cm}T\in\left\{1_{s},8_{s},27_{s}\right\}\,,
  \\
  \mathcal{H}_{gq}^{[8_{a}]}(z)&=&\mathcal{N}_{q\overline{q}}^{[8_a]}\biggl\{-\frac{1}{2}\mathcal{P}_{qg}(z)\ln\left(\frac{\mu_F^2z}{M^2(1-z)^2}\right)+\frac{9(z+1)}{8}\ln(z)
  \nonumber\\
  &&\hspace{1.2cm}+\frac{(1-z)(32z^2+11z+18)}{16z}\biggr\}\,,
  \\
  \mathcal{H}_{gq}^{[10]}(z)&=&0
  \,,
\end{eqnarray}
  \label{realGQ}
\end{subequations}

and $\mathcal{H}_{gq}^{[1_{s}]}$ of course coincides with the result of
Eq.~(12) of Ref.~\cite{Kauth:2009ud}. The functions
$\mathcal{F}_{gq}^{[T]}$ are listed in Appendix~\ref{app:results}, see
Eqs.~(\ref{Fgq1})-(\ref{Fgq27}). A similar decomposition can be made for
the $q\overline{q}$ initiated processes (cf. Fig. \ref{realqq})
\begin{subequations}
\begin{eqnarray}
  \mathcal{R}_{q\overline{q}}^{[T]}(z)&=&\mathcal{H}_{q\overline{q}}^{[T]}(z)+\mathcal{F}_{q\overline{q}}^{[T]}(z,r)\,,
  \\
  \mathcal{H}_{q\overline{q}}^{[T]}(z)&=&\mathcal{N}_{gg}^{[T]}\frac{32}{27}z(1-z)
  \nonumber\\
  &&\mbox{for}\hspace{0.6cm}T\in\left\{1_{s},8_{s},27_{s}\right\}\,,
  \\
  \mathcal{H}_{q\overline{q}}^{[8_{a}]}(z)&=&\left[\frac{4}{3\varepsilon_{\mbox{\tiny{IR}}}^2}+\frac{7}{2\varepsilon_{\mbox{\tiny{IR}}}}-2\ln\left(\frac{\mu_F^2}{M^2}\right)+3-\frac{4\pi^2}{9}\right]f_{\varepsilon}(M^2)\,\delta(1-z)
  \nonumber\\
  &&\hspace{0.05cm}+(1-z)\mathcal{P}_{qq}(z)\left(2\left[\frac{\ln(1-z)}{1-z}\right]_{+}-\ln\left(\frac{z\mu_F^2}{M^2}\right)\left[\frac{1}{1-z}\right]_{+}\right)
  \nonumber\\
  &&\hspace{.05cm}+z+2-3\left[\frac{1}{1-z}\right]_{+}\,,
  \\
  \mathcal{H}_{q\overline{q}}^{[10]}(z)&=&0
  \,,
\end{eqnarray}
  \label{realQQ}
\end{subequations}

and $\mathcal{F}_{q\overline{q}}^{[T]}$ is again listed in
Appendix~\ref{app:results}. The splitting functions entering the
previous expressions are defined as
{\allowdisplaybreaks
\begin{align}
  \mathcal{P}_{gg}(z)&=6\left(\frac{1}{z}+\frac{1}{1-z}+z(1-z)-2\right)\,,
  &&\mathcal{P}_{gq}(z)=\frac{4\left[1+(1-z)^2\right]}{3z}\,,
  \nonumber\\
  \mathcal{P}_{qg}(z)&=\frac{z^2+(1-z)^2}{2}\,,
  &&\mathcal{P}_{qq}(z)=\frac{8}{3}\left(\frac{1}{1-z}-\frac{1+z}{2}\right)
  \,.
  \label{split}
\end{align}
}
\section{\label{sec:numerics}Hadronic production}
As described in Eq.~(\ref{master}), the luminosity function
Eq.~(\ref{master2}) is convoluted with the partonic cross section, which in
turn is composed of the Green's function (Eq.~(\ref{Green})) and the
short distance corrections (Eqs.~(\ref{formulae}) and
(\ref{formulae2})). For the numerical evaluation we use the PDF set
MSTW2008NLO~\cite{Martin:2009iq} which corresponds to
$\alpha_s(M_Z)=0.1202$. (Note, that for the LO study a value
$\alpha_s(M_Z)=0.1394$ was adopted.)  From this value as starting point
we compute $\alpha_s^{\rm SQCD}$ employing the same procedure as
described before Eq.~(\ref{eq::FqqLO}). We thus arrive at the cross section
for proton-proton collisions, which is still differential in the mass of
the gluino pair.
\begin{figure}[tbp]
\begin{center}
\begin{tabular}{cc}
\mbox{(a)}&
\includegraphics[angle=270,width=0.8\textwidth]{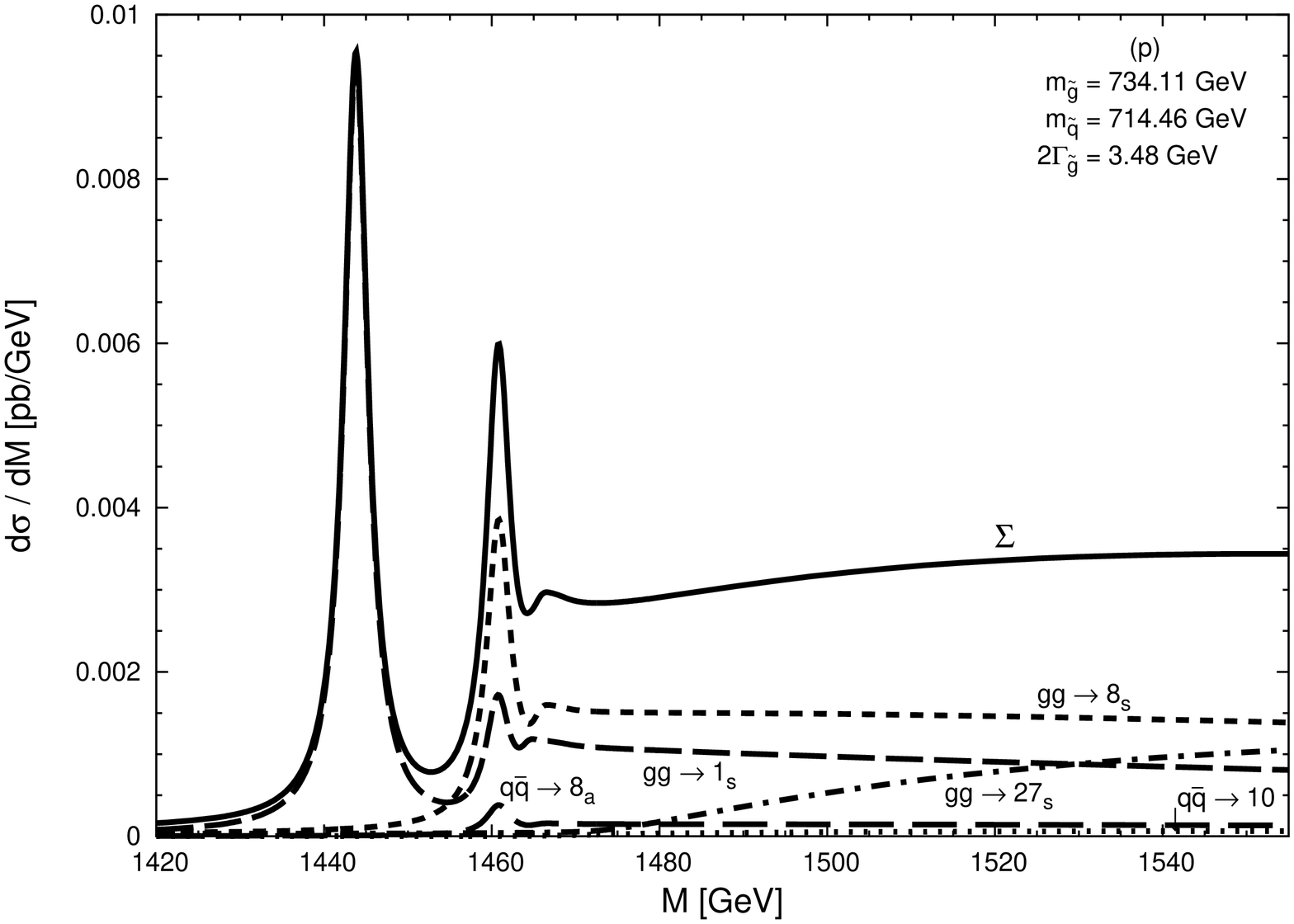}
\\
\mbox{(b)}&
\includegraphics[angle=270,width=0.8\textwidth]{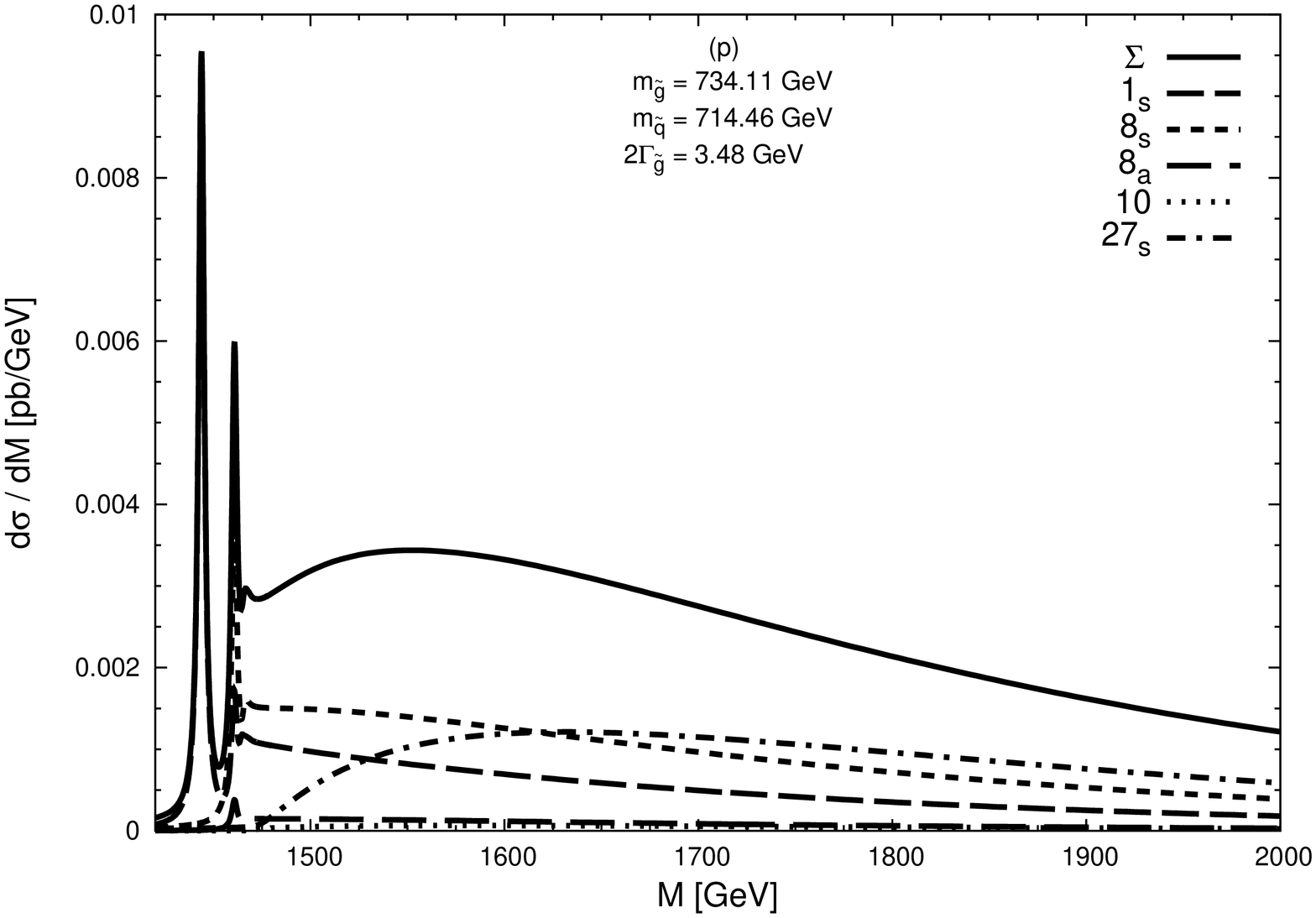}
\end{tabular}
\caption{NLO prediction for the differential cross section for scenario
  (p) for two different regions of $M$.}
\label{NLOp}
\end{center}
\end{figure}
In a first step we restrict ourselves to scenario (p) and display the
region very close to threshold (Fig.~\ref{NLOp}(a)) and a wider range
with $M_{\mbox{\tiny max}}=2000\,\mbox{GeV}$ corresponding to
$v_{\mbox{\tiny max}}^2=0.72$ (Fig.~\ref{NLOp}(b)). The contributions
from the different colour configurations are shown separately. Close to
threshold the strong enhancement of the singlet is contrasted with the
strong suppression of the twenty-seven representation. For larger $M$
and correspondingly larger relative velocities the final state
interaction becomes less relevant and about $200\,\mbox{GeV}$ above
threshold the twenty-seven representation starts to dominate. (This
behaviour is quite similar to the one of top pair production with
dominant singlet close to and dominant octet far above threshold.). The
process $q\overline{q}\rightarrow [8_a]$ is non-vanishing in Born
approximation. However, for squark and gluino masses being roughly equal
it is strongly suppressed. This suppression is also present at NLO. The
contribution from the decuplet, which is absent in Born approximation,
remains small throughout.
\begin{figure}[tbp]
\begin{center}
\includegraphics[angle=270,width=0.8\textwidth]{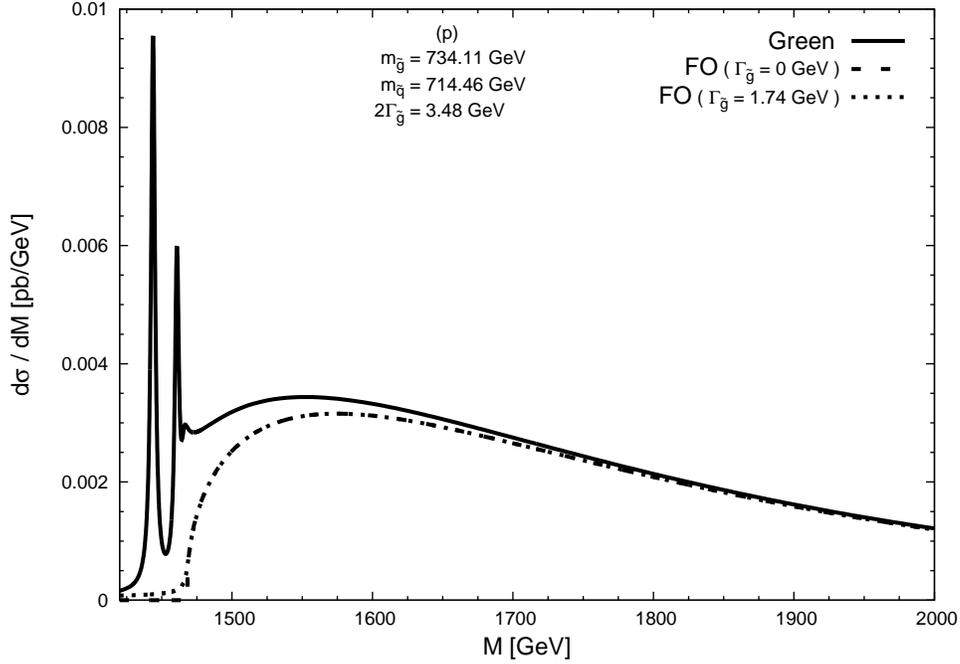}
\caption{Prediction for the differential cross section
  in NLO using the Green's function, in comparison with the fixed order
  cross section without and with vanishing single decay width for
  scenario (p).}
\label{NLOpFix}
\end{center}
\end{figure}
In Fig.~\ref{NLOpFix} we compare this result (solid curve) with the NLO
result using a fixed order treatment for the Green's function, i.e. we
replace the imaginary part of the Green's function by its expansion in
$\alpha_s$, first keeping $\Gamma_{\tilde{g}}$ non-vanishing (dotted
curve) and then in the limit $\Gamma_{\tilde{g}}\rightarrow 0$ (dashed
curve)
\begin{eqnarray}
\mbox{Im}\,G^{[R]}&\rightarrow&\frac{m_{\tilde{g}}^2}{4\pi}\,v\,\left(1+C^{[R]}\frac{\alpha_s\pi}{2v}\right)
\,,
\end{eqnarray}
leaving hard correction and PDFs unchanged. In the threshold region the
latter choice is a valid approximation to the complete fixed order NLO
prediction for the cross section. For invariant masses above
$1700\,\mbox{GeV}$ fixed order and Green's function modulated approach
agree reasonably well, between threshold and $1600\,\mbox{GeV}$ they
differ significantly. (Inclusion of finite width effects is quite
irrelevant in the fixed order treatment.) The integrated difference
between dashed and solid curves amounts to $0.183\,\mbox{pb}$ and is a
measure of the threshold enhancement, that would escape the strict fixed
order treatment. Relative to the total cross section\footnote{This
result has been obtained with the code {\tt Prospino2}
\cite{Beenakker:1996ch} in NLO, using CTEQ5 PDFs.} for
hadro-production of two gluinos, which, within scenario (p), amounts to
$\sigma_{\mbox{\tiny tot}}=2.59\,\mbox{pb}$, this corresponds to an
enhancement of $7.1\%$.
\par
\begin{figure}[tbp]
\begin{center}
\begin{tabular}{cc}
\mbox{(a)}&
\includegraphics[angle=270,width=0.8\textwidth]{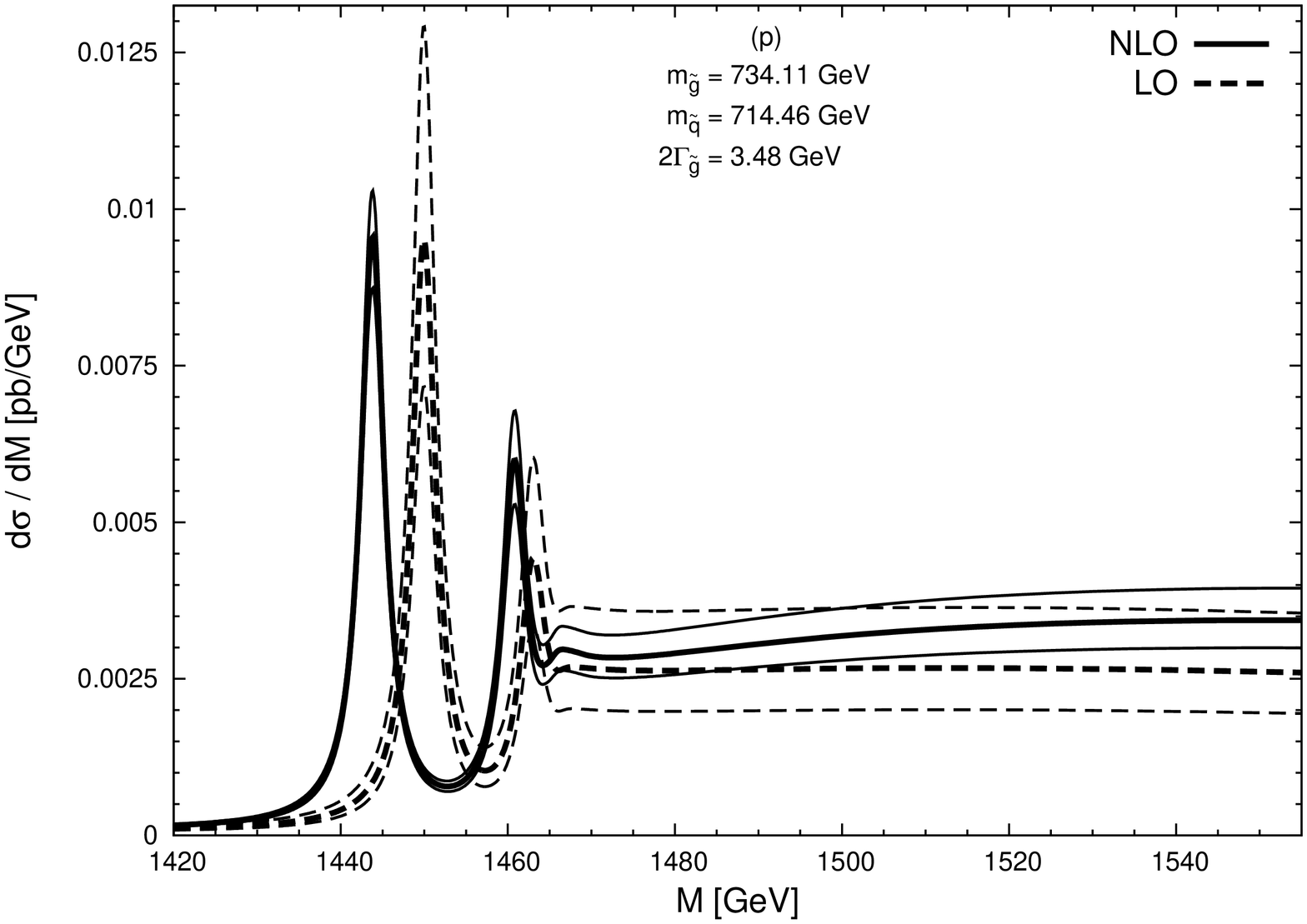}
\\
\mbox{(b)}&
\includegraphics[angle=270,width=0.8\textwidth]{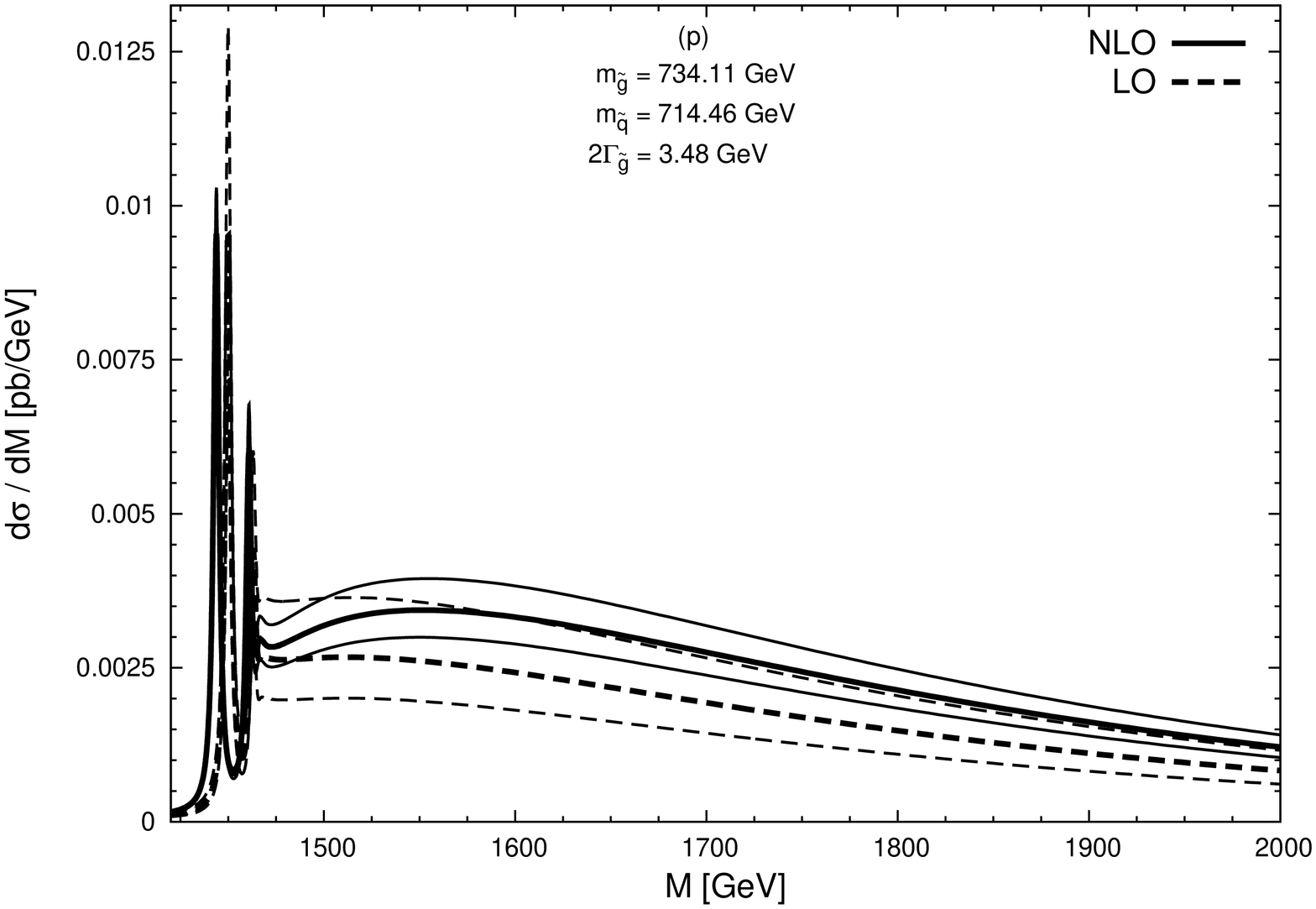}
\end{tabular}
\caption{Renormalization- and factorization-scale dependence of the
  differential cross section for scenario (p) for two different regions
  of $M$.}
\label{NLOscale}
\end{center}
\end{figure}
The renormalization ($\mu_R$) and factorization ($\mu_F$) scale
dependence of the differential cross section is shown in
Fig.~\ref{NLOscale} for three different choices, namely
$\mu_R=\mu_F=m_{\tilde{g}},2m_{\tilde{g}},4m_{\tilde{g}}$ and again for the
two regions in $M$ and for the LO and the NLO prediction. Note, that we
keep the Green's function unchanged and identify factorization and
renormalization scales. For the LO prediction we use the parameters
described at the end of Section\,2. Between LO and NLO we observe a
slight shift of the location of the resonance peaks by about
$10\,\mbox{GeV}$, a slight increase of the cross section by about $15\%$
and a reduction of the scale dependence. For the NLO prediction this
residual $\mu$ dependence amounts to $15\%$.
\par
\begin{figure}[tbp]
\begin{center}
\begin{tabular}{cc}
\mbox{(a)}&
\includegraphics[angle=270,width=0.8\textwidth]{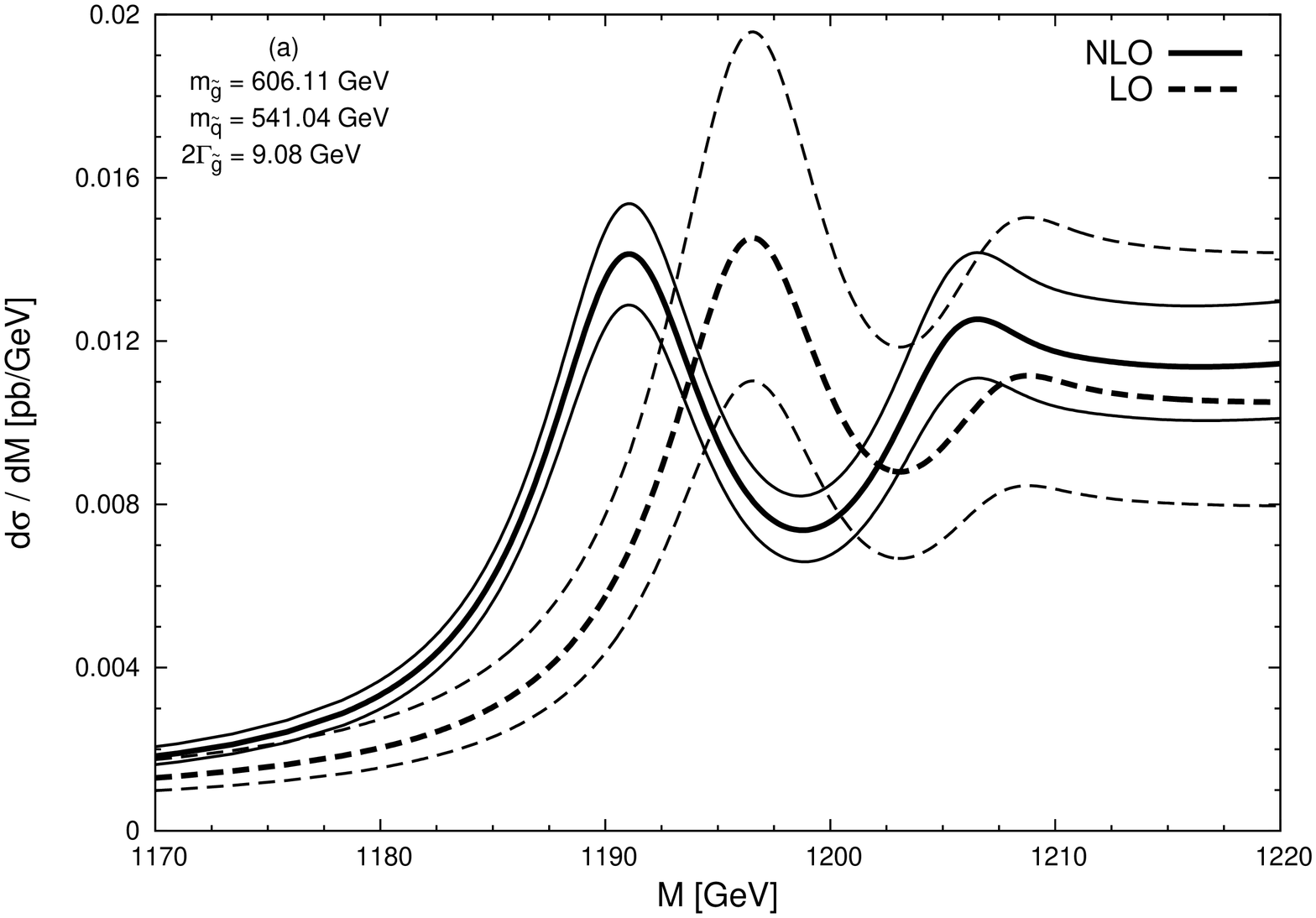}
\\
\mbox{(b)}&
\includegraphics[angle=270,width=0.8\textwidth]{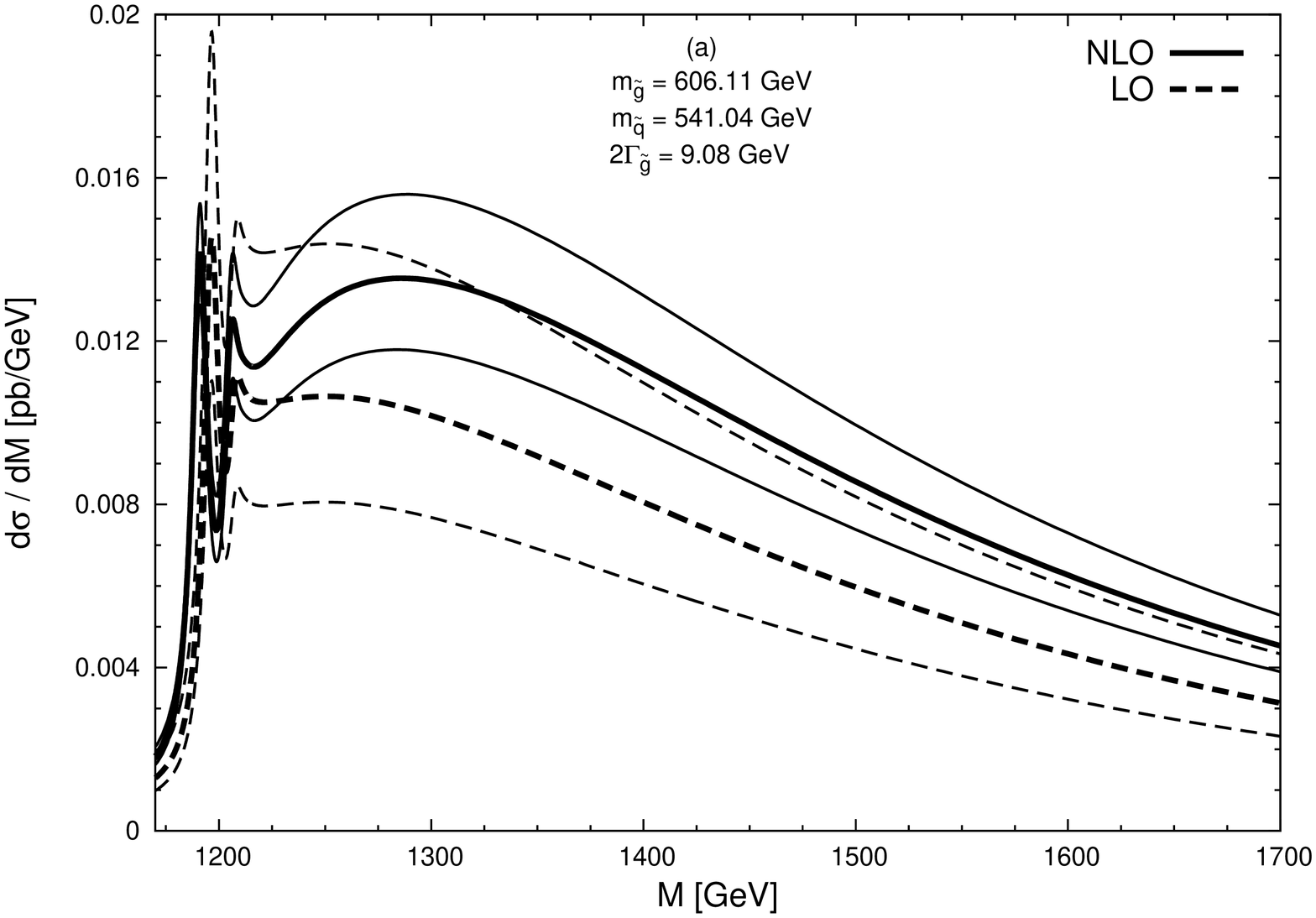}
\end{tabular}
\caption{Renormalization- and factorization-scale dependence of the
  differential cross section of scenario (a) for two different regions
  of $M$.}
\label{NLOa}
\end{center}
\end{figure}
\begin{figure}[tbp]
\begin{center}
\begin{tabular}{cc}
\mbox{(a)}&
\includegraphics[angle=270,width=0.8\textwidth]{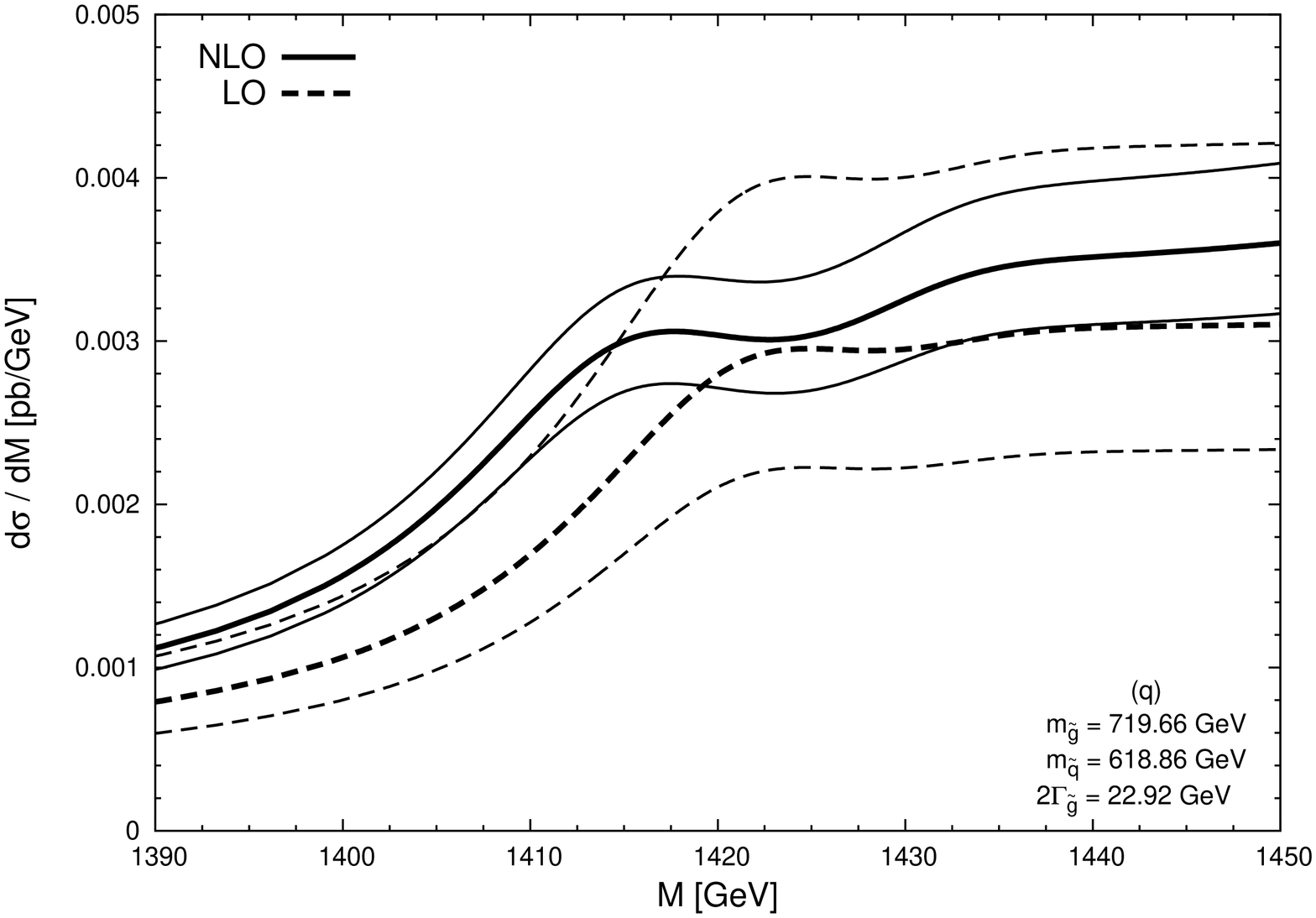}
\\
\mbox{(b)}&
\includegraphics[angle=270,width=0.8\textwidth]{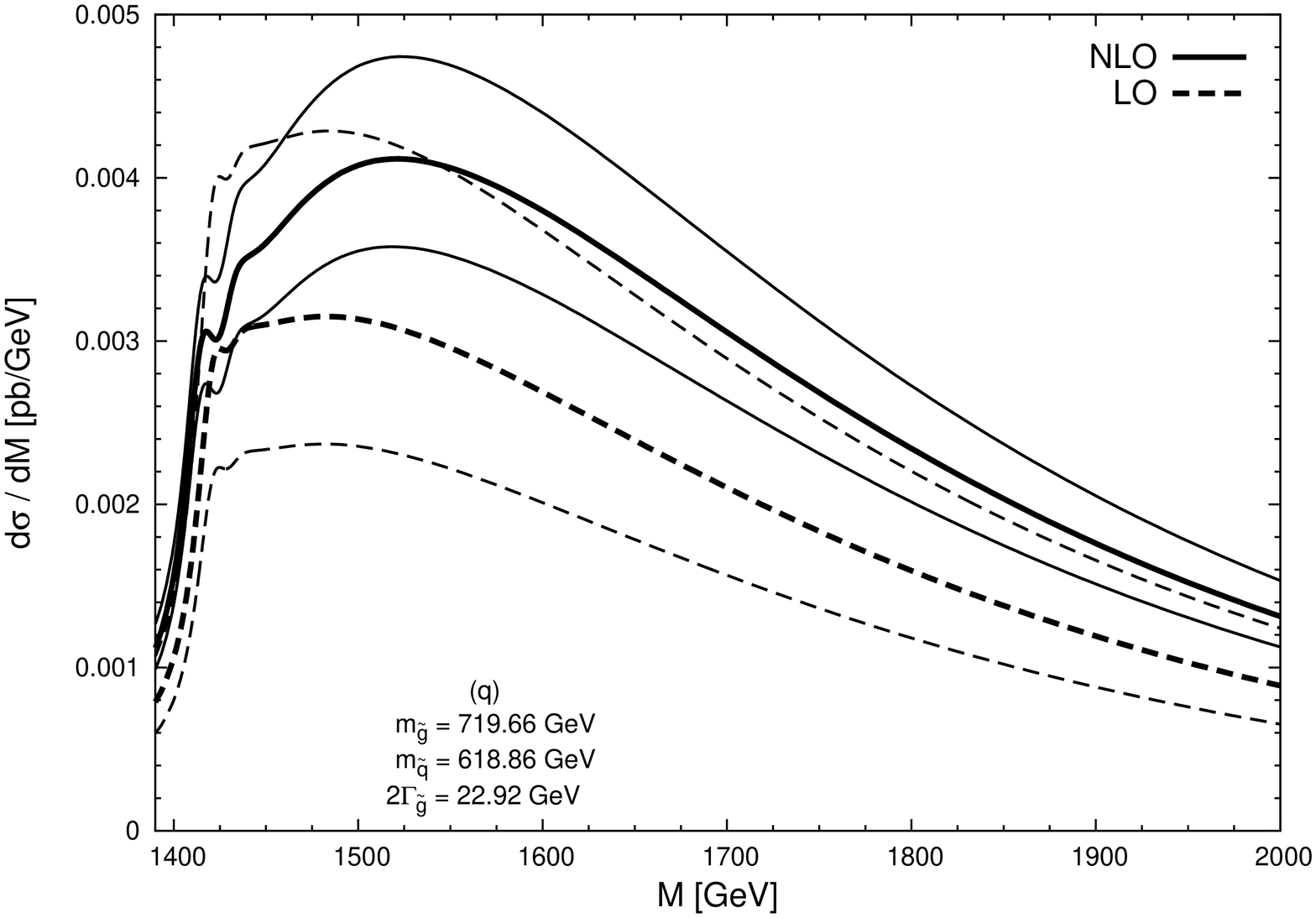}
\end{tabular}
\caption{Renormalization- and factorization-scale dependence of the
  differential cross sections of scenario (q) for two different regions
  of $M$.}
\label{NLOq}
\end{center}
\end{figure}
\begin{figure}[tbp]
\begin{center}
\begin{tabular}{cc}
\mbox{(a)}&
\includegraphics[angle=270,width=0.8\textwidth]{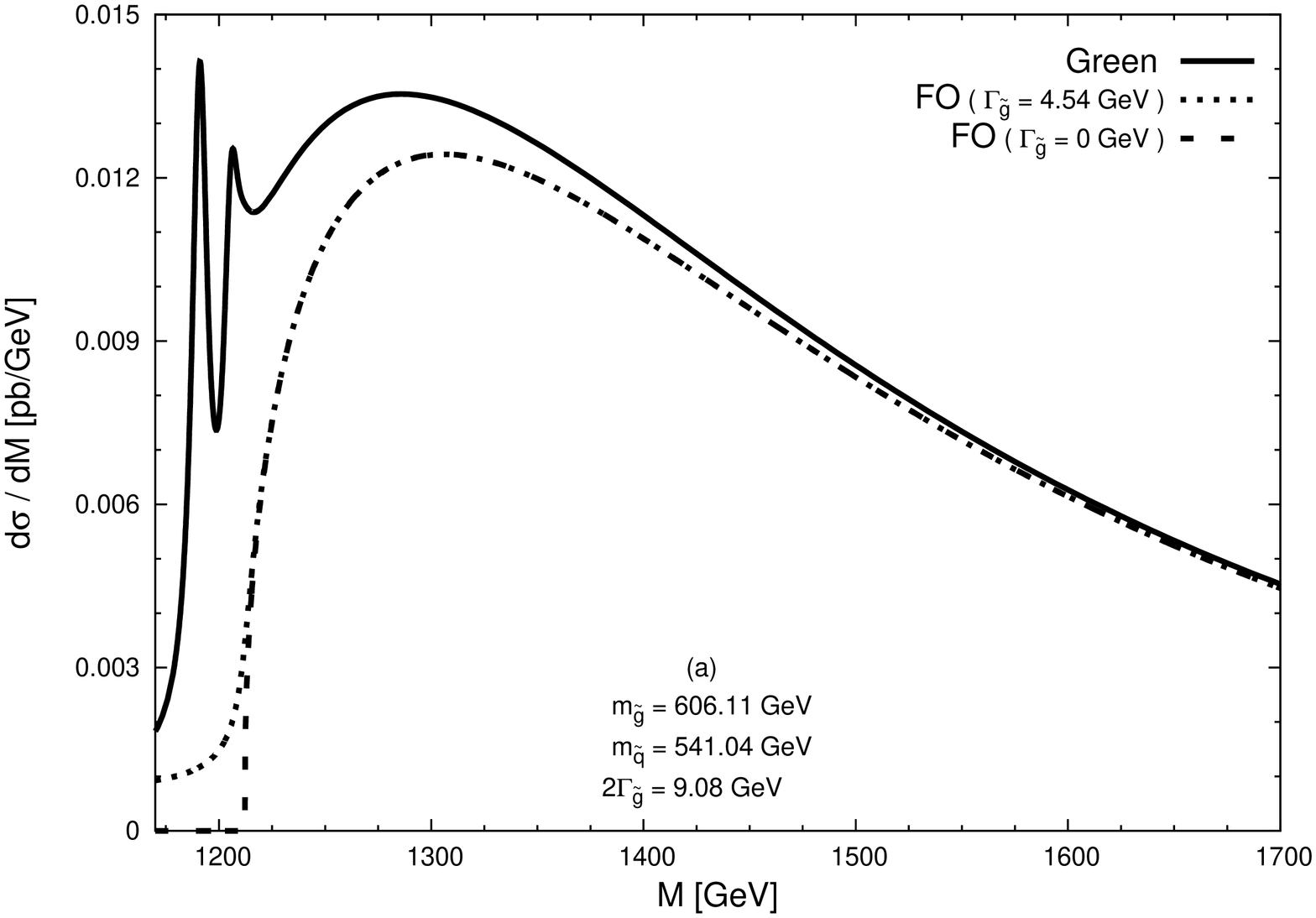}
\\
\mbox{(b)}&
\includegraphics[angle=270,width=0.8\textwidth]{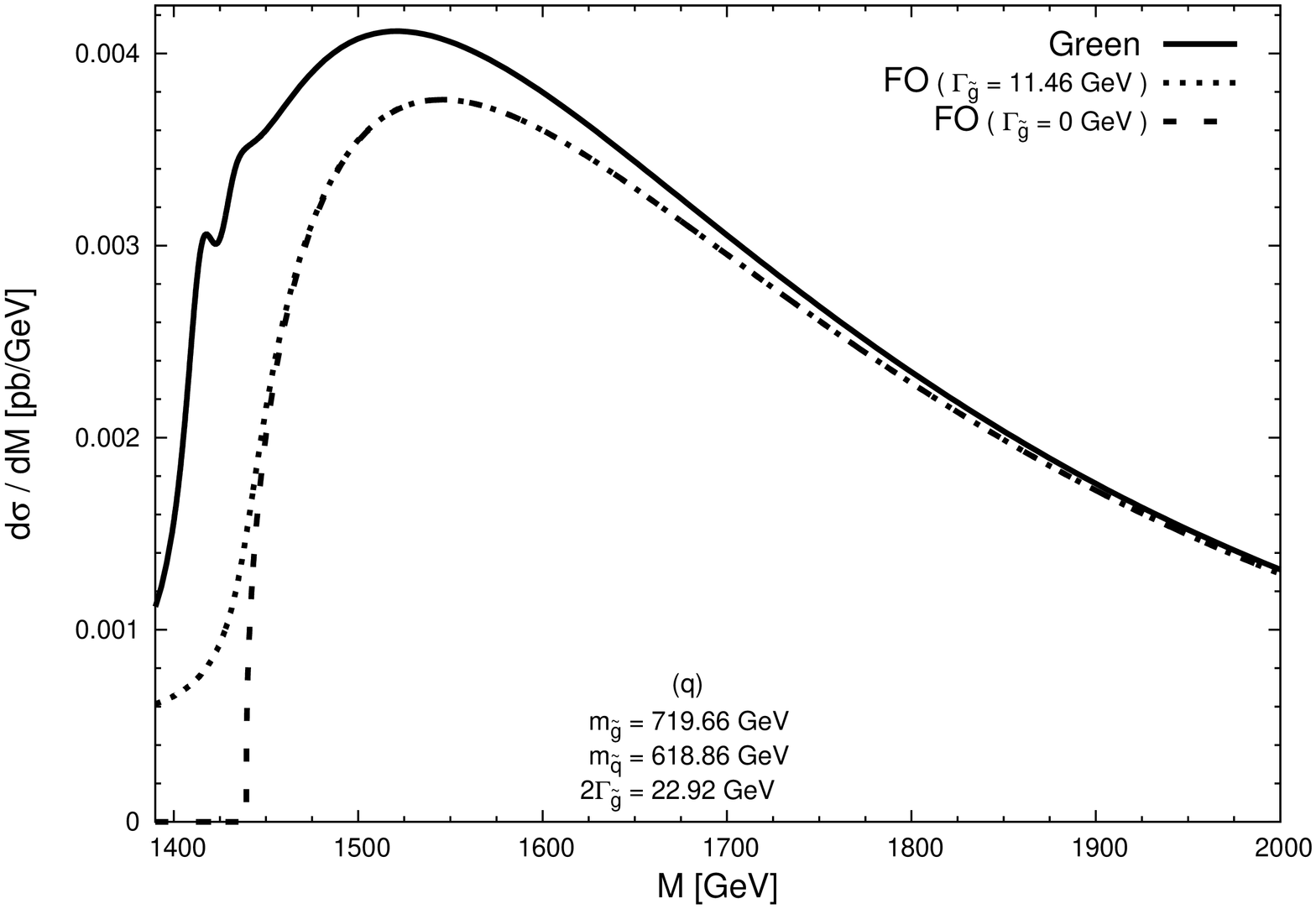}
\end{tabular}
\caption{Prediction for the differential cross section
  in NLO using the Green's function in comparison with the fixed order
  cross section without and with vanishing single decay width for
  scenarios (a) and (q).}
\label{NLOfixAQ}
\end{center}
\end{figure}

In Figs.~\ref{NLOa} and \ref{NLOq} the corresponding results are shown
for scenarios (a) and (q).
The gluino masses of (p), (a) and (q) are comparable, the difference
between gluino and squark mass, however, increases and, correspondingly,
the gluino decay rate. For scenario (a) with $\Delta
M=13.89\,\mbox{GeV}$ and $2\Gamma_{\tilde{g}}=9.08\,\mbox{GeV}$ the $1S$
peak is still clearly visible, for scenario (q), with $\Delta
M=15.79\,\mbox{GeV}$ and $2\Gamma_{\tilde{g}}=22.92\,\mbox{GeV}$, the
resonant structures have essentially disappeared. Nevertheless, final
state interaction leads to a significantly enhanced cross section in the
threshold region also in these two cases (Fig.~\ref{NLOfixAQ}(a) and (b)).
\par
\begin{figure}[tbp]
\begin{center}
\begin{tabular}{cc}
\mbox{(a)}&
\includegraphics[angle=270,width=0.8\textwidth]{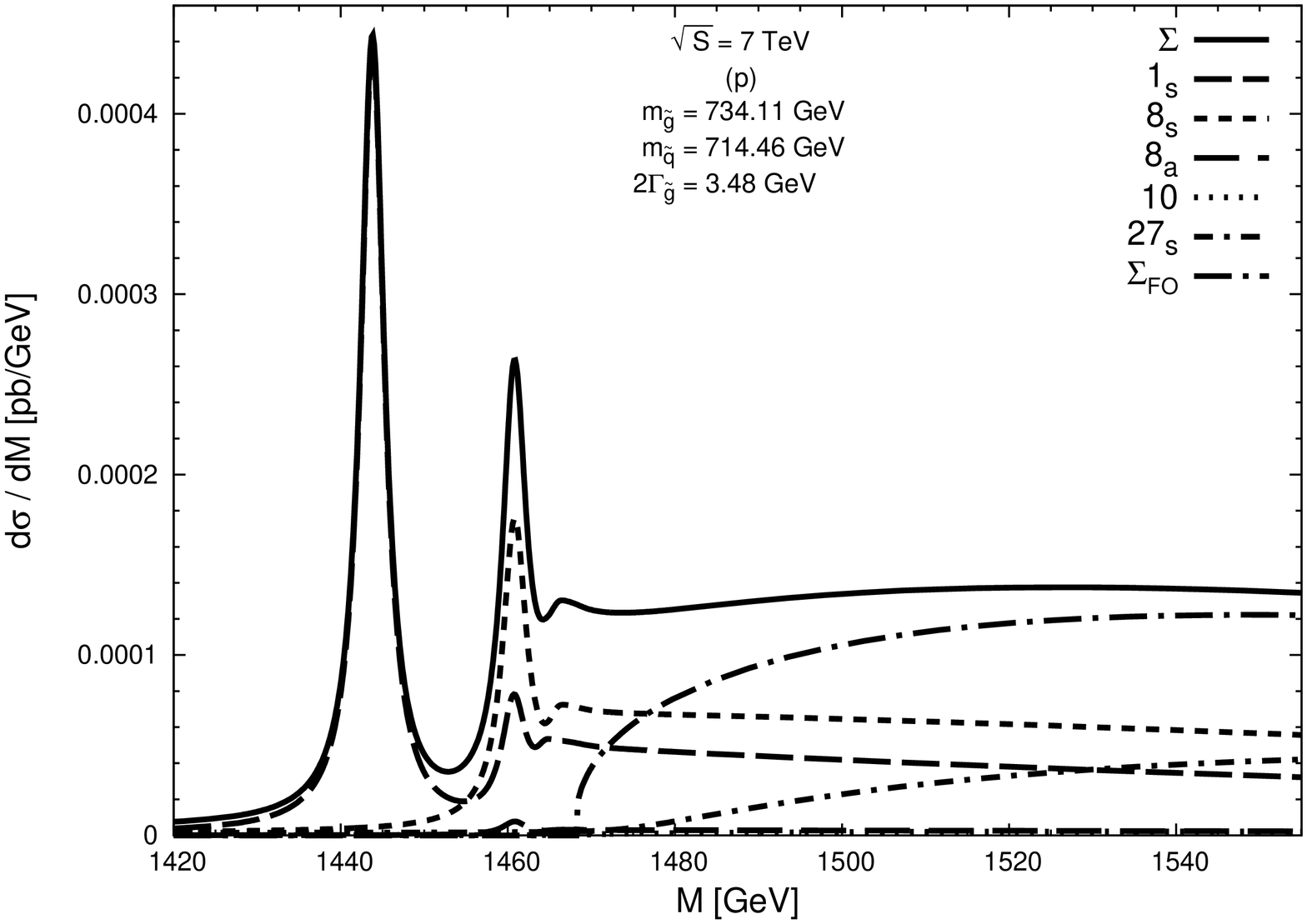}
\\
\mbox{(b)}&
\includegraphics[angle=270,width=0.8\textwidth]{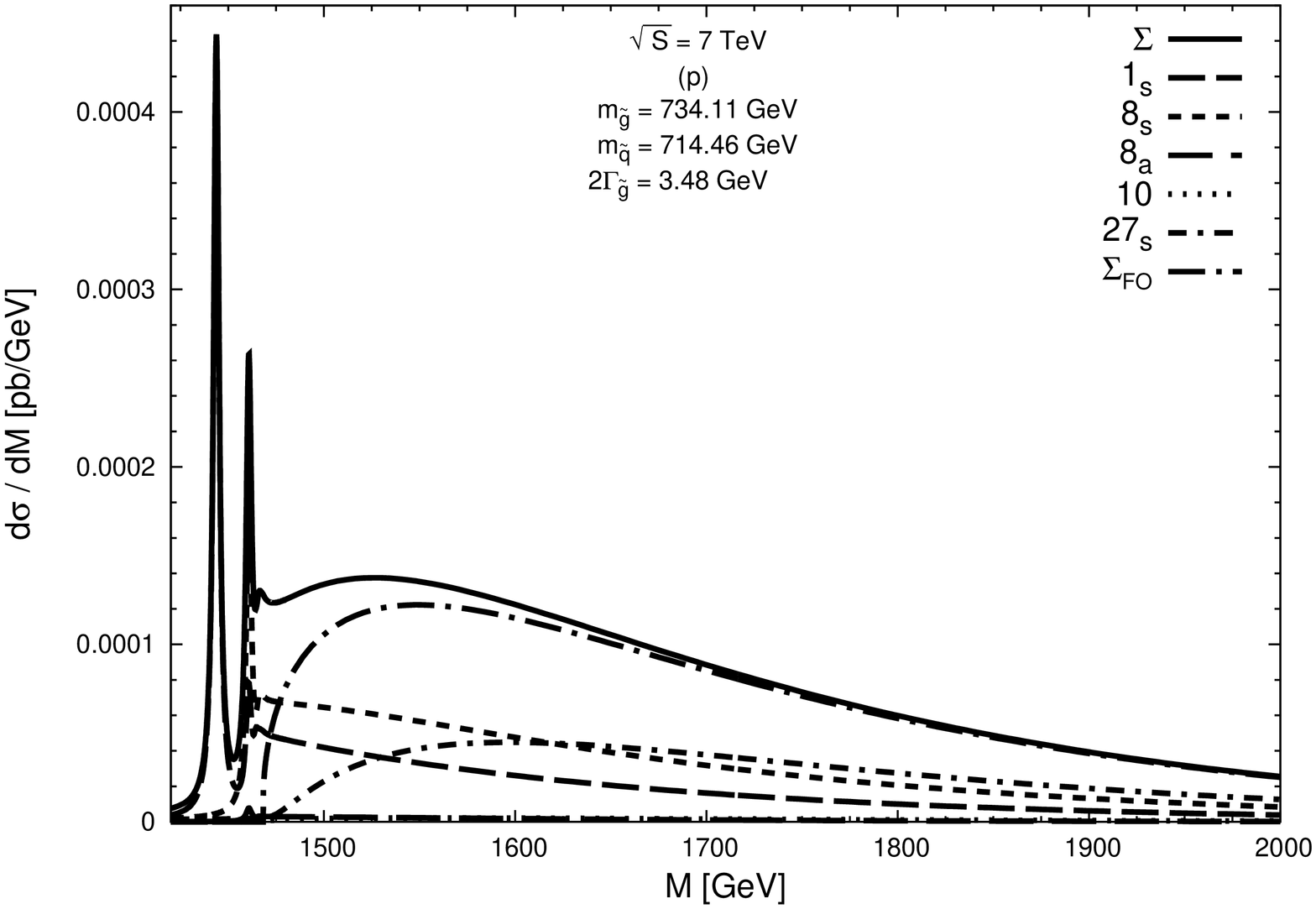}
\end{tabular}
\caption{NLO prediction for the differential cross section at
  $\sqrt{s}=7\,\mbox{TeV}$ for scenario (p) and comparison to
  the fixed order cross section with vanishing single decay width.}
\label{NLOp7}
\end{center}
\end{figure}

The results for the invariant mass distribution with
$\sqrt{s}=7\,\mbox{TeV}$ are shown in Fig.~\ref{NLOp7}(a) and (b),
restricting ourselves again to scenario (p). These results are
qualitatively similar to those of Fig.~\ref{NLOp}. A reduction of the
cross section by a factor $20$ is observed.
\par
\begin{table}
\begin{center}
\begin{tabular}{c||c||c|c|c|c|c}
$\hspace{3.3cm}$&$\hspace{0.2cm}\alpha_s(2m_{\tilde{g}})\hspace{0.2cm}$&$\hspace{0.6cm}1_s\hspace{0.6cm}$&$\hspace{0.6cm}8_s\hspace{0.6cm}$&$\hspace{0.6cm}8_a\hspace{0.6cm}$&$\hspace{0.6cm}10\hspace{0.6cm}$&$\hspace{0.6cm}27_s\hspace{0.6cm}$
\\\hline\hline
&\multicolumn{6}{c}{NLO calculation}
\\\hline\hline
$s^{\mbox{\tiny{LO}}}\,[\mbox{fb}/\mbox{GeV}^2]$&$0.1031$&
$5.00$&$10.0$&$0.137$&$-$&$15.0$
\\\cline{1-7}
$s^{\mbox{\tiny{NLO}}}\,[\mbox{fb}/\mbox{GeV}^2]$&$0.0923$&
$\hspace{0.1cm}4.98\hspace{0.1cm}$&$\hspace{0.1cm}13.3\hspace{0.1cm}$&$\hspace{0.1cm}1.40\hspace{0.1cm}$&$\hspace{0.1cm}0.984\hspace{0.1cm}$&$\hspace{0.1cm}29.5\hspace{0.1cm}$
\\\cline{1-7}
$\hspace{0.1cm}s^{\mbox{\tiny{NLO}}}/s^{\mbox{\tiny{LO}}}-1\,[\%]\hspace{0.1cm}$&$-$&
$-0.4$&$33$&$922$&$-$&$96.7$
\\\hline\hline
&\multicolumn{6}{c}{approximation}
\\\hline\hline
$s^{\mbox{\tiny{LO}}}\,[\mbox{fb}/\mbox{GeV}^2]$&$0.0954$&$4.29$&$8.57$&$0.117$&$-$&$12.9$
\\\cline{1-7}
$s^{\mbox{\tiny{NLO}}}\,[\mbox{fb}/\mbox{GeV}^2]$&$0.0862$&$6.50$&$14.0$&$0.136$&$-$&$23.5$
\\\cline{1-7}
$s^{\mbox{\tiny{NLO}}}/s^{\mbox{\tiny{LO}}}-1\,[\%]$&$-$&$51.5$&$63.4$&$16.2$&$-$&$82.2$
\end{tabular}
\caption{Comparison of full and approximate NLO results (see text).}
\label{tab:hag}
\end{center}
\end{table}
In Tab.~\ref{tab:hag} we investigate the relative size of the
corrections for the different states and contrast our result with those
based on an approximation discussed in \cite{Hagiwara:2009hq}. The
latter employs a hard correction factor, which is different for
quark-anti-quark annihilation and gluon fusion, does not distinguish
between different colour representations and has been numerically extracted
from the continuum result \cite{Beenakker:1996ch}. Furthermore, it
includes the logarithmically enhanced terms from soft radiation, which
are determined by the colour charge of initial and final states. To
allow for a consistent comparison, we employ for both cases MSTW2008LO
PDFs with $\alpha_s(M_Z)=0.1394$ in LO and MSTW2008NLO PDFs with
$\alpha_s(M_Z)=0.1202$ for the (approximate) NLO result. However,
following \cite{Hagiwara:2009hq} we use for the approximate treatment
two loop running with $n_f=5$ both for LO and NLO. The corresponding
$\alpha_s$ values are listed in the table. In Tab.~\ref{tab:hag} we also
show the results for the quantities $s^{\mbox{\tiny LO}}$ and
$s^{\mbox{\tiny NLO}}$ which were obtained from the corresponding
results for $\mbox{d}\sigma/\mbox{d}M$ with $G\equiv
m_{\tilde{g}}^2/(4\pi)$. In this way we remove the dependence on the
Green's function. The hard corrections are only weakly dependent on $M$
and we adopt $M=2m_{\tilde{g}}$ as reference point. Furthermore, we
adopt benchmark point (a), corresponding to
$m_{\tilde{g}}=606.11\,\mbox{GeV}$ and
$m_{\tilde{q}}=541.04\,\mbox{GeV}$.
\par
The difference between the two LO results can be traced to the different
values of $\alpha_s$. The difference between the relative size of the
corrections is due to additional subprocesses present in the full
calculation and the appearance of additional contributions with virtual squarks.

{
As stated in Section 2, the three scenarios (p), (a) and (q) serve to
illustrate the change of the excitation curve by moving from relatively
smaller gluino decay rate up to a situation with $2\Gamma_{\tilde g}$
even somewhat larger than the excitation energy $\Delta M$. Although
already excluded by recent LHC results, their
detailed discussion serves to illustrate the impact of NLO corrections
and threshold enhancement. From the phenomenological side,
however, it seems appropriate to also present results for scenarios
still consistent with the LHC limits. Indeed, the behaviour is quite
similar for the two scenarios (X) and (Y) (see Tabs. 3 and 5). 
In view of the smallness of the respective cross section for $\sqrt{s} =
7\,\mbox{TeV}$ only the results for $\sqrt{s} =14\,{\rm TeV}$ are presented. 
We separate the cross section according the the different colour and
spin configurations (Figs. \ref{fig:X14} and \ref{fig:Y14}) and compare the
results with the corresponding ones based on NLO fixed order calculations
with vanishing and non-vanishing gluino decay rates
(Figs. \ref{fig:X14fix} and \ref{fig:Y14fix}). As anticipated, the qualitative features of the results
are quite similar to those for scenarios  (p), (a) and (q).
\begin{figure}
  \centering
  \includegraphics[width=0.8\linewidth]{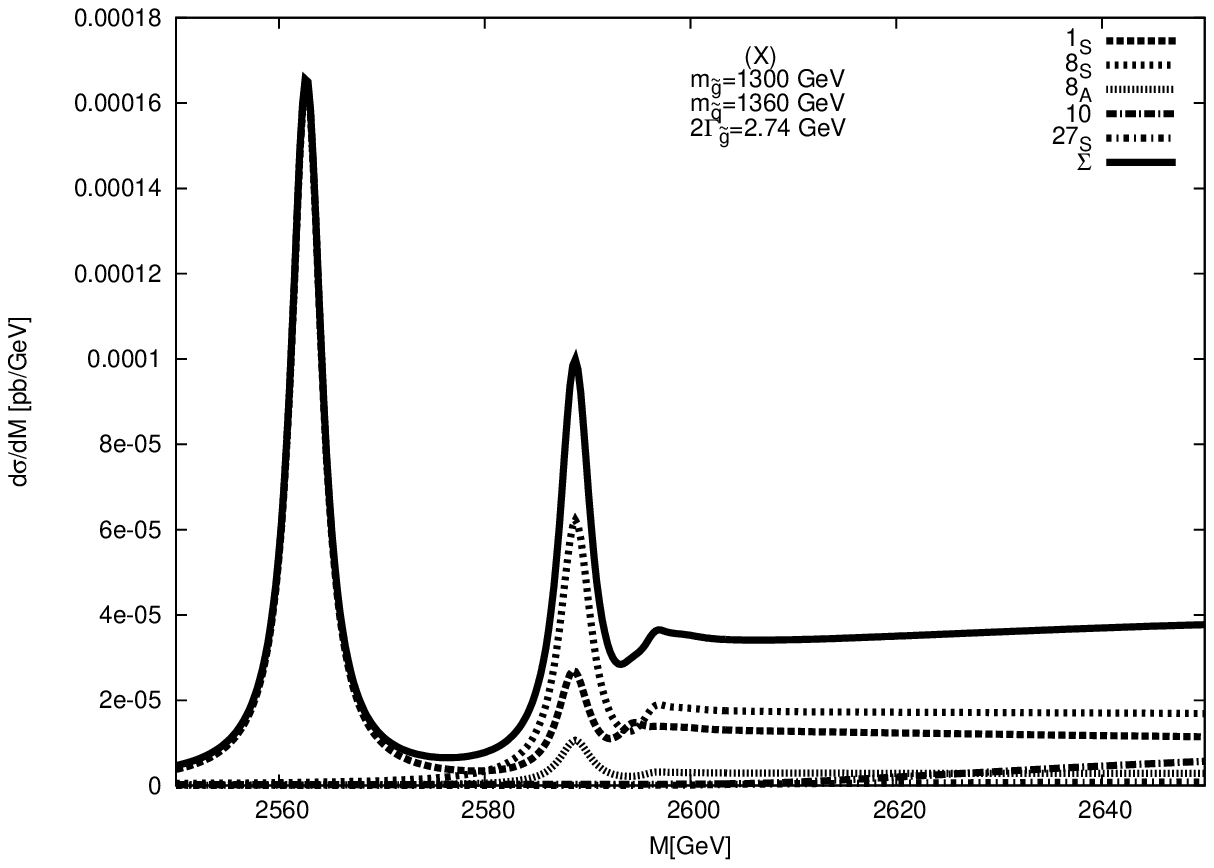}\hfill
  \includegraphics[width=0.8\linewidth]{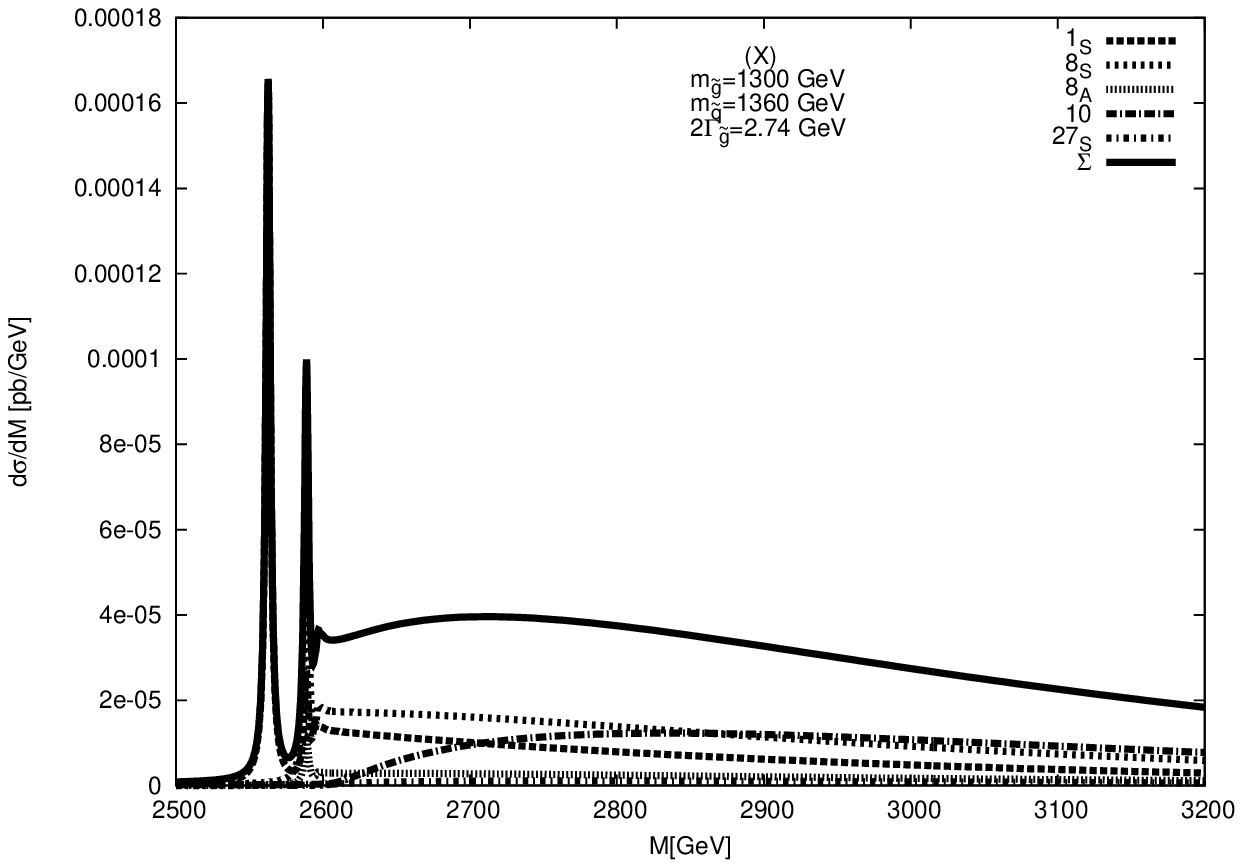}  
\caption{NLO prediction for the differential cross section for scenario
  (X) at $\sqrt{s} = 14\,\mbox{TeV}$ for two different regions of $M$.}
  \label{fig:X14}
\end{figure}
\begin{figure}
  \centering
    \includegraphics[width=0.8\linewidth]{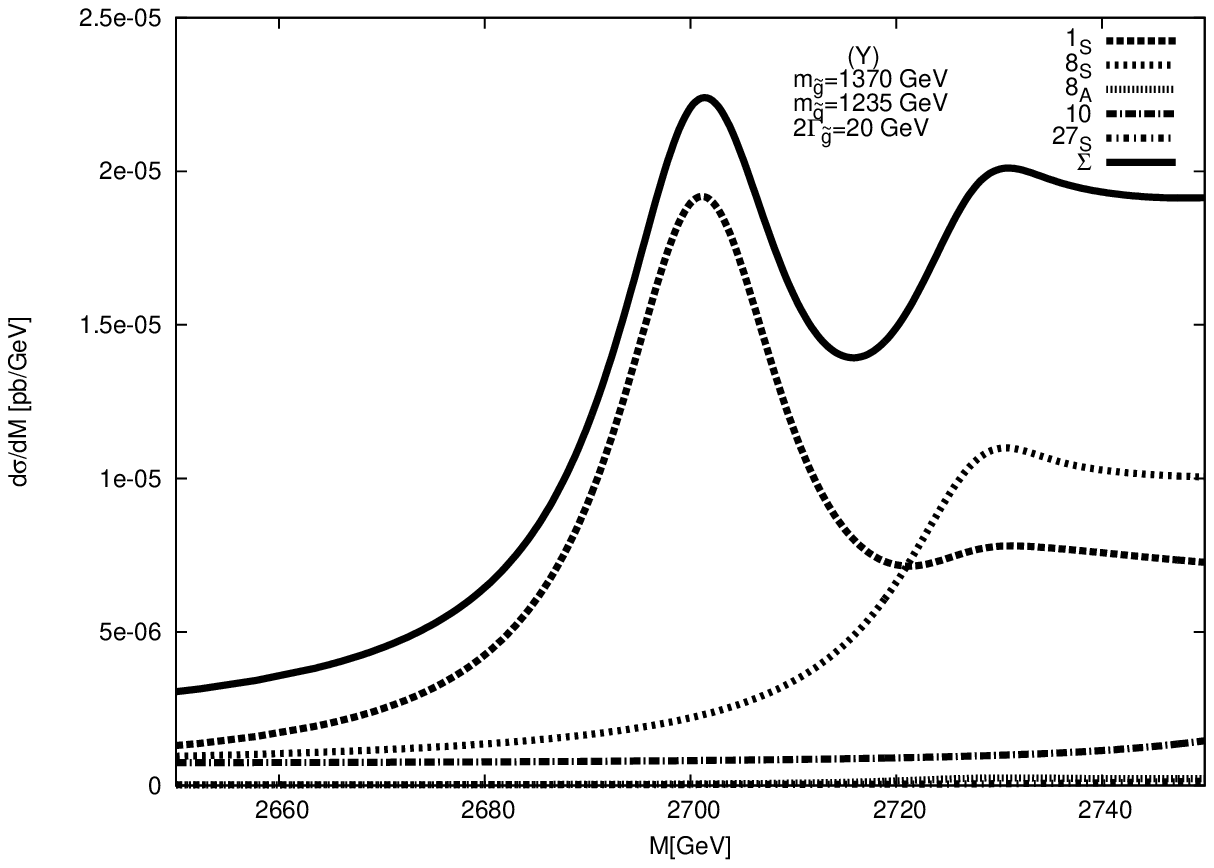}\hfill
  \includegraphics[width=0.8\linewidth]{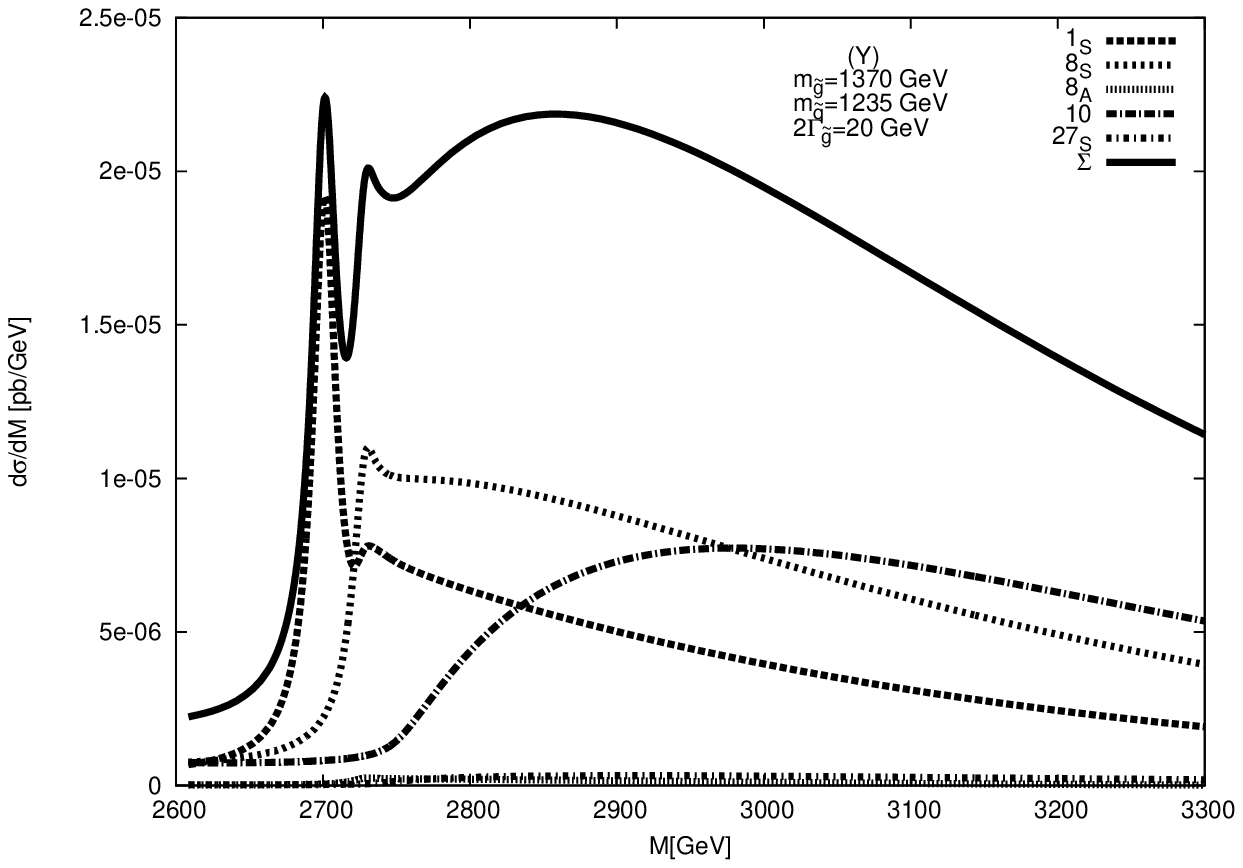}
\caption{NLO prediction for the differential cross section for scenario
  (Y) at $\sqrt{s} = 14\,\mbox{TeV}$ for two different regions of $M$.}
  \label{fig:Y14}
\end{figure}
\begin{figure}
  \centering
  \includegraphics[width=0.8\linewidth]{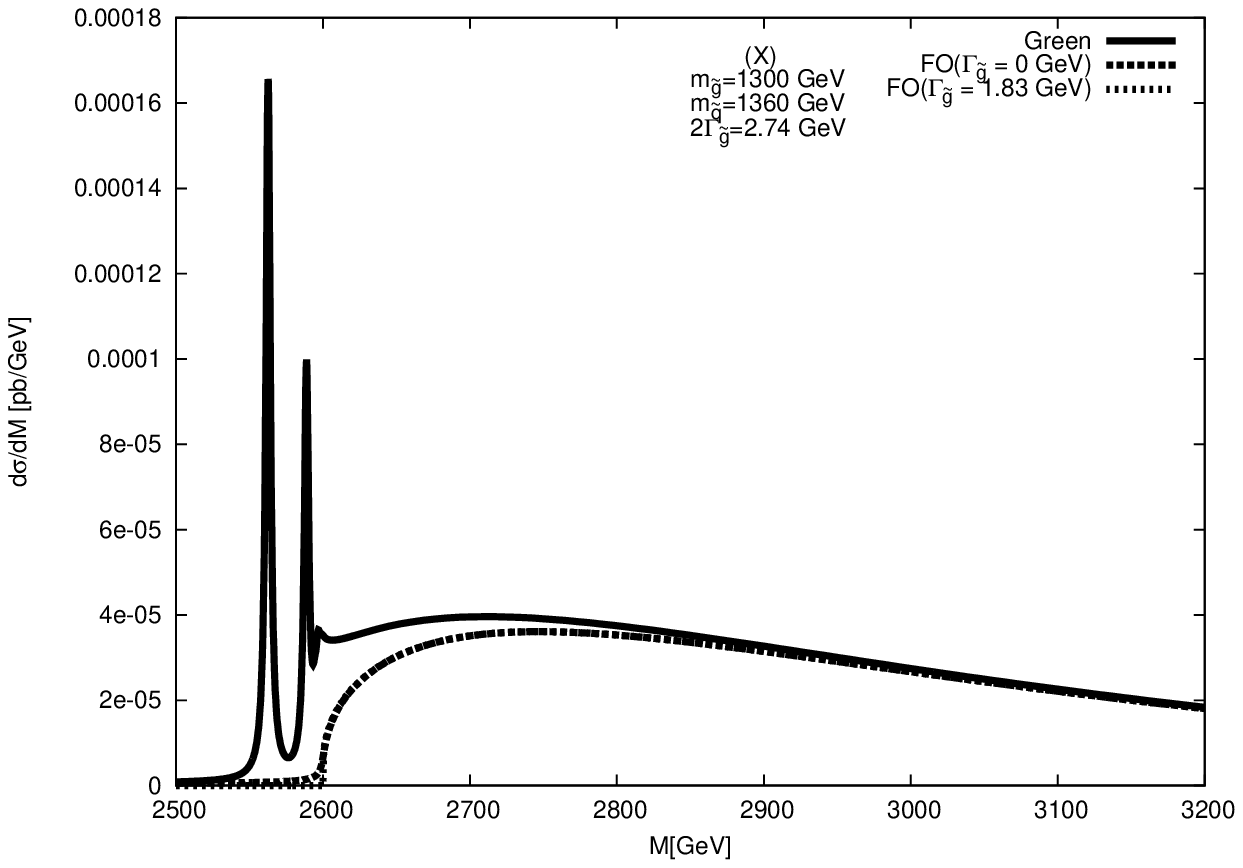}    
\caption{Prediction for the differential cross section
  in NLO using the Green's function, in comparison with the fixed order
  cross section without and with vanishing single decay width for
  scenario (X) at $\sqrt{s}=14\,\mbox{TeV}$.}
  \label{fig:X14fix}
\end{figure}
\begin{figure}
  \centering
    \includegraphics[width=0.8\linewidth]{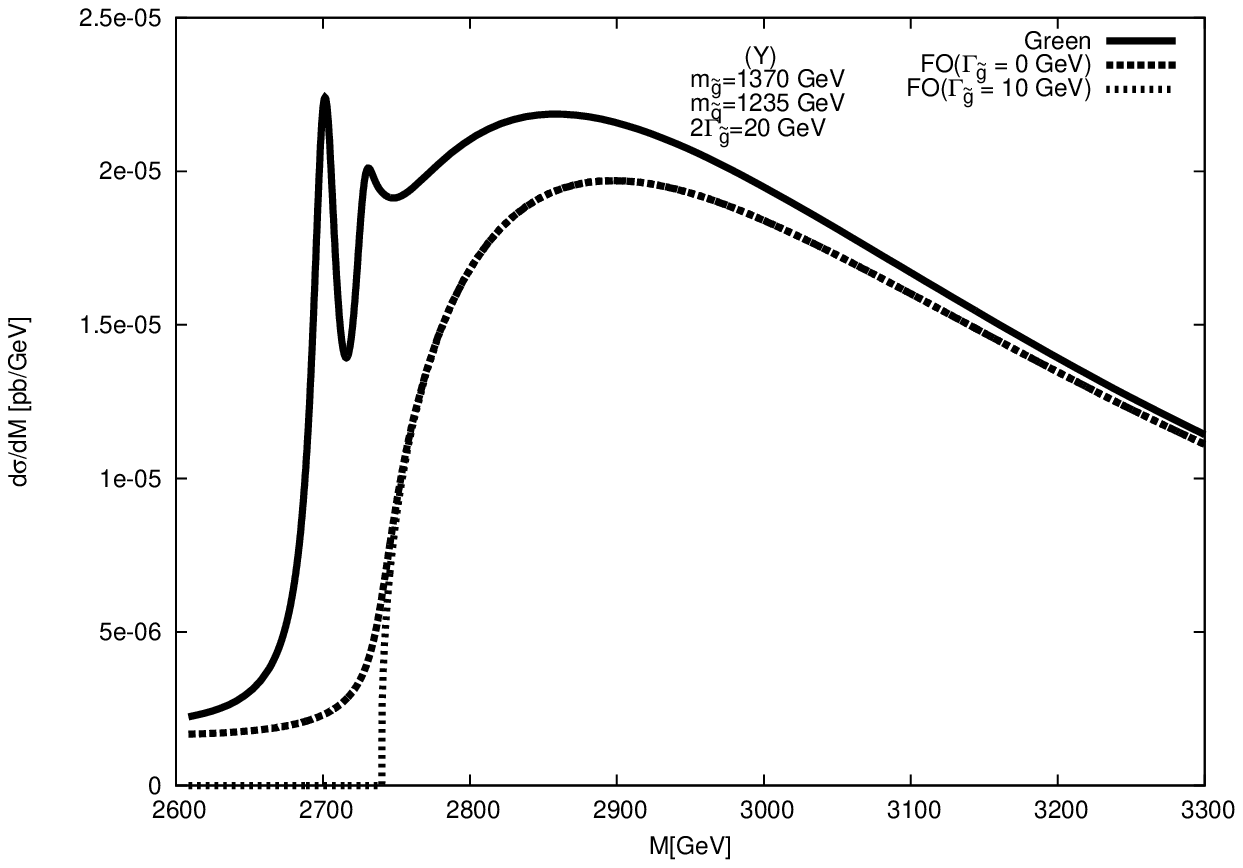}  
\caption{Prediction for the differential cross section
  in NLO using the Green's function, in comparison with the fixed order
  cross section without and with vanishing single decay width for
  scenario (Y) at $\sqrt{s}=14\,\mbox{TeV}$.}
  \label{fig:Y14fix}
\end{figure}
}
\section{\label{sec:conclusions}Conclusions}
The next-to-leading order analysis for hadronic production of gluino
pairs close to thres\-hold has been presented. The matching coefficients
were evaluated separately for the different colour configurations of
gluino pairs with relative angular momentum zero. The cross section is
strongly affected by final state interaction which is encoded in the NLO
Green's function and which depends on the gluino decay rate, the colour
configuration and the invariant mass of the pair. In a first step we
have investigated three different SUSY-scenarios covering a wide range
of gluino and squark masses, and studied the renormalization and
factorization scale dependence of the result. Compared to the leading
order prediction the result is more stable, and, for $\mu$ between
$m_{\tilde{g}}$ and $4m_{\tilde{g}}$, varies by $\pm15\%$. The complete
NLO threshold production of $(\tilde{g}\tilde{g})$ boundstates
considered in this paper enhances the fixed order prediction by
typically $7\%$ to $9\%$. {To accomondate the most recent limits on squark
and gluino masses, we also presented predictions for two additional
scenarios with squark and gluino masses above 1~TeV. The qualitative
features of our results remain unchanged.}
\begin{appendix}
\section{\label{app:Hyp}The Generalized Hypergeometric Function}
The Generalized Hypergeometric Function (GHF) is defined by the
series
\begin{eqnarray}
_{p}F_{q}(a_{1},a_{2},\ldots,a_{p};b_{1},\ldots,b_{q};z)&=&\sum_{n=0}^{\infty}\frac{(a_{1})_{n}(a_{2})_{n}\cdots(a_{p})_{n}}{(b_{1})_{n}\cdots(b_{q})_{n}}\frac{z^n}{n!}
\,,
\label{hyp}
\end{eqnarray}
with the Pochhammer symbols $(x)_{n}=\Gamma(x+n)/\Gamma(x)$ and the
restriction $b_{i}\neq0,-1,\ldots$ for $i=1,2,\ldots,q$. The series
converges if one of the following conditions holds
\begin{itemize}
\item[(1)] $p\leq q$, $\left|z\right|<\infty$\,,
\item[(2)] $p=q+1$, $\left|z\right|<1$\,,
\item[(3)] $p=q+1$, $\left|z\right|=1$, $\mbox{Re}\left(\sum_{n=1}^{q}b_{n}-\sum_{n=1}^{q+1}a_{n}\right)>0$\,,
\item[(4)] $p=q+1$, $\left|z\right|=1$, $z\neq 1$, $-1<\mbox{Re}\left(\sum_{n=1}^{q}b_{n}-\sum_{n=1}^{q+1}a_{n}\right)\leq  0$\,.
\end{itemize}
For the numerical evaluation of the Green's function of Eq.\,(\ref{Green})
it is necessary to evaluate the GHF $_{4}F_{3}(1,1,1,1;2,2,x;1)$ for $x\in
\mathbb{C}$, hence the remaining arguments have to fulfill condition
(3). For $\mbox{Re}(x)<0$ the convergence of the series is not
guarantied.
\par
In Ref.~\cite{Huber:2005yg} the following algorithm is introduced which
allows to decompose this function via partial fractioning in GHFs with
convergent series representation. Following Ref.~\cite{Huber:2005yg} we
employ the following identity
\begin{eqnarray}
_{4}F_{3}(1,1,1,1;2,2,x;1)&=&\frac{1}{4x^2(2-x)^2}\Bigl[\,4(x-1)^4\,_{4}F_{3}(1,1,1,1;2,2,x+1;1)
\nonumber\\
&&\hspace{2.4cm}+2x(7-4x)\,_{4}F_{3}(1,1,1,1;3,2,x;1)
\nonumber\\
&&\hspace{2.4cm}+x(x-2)\,_{4}F_{3}(1,1,1,1;3,3,x;1)\Bigr]
\,.
\label{HypRep1}
\end{eqnarray}
For arbitrary values $\mbox{Re}(x)<0$ with $x\neq -1,-2,\ldots$ this
relation is applied repeatedly, leading to the following results
\begin{eqnarray}
_{4}F_{3}(1,1,1,1;a,a,x;1)&=&\frac{1}{a^2x\bigl(x-2(2-a)\bigr)(a-x)^2}
\nonumber\\
&&\hspace{0.cm}\times\Bigl[\,a^2(x-1)^4\,_{4}F_{3}(1,1,1,1;a,a,x+1;1)
\nonumber\\
&&\hspace{0.5cm}+a(a-1)^3x(3a+1-4x)\,_{4}F_{3}(1,1,1,1;a+1,a,x;1)
\nonumber\\
&&\hspace{0.5cm}+(a-1)^4x(x-a)\,_{4}F_{3}(1,1,1,1;a+1,a+1,x;1)\Bigr]\,,
\nonumber\\
_{4}F_{3}(1,1,1,1;a,b,x;1)&=&\frac{1}{a+b+x-4}\Bigl[\,\frac{(a-1)^4}{a(a-b)(a-x)}\,_{4}F_{3}(1,1,1,1;a+1,b,x;1)
\nonumber\\
&&\hspace{2.65cm}+\frac{(b-1)^4}{b(b-a)(b-x)}\,_{4}F_{3}(1,1,1,1;a,b+1,x;1)
\nonumber\\
&&\hspace{2.65cm}+\frac{(x-1)^4}{x(x-a)(x-b)}\,_{4}F_{3}(1,1,1,1;a,b,x+1;1)\Bigr]
\,,
\nonumber\\
\label{HypRep2}
\end{eqnarray}
with $a,b\in \mathbb{N}\backslash\{0,1\}$ and $a\neq b$.
\section{\label{app:results}Corrections from virtual and real emission}
The missing piece for the virtual corrections to the hard part of the
partonic cross section for the $[8_a]$ configuration (see
Eq.(\ref{virtual})) is
\begin{eqnarray}
\mathcal{A}_{q\overline{q}\rightarrow8_{a}}(r)&=&\frac{3a_{1}(r)}{4(r-1)}-
\frac{(r^2-5)b_{1}(r)}{12(r-1)^2}+
\frac{(r-3)(r+1)^2b_{2}(r)}{24(r-1)^2}+
\frac{(4r-13)b_{3}(r)}{12}
\nonumber\\&&-\frac{2(r^2-2r+5)b_{4}(r)}{3(r^2-1)}-
\frac{16rb_{5}(r)}{3(r^2-1)}+
\frac{8b_{6}(r)}{3(r+1)}+
\frac{2(r-1)b'_{2}(r)}{3}
\nonumber\\&&-
\frac{(r+1)^2(r^2-2r+5)c_{1}(r)}{24(r-1)^2}-
\frac{(r^3-5r^2+11r-15)c_{2}(r)}{12(r-1)^2}
\nonumber\\&&-
\frac{3(r+1)(r^2-6r+17)c_{3}(r)}{8(r-1)}-
\frac{3(r+1)(r^2-2r+5)c_{4}(r)}{4(r-1)}
\nonumber\\&&-
\frac{(3r^2-4r-17)\ln(2)}{4(r-1)}-
\frac{18r^3+14r^2+23r-101}{12(r^2-1)}
\nonumber\\&&+
n_{f}\biggl[\,
\frac{(r-5)a_{1}(r)}{6(r-1)}-
\frac{rb_{1}(r)}{3}+
\frac{(r+1)^2b_{2}(r)}{6(r-1)}
\nonumber\\&&\hspace{1.cm}+
(r-1)b'_{1}(r)+
\frac{3r^2+r+2}{18(r-1)}
\biggr]
\,,
\end{eqnarray}
where the scalar one-, two- and three-point functions are 
\begin{eqnarray}
A_{0}(m_{\tilde{q}}^2)&=&m_{\tilde{q}}^2\,\Delta+m_{\tilde{g}}^2\,a_{1}\left(r\right)\,,
\nonumber\\
\mbox{Re}\left\{B_{0}(m_{\tilde{g}}^2;m_{\tilde{q}}^2,0)\right\}&=&\Delta+b_{1}\left(r\right)\,,
\nonumber\\
\mbox{Re}\left\{B_{0}(4m_{\tilde{g}}^2;m_{\tilde{q}}^2,m_{\tilde{q}}^2)\right\}&=&\Delta+b_{2}\left(r\right)\,,
\nonumber\\
B_{0}(0;m_{\tilde{q}}^2,m_{\tilde{g}}^2)&=&\Delta+b_{3}\left(r\right)\,,
\nonumber\\
B_{0}(-m_{\tilde{g}}^2;m_{\tilde{q}}^2,0)&=&\Delta+b_{4}\left(r\right)\,,
\nonumber\\
B_{0}(m_{\tilde{q}}^2;m_{\tilde{q}}^2,0)&=&\Delta+b_{5}\left(r\right)\,,
\nonumber\\
\mbox{Re}\left\{B_{0}(m_{\tilde{q}}^2;m_{\tilde{g}}^2,0)\right\}&=&\Delta+b_{6}\left(r\right)\,,
\nonumber\\
m_{\tilde{g}}^2\,\mbox{Re}\left\{B'_{0}(m_{\tilde{g}}^2;m_{\tilde{q}}^2,0)\right\}&=&b'_{1}\left(r\right)\,,
\nonumber\\
m_{\tilde{g}}^2B'_{0}(0\,;m_{\tilde{g}}^2,m_{\tilde{q}}^2)&=&b'_{2}\left(r\right)\,,
\nonumber\\
m_{\tilde{g}}^2\,\mbox{Re}\left\{C_{0}(4m_{\tilde{g}}^2,0,0\,;m_{\tilde{q}}^2,m_{\tilde{q}}^2,m_{\tilde{g}}^2)\right\}&=&c_{1}\left(r\right)\,,
\nonumber\\
m_{\tilde{g}}^2\,\mbox{Re}\left\{C_{0}(0,-m_{\tilde{g}}^2,m_{\tilde{g}}^2;m_{\tilde{q}}^2,m_{\tilde{g}}^2,0)\right\}&=&c_{2}\left(r\right)\,,
\nonumber\\
m_{\tilde{g}}^2C_{0}(0,-m_{\tilde{g}}^2,m_{\tilde{g}}^2;m_{\tilde{g}}^2,m_{\tilde{q}}^2,0)&=&m_{\tilde{g}}^2C_{0}(4m_{\tilde{g}}^2,0,0\,;m_{\tilde{g}}^2,m_{\tilde{g}}^2,m_{\tilde{q}}^2)\hspace{0.2cm}=\hspace{0.2cm}c_{3}\left(r\right)\,,
\nonumber\\
c_{4}(r)&=&-\frac{1}{r+1}\left[1+\ln\left(\frac{r+1}{2}\right)-\frac{r}{r+1}\ln\left(r\right)\right]\,,
\nonumber\\
m_{\tilde{g}}^2\,\mbox{Re}\left\{C_{0}(0,-m_{\tilde{g}}^2,m_{\tilde{g}}^2;m_{\tilde{q}}^2,m_{\tilde{q}}^2,0)\right\}&=&c_{5}(r)\,,
\end{eqnarray}
with the conventions of \cite{'tHooft:1978xw} and
$\Delta=1/\varepsilon_{\mbox{\tiny{UV}}}-\gamma_{E}
+\ln\left[4\pi\mu_R^2/(2m_{\tilde{g}})^2\right]$. The remaining
functions complete the real corrections
\begin{eqnarray}
\mathcal{F}_{gq}^{[1_s]}(z,r)&=&\Bigl[\,9r(r+1)^3z^4-(r-1)(r+1)^2z^3-2(r+1)(19r+35)z^2
\nonumber\\
&&\hspace{0.2cm}-32(r+1)z-64\Bigr]\frac{16z}{27\left[(r+1)z-2\right]^2\left[(r+1)z+2\right]^3}\ln(z)
\nonumber\\
&&\hspace{-0.3cm}+\Bigl[\,(r+1)^3(21r+5)z^4+2(r+1)^2(5r+9)z^3+4(r+1)(23r-29)z^2
\nonumber\\
&&\hspace{0.2cm}+200(r+1)z-224\Bigr]\frac{2z\left[5z(r+1)-8\right]\ln\left(1+2\frac{1-z}{z(r+1)}\right)}{243\left[(r+1)z-2\right]^2\left[(r+1)z+2\right]^3}
\nonumber\\
&&\hspace{-0.3cm}+\Bigl[\,(r+1)^3(57r+41)z^4+4(r+1)^2(81r^2+61r-36)z^3
\nonumber\\
&&\hspace{0.2cm}-4(r+1)(55r+71)z^2-16(r+1)(81r+133)z+256\Bigr]
\nonumber\\
&&\hspace{0.2cm}\times\frac{4(1-z)}{243(r+1)\left[z^2(r+1)^2-4\right]^2}\,,
\label{Fgq1}
\end{eqnarray}

\begin{eqnarray}
\mathcal{F}_{gq}^{[8_s]}(z,r)&=&\Bigl[\,9r(r+1)^3z^4-4(r-1)(r+1)^2z^3-4(r+1)(11r+26)z^2
\nonumber\\
&&\hspace{0.2cm}-40(r+1)z-80\Bigr]\frac{32z\ln(z)}{27\left[(r+1)z-2\right]^2\left[(r+1)z+2\right]^3}
\nonumber\\
&&\hspace{-0.3cm}+\Bigl[\,(r+1)^4(33r+28)z^5-4(r+1)^3(37r-14)z^4
\nonumber\\
&&\hspace{0.2cm}+4(r+1)^2(47r-175)z^3+8(r+1)(33r+163)z^2
\nonumber\\
&&\hspace{0.2cm}-16(86r+131)z+1120\Bigr]\frac{8z\ln\left(1+2\frac{1-z}{z(r+1)}\right)}{243\left[(r+1)z-2\right]^2\left[(r+1)z+2\right]^3}
\nonumber\\
&&\hspace{-0.3cm}+\Bigl[\,(r+1)^3(24r+19)z^4+(r+1)^2(81r^2+56r-45)z^3
\nonumber\\
&&\hspace{0.2cm}-4(r+1)(11r+16)z^2-4(r+1)(81r+191)z+80\Bigr]
\nonumber\\
&&\hspace{0.2cm}\times\frac{32(1-z)}{243(r+1)\left[z^2(r+1)^2-4\right]^2}
\,,
\nonumber\\
\label{Fgq8s}
\end{eqnarray}

\begin{eqnarray}
\mathcal{F}_{gq}^{[8_a]}(z,r)&=&\Bigl[\,9(r-1)^2(r+1)^4z^6+(r+1)^3(67r^2-186r+67)z^5
\nonumber\\
&&\hspace{0.2cm}-(r+1)^2(13r^3+207r^2-81r-67)z^4
\nonumber\\
&&\hspace{0.2cm}-2(r+1)^2(85r^2+100r-233)z^3-36(r+1)(5r^2-6r-3)z^2
\nonumber\\
&&\hspace{0.2cm}+72(7r^2\hspace{-0.05cm}+\hspace{-0.05cm}18r+3)z\hspace{-0.05cm}+\hspace{-0.05cm}576r\Bigr]\frac{32\ln(z)}{27(r+1)^2\left[(r+1)z\hspace{-0.05cm}-\hspace{-0.05cm}2\right]^2\left[(r+1)z\hspace{-0.05cm}+\hspace{-0.05cm}2\right]^3}
\nonumber\\
&&\hspace{-0.3cm}-\Bigl[\,(r+1)^4(3r^3+35r^2+13r+13)z^6
\nonumber\\
&&\hspace{0.2cm}-2(r+1)^3(35r^3-195r^2+353r-57)z^5
\nonumber\\
&&\hspace{0.2cm}-8(r+1)^2(5r^3+125r^2-189r+91)z^4
\nonumber\\
&&\hspace{0.2cm}+16(r+1)(15r^3+7r^2+25r+97)z^3
\nonumber\\
&&\hspace{0.2cm}+16(19r^3+11r^2-227r-91)z^2+32(r^2+82r+17)z-1024r\Bigr]
\nonumber\\
&&\hspace{0.2cm}\times\frac{4\ln\left(1+2\frac{1-z}{z(r+1)}\right)}{27(r+1)^2\left[(r+1)z-2\right]^2\left[(r+1)z+2\right]^3}
\nonumber\\
&&\hspace{-0.3cm}+\Bigl[\,(r-1)(r+1)^3(21r^2-38r+37)z^6
\nonumber\\
&&\hspace{0.2cm}+4(r+1)^2(35r^3-324r^2+327r-82)z^5
\nonumber\\
&&\hspace{0.2cm}-8(r+1)(6r^3+75r^2+136r-93)z^4
\nonumber\\
&&\hspace{0.2cm}-16(18r^4+25r^3-104r^2+61r+44)z^3
\nonumber\\
&&\hspace{0.2cm}-16(36r^3-23r^2-214r-115)z^2+384(3r^2+r-1)z+2304r\Bigr]
\nonumber\\
&&\hspace{0.2cm}\times\frac{8(1-z)}{27(r+1)^2z\left[z(r-1)+2\right]\left[z^2(r+1)^2-4\right]^2}\,,
\label{Fgq8a}
\end{eqnarray}

\begin{eqnarray}
\mathcal{F}_{gq}^{[10]}(z,r)&=&\Bigl[\,(r+1)(r^2-6r+1)z^2-(r-1)(r^2+10r+1)z
\nonumber\\
&&\hspace{0.2cm}-2(r^2-2r-11)\Bigr]\frac{320z^3\ln(z)}{27\left[(r+1)z-2\right]^2\left[(r+1)z+2\right]^3}
\nonumber\\
&&\hspace{-0.3cm}-\Bigl[\,(r+1)^5z^4+2(r+1)(3r^3+9r^2-23r+3)z^3
\nonumber\\
&&\hspace{0.2cm}+4(r^3-21r^2+3r-7)z^2-8(r^2+6r-11)z-64\Bigr]
\nonumber\\
&&\hspace{0.2cm}\times\frac{40z^2\ln\left(1+2\frac{1-z}{z(r+1)}\right)}{27\left[(r+1)z-2\right]^2\left[(r+1)z+2\right]^3}
\nonumber\\
&&\hspace{-0.3cm}+\Bigl[\,(r-1)(r+1)^3z^4+2(3r-1)(r^2\hspace{-0.05cm}-\hspace{-0.05cm}10r+5)z^3+4(3r^2\hspace{-0.05cm}-\hspace{-0.05cm}20r+9)z^2
\nonumber\\
&&\hspace{0.2cm}+8(5r-7)z+64\Bigr]\frac{80z(1-z)}{27\left[z(r-1)+2\right]\left[z^2(r+1)^2-4\right]^2}
\,,
\label{Fgq10}
\end{eqnarray}

\begin{eqnarray}
\mathcal{F}_{gq}^{[27_s]}(z,r)&=&\Bigl[\,r(r+1)z^2-(r-1)z-6\Bigr]\frac{16z^3(r+1)^2\ln(z)}{\left[(r+1)z-2\right]^2\left[(r+1)z+2\right]^3}
\nonumber\\
&&\hspace{-0.3cm}+\Bigl[\,(r+1)^4z^4-2(r+1)^2(3r-1)z^3-4(r+1)(r+5)z^2
\nonumber\\
&&\hspace{0.2cm}+8(r+1)z-32\Bigr]\frac{2z^2(r+1)\ln\left(1+2\frac{1-z}{z(r+1)}\right)}{\left[(r+1)z-2\right]^2\left[(r+1)z+2\right]^3}
\nonumber\\
&&\hspace{-0.3cm}+\Bigl[\,(r+1)z^2+2(2r+1)z+8\Bigr]\frac{4z(1-z)(r+1)}{\left[z(r+1)+2\right]\left[z^2(r+1)^2-4\right]}
\,,
\label{Fgq27}
\end{eqnarray}

\begin{eqnarray}
\mathcal{F}_{q\overline{q}}^{[1_s]}(z,r)&=&-\Bigl[\,18r^2(r^2+1)z^4+(81r^3\hspace{-0.05cm}-\hspace{-0.05cm}35r^2+27r\hspace{-0.05cm}-\hspace{-0.05cm}1)z^3+4(38r^2\hspace{-0.05cm}-\hspace{-0.05cm}13r+3)z^2
\nonumber\\
&&\hspace{0.6cm}+4(31r-5)z+36\Bigr]\frac{8z\left[(9r+1)z+8\right]\ln\left(1+2\frac{1-z}{z(r+1)}\right)}{729(1-z)^2(rz+1)^3}
\nonumber\\
&&\hspace{-0.0cm}+\Bigl[\,18r^2(r^2\hspace{-0.05cm}-\hspace{-0.05cm}1)z^4+(81r^3+37r^2\hspace{-0.05cm}-\hspace{-0.05cm}27r+1)z^3+4(29r^2+14r-3)z^2
\nonumber\\
&&\hspace{0.6cm}+2(35r+11)z+16\Bigr]\frac{16\left[(9r+1)z+8\right]}{729(r+1)(1-z)\left[(r-1)z+2\right](rz+1)^2}\,,
\label{Fqq1}
\end{eqnarray}

\begin{eqnarray}
\mathcal{F}_{q\overline{q}}^{[8_s]}(z,r)&=&-\Bigl[\,9r^2(9r+4)(r^2+1)z^5+(477r^4+18r^3+130r^2+36r-13)z^4
\nonumber\\
&&\hspace{0.6cm}+(1143r^3-44r^2+125r+52)z^3+2(668r^2-53r-1)z^2
\nonumber\\
&&\hspace{0.6cm}+2(391r-14)z+180\Bigr]\frac{32z\ln\left(1+2\frac{1-z}{z(r+1)}\right)}{729(1-z)^2(rz+1)^3}
\nonumber\\
&&\hspace{-0.0cm}+\Bigl[\,9r^2(r^2-1)(9r+4)z^5+(477r^4+342r^3-104r^2-36r+13)z^4
\nonumber\\
&&\hspace{0.6cm}+(981r^3+640r^2-73r-52)z^3+2(479r^2+253r+14)z^2
\nonumber\\
&&\hspace{0.6cm}+40(11r+3)z+80\Bigr]\frac{64}{729(r+1)(1-z)\left[(r-1)z+2\right](rz+1)^2}\,,
\label{Fqq8s}
\end{eqnarray}

\begin{eqnarray}
\mathcal{F}_{q\overline{q}}^{[8_a]}(z,r)&=&\Bigl[\,2r^3(9r^4+35r^3+41r^2-19r-2)z^6
\nonumber\\
&&\hspace{0.2cm}+(88r^6+264r^5+309r^4-146r^3-26r^2-2r+1)z^5
\nonumber\\
&&\hspace{0.2cm}+2r(104r^4+213r^3+271r^2-65r-3)z^4
\nonumber\\
&&\hspace{0.2cm}+(349r^4+386r^3+492r^2\hspace{-0.05cm}-\hspace{-0.05cm}58r\hspace{-0.05cm}-\hspace{-0.05cm}1)z^3+4(99r^3+52r^2+63r\hspace{-0.05cm}-\hspace{-0.05cm}2)z^2
\nonumber\\
&&\hspace{0.2cm}+2(125r^2+26r+29)z+64r\Bigr]\frac{\ln\left(1+2\frac{1-z}{z(r+1)}\right)}{12(r-1)^2(1-z)^2(rz+1)^3}
\nonumber\\
&&\hspace{-0.3cm}-\Bigl[\,2r^2(r-1)(9r^3+35r^2+35r-3)z^5
\nonumber\\
&&\hspace{0.2cm}+(88r^5+200r^4+129r^3-191r^2+15r-1)z^4
\nonumber\\
&&\hspace{0.2cm}+2(107r^4+125r^3+140r^2-91r+3)z^3
\nonumber\\
&&\hspace{0.2cm}+(329r^3+165r^2+215r-53)z^2+2(141r^2+54r+25)z
\nonumber\\
&&\hspace{0.2cm}+48(2r+1)\Bigr]\frac{1}{6(r-1)^2(1-z)\left[(r-1)z+2\right](rz+1)^2}\,,
\label{Fqq8a}
\end{eqnarray}

\begin{eqnarray}
\mathcal{F}_{q\overline{q}}^{[10]}(z,r)&=&\Bigl[\,8r^4z^4-(r^4-22r^3+6r-1)z^3-2(r^3-13r^2-5r+1)z^2
\nonumber\\
&&\hspace{0.2cm}+16rz+8\Bigr]\frac{160z^2\ln\left(1+2\frac{1-z}{z(r+1)}\right)}{81(1-z)^2(rz+1)^3}
\nonumber\\
&&\hspace{-0.3cm}-\Bigl[\,8r^2(r-1)z^3-(r^3-23r^2+9r+1)z^2-2(r^2-12r+3)z
\nonumber\\
&&\hspace{0.2cm}-2(r-7)\Bigr]\frac{320z^2}{81(1-z)\left[(r-1)z+2\right](rz+1)^2}\,,
\label{Fqq10}
\end{eqnarray}

\begin{eqnarray}
\mathcal{F}_{q\overline{q}}^{[27_s]}(z,r)&=&-\Bigl[\,2r^2(r^2+1)z^4+(3r^2+1)(3r-1)z^3+4(4r^2-r+1)z^2
\nonumber\\
&&\hspace{0.6cm}+4(3r-1)z+4\Bigr]\frac{8z^2(r+1)\ln\left(1+2\frac{1-z}{z(r+1)}\right)}{3(1-z)^2(rz+1)^3}
\nonumber\\
&&\hspace{-0.0cm}+\Bigl[\,2r^2(r-1)z^3+(9r^2-4r+1)z^2+4(3r-1)z+6\Bigr]
\nonumber\\
&&\hspace{0.6cm}\times\frac{16z^2(r+1)}{3(1-z)\left[(r-1)z+2\right](rz+1)^2}\,.
\label{Fqq27}
\end{eqnarray}
\section{\label{app:3}Benchmark scenarios}
This Appendix contains detailed information about the scenarios
(a)-(q), (X) and (Y). The values of the {\tt msugra} parameters defining the
scenarios are listed in Tab.~\ref{SPSa}. In Tabs.~\ref{SPSb}
and~\ref{SPSc} we list the values for the squark masses as provided by
{\tt SuSpect}~\cite{Djouadi:2002ze}. Note that for our analysis the
averaged values as provided in Tab.~\ref{SPSmain} are used.
\begin{table}
\begin{center}
\footnotesize
\begin{tabular}{c|c||c|c|c|c|c}
benchmark&SPS&$m_{0}\,\left[\mbox{GeV}\right]$&$m_{1/2}\,\left[\mbox{GeV}\right]$&$A_{0}\,\left[\mbox{GeV}\right]$&$\tan(\beta)$&$\mbox{sign}(\mu)$
\\
point&scenario&&&&&
\\\hline\hline
(a)&SPS1a-point&$100$&$250$&$-100$&$10$&$1$
\\\hline
(b)&SPS1a-slope1&$80$&$200$&$-80$&$10$&$1$
\\\hline
(c)&SPS1a-slope2&$60$&$150$&$-60$&$10$&$1$
\\\hline
(d)&SPS1a-slope3&$120$&$300$&$-120$&$10$&$1$
\\\hline
(e)&SPS1a-slope4&$140$&$350$&$-140$&$10$&$1$
\\\hline
(f)&SPS1a-slope5&$160$&$400$&$-160$&$10$&$1$
\\\hline
(g)&SPS1a-slope6&$180$&$450$&$-180$&$10$&$1$
\\\hline
(h)&SPS1a-slope7&$200$&$500$&$-200$&$10$&$1$
\\\hline
(i)&SPS1b-point&$200$&$400$&$0$&$30$&$1$
\\\hline
(j)&SPS2-point&$1450$&$300$&$0$&$10$&$1$
\\\hline
(k)&SPS2-slope1&$1250$&$200$&$0$&$10$&$1$
\\\hline
(l)&SPS2-slope2&$1050$&$100$&$0$&$10$&$1$
\\\hline
(m)&SPS2-slope3&$1650$&$400$&$0$&$10$&$1$
\\\hline
(n)&SPS2-slope4&$1850$&$500$&$0$&$10$&$1$
\\\hline
(o)&SPS3-point&$90$&$400$&$0$&$10$&$1$
\\\hline
(p)&SPS4-point&$400$&$300$&$0$&$50$&$1$
\\\hline
(q)&SPS5-point&$150$&$300$&$-1000$&$5$&$1$\\\hline
(X)&--&900&550&0&10&1
\\\hline
(Y)&--&400&600&0&10&1
\end{tabular}
\caption{Initial parameters for the SPS scenarios and scenarios (X) and
  (Y) to obtain the benchmark points.}
\label{SPSa}
\end{center}
\end{table}
\begin{table}
\begin{center}
\footnotesize
\begin{tabular}{c||c|c|c|c|c|c}
benchmark&$m_{\tilde{d}_{L}}\,\left[\mbox{GeV}\right]$&$m_{\tilde{d}_{R}}\,\left[\mbox{GeV}\right]$&$m_{\tilde{u}_{L}}\,\left[\mbox{GeV}\right]$&$m_{\tilde{u}_{R}}\,\left[\mbox{GeV}\right]$&$m_{\tilde{s}_{L}}\,\left[\mbox{GeV}\right]$&$m_{\tilde{s}_{R}}\,\left[\mbox{GeV}\right]$
\\
point&&&&&&
\\\hline\hline
(a)&$567.77$&$545.62$&$562.26$&$545.89$&$567.77$&$545.62$
\\\hline
(b)&$464.72$&$446.20$&$457.94$&$445.62$&$464.72$&$446.20$
\\\hline
(c)&$360.24$&$344.94$&$351.37$&$343.26$&$360.24$&$344.94$
\\\hline
(d)&$669.23$&$643.23$&$664.59$&$644.23$&$669.23$&$643.23$
\\\hline
(e)&$769.54$&$739.52$&$765.52$&$741.20$&$769.54$&$739.52$
\\\hline
(f)&$868.90$&$834.78$&$865.36$&$837.09$&$868.90$&$834.78$
\\\hline
(g)&$967.33$&$929.03$&$964.15$&$931.94$&$967.33$&$929.03$
\\\hline
(h)&$1065.22$&$1022.71$&$1062.35$&$1026.20$&$1065.23$&$1022.71$
\\\hline
(i)&$876.84$&$843.16$&$873.27$&$845.41$&$876.84$&$843.16$
\\\hline
(j)&$1556.70$&$1552.46$&$1554.76$&$1552.07$&$1556.70$&$1552.46$
\\\hline
(k)&$1300.76$&$1300.38$&$1298.41$&$1299.46$&$1300.76$&$1300.38$
\\\hline
(l)&$1055.59$&$1058.08$&$1052.65$&$1056.68$&$1055.59$&$1058.08$
\\\hline
(m)&$1817.49$&$1808.85$&$1815.83$&$1809.03$&$1817.49$&$1808.85$
\\\hline
(n)&$2080.64$&$2067.30$&$2079.20$&$2068.06$&$2080.64$&$2067.30$
\\\hline
(o)&$858.66$&$824.01$&$855.07$&$826.36$&$858.66$&$824.01$
\\\hline
(p)&$764.91$&$743.68$&$760.88$&$744.40$&$764.91$&$743.68$
\\\hline
(q)&$678.15$&$652.40$&$673.86$&$653.45$&$678.15$&$652.40$\\\hline
(X)&1437.13&1403.17&1435.06&1406.01&1437.13&1403.17
\\\hline
(Y)&1296.51&1247.59&1294.17&1252.01&1296.51&1247.59
\end{tabular}
\caption{$\tilde{d}$, $\tilde{u}$, $\tilde{s}$ masses for the various
  scenarios. The numbers have been obtained with {\tt
SuSpect}~\cite{Djouadi:2002ze}}
\label{SPSb}
\end{center}
\end{table}
\begin{table}
\begin{center}
\footnotesize
\begin{tabular}{c||c|c|c|c|c|c}
benchmark&$m_{\tilde{c}_{L}}\,\left[\mbox{GeV}\right]$&$m_{\tilde{c}_{R}}\,\left[\mbox{GeV}\right]$&$m_{\tilde{b}_{1}}\,\left[\mbox{GeV}\right]$&$m_{\tilde{b}_{2}}\,\left[\mbox{GeV}\right]$&$m_{\tilde{t}_{1}}\,\left[\mbox{GeV}\right]$&$m_{\tilde{t}_{2}}\,\left[\mbox{GeV}\right]$
\\
point&&&&&&
\\\hline\hline
(a)&$562.26$&$545.89$&$516.91$&$546.24$&$399.73$&$586.53$
\\\hline
(b)&$457.94$&$445.62$&$421.76$&$448.10$&$319.81$&$499.59$
\\\hline
(c)&$351.37$&$343.26$&$325.09$&$348.45$&$242.07$&$413.19$
\\\hline
(d)&$664.59$&$644.23$&$610.48$&$642.77$&$479.61$&$673.26$
\\\hline
(e)&$765.52$&$741.20$&$702.91$&$738.14$&$559.00$&$759.85$
\\\hline
(f)&$865.36$&$837.09$&$794.42$&$832.56$&$637.74$&$846.32$
\\\hline
(g)&$964.15$&$931.94$&$885.04$&$926.03$&$715.69$&$932.52$
\\\hline
(h)&$1062.35$&$1026.20$&$975.15$&$1018.98$&$793.11$&$1018.75$
\\\hline
(i)&$873.27$&$845.41$&$777.67$&$826.04$&$661.71$&$839.81$
\\\hline
(j)&$1554.76$&$1552.07$&$1298.73$&$1539.74$&$970.87$&$1307.45$
\\\hline
(k)&$1298.41$&$1299.46$&$1073.69$&$1289.26$&$789.69$&$1083.38$
\\\hline
(l)&$1052.65$&$1056.68$&$860.37$&$1048.57$&$622.94$&$870.89$
\\\hline
(m)&$1815.83$&$1809.03$&$1528.82$&$1794.54$&$1157.84$&$1536.48$
\\\hline
(n)&$2079.20$&$2068.06$&$1761.18$&$2051.40$&$1347.10$&$1767.83$
\\\hline
(o)&$855.07$&$826.36$&$792.13$&$822.99$&$649.29$&$842.40$
\\\hline
(p)&$760.88$&$744.40$&$617.79$&$685.28$&$546.52$&$696.19$
\\\hline
(q)&$673.86$&$653.45$&$561.85$&$650.48$&$248.72$&$649.57$\\\hline
(X)&1435.06&1406.01&1279.28&1394.96&1035.19&1300.47
\\\hline
(Y)&1294.17&1252.01&1189.74&1242.61&9878.50&1221.42
\end{tabular}
\caption{$\tilde{c}$, $\tilde{b}$, $\tilde{t}$ masses for the various
  scenarios. The numbers have been obtained with {\tt
SuSpect}~\cite{Djouadi:2002ze}}
\label{SPSc}
\end{center}
\end{table}
\end{appendix}


\end{document}